%% file: main.tex
\newif\ifpdflatex    
\pdflatextrue           

\documentclass[fleqn,usenatbib,useAMS,twocolumn,numberedappendix]{aastex}
\usepackage{amsmath,esint}
\usepackage{url,aasmacros}
\usepackage{natbib}
\usepackage{soul, CJK}
\usepackage{multirow}
\usepackage{tablefootnote}
\usepackage{appendix}
\usepackage{bm}
\usepackage{xspace}
\usepackage{color}
\usepackage{enumitem}
\usepackage{booktabs}

\usepackage{graphicx}	
\usepackage{amsmath}	
\usepackage{amssymb}	
\usepackage{bm}		
\usepackage{url}
\usepackage{amsbsy}
\usepackage{xspace}
\usepackage{hyperref}
\usepackage{longtable}
\usepackage[colorinlistoftodos]{todonotes}
\usepackage[flushleft]{threeparttable}

\usepackage[T1]{fontenc}
\usepackage{ae,aecompl}

\def\lesssim{\mathrel{\hbox{\rlap{\hbox{\lower5pt\hbox{$\sim$}}}\hbox{$<$}}}}
\def\gtrsim{\mathrel{\hbox{\rlap{\hbox{\lower5pt\hbox{$\sim$}}}\hbox{$>$}}}}

\usepackage{upgreek}                                                  
\usepackage{xspace}                                                       
\newcommand{\um}{$\upmu$m\xspace}            
\input{journals.tex}

\shorttitle{LRN/ILRT in ZTF}
\shortauthors{Karambelkar et al.}

\begin{document}
\title{Volumetric rates of Luminous Red Novae and Intermediate Luminosity Red Transients with the Zwicky Transient Facility}

\input{authors.tex}


\begin{abstract}
    Luminous red novae (LRNe) are transients characterized by low luminosities and expansion velocities, and are associated with mergers or common envelope ejections in stellar binaries. Intermediate-luminosity red transients (ILRTs) are an observationally similar class with unknown origins, but generally believed to either be electron capture supernovae (ECSN) in super-AGB stars, or outbursts in dusty luminous blue variables (LBVs). In this paper, we present a systematic sample of 8 LRNe and 8 ILRTs detected as part of the Census of the Local Universe (CLU) experiment on the Zwicky Transient Facility (ZTF). The CLU experiment spectroscopically classifies ZTF transients associated with nearby ($<150$ Mpc) galaxies, achieving 80\% completeness for m$_{r}<20$\,mag. Using the ZTF-CLU sample, we derive the first systematic LRNe volumetric-rate of 7.8$^{+6.5}_{-3.7}\times10^{-5}$\,Mpc$^{-3}$\,yr$^{-1}$ in the luminosity range $-16\leq$\,M$_{\rm{r}}$\,$\leq -11$ mag. We find that in this luminosity range, the LRN rate scales as dN/dL $\propto L^{-2.5\pm0.3}$ -- significantly steeper than the previously derived scaling of $L^{-1.4\pm0.3}$ for lower luminosity LRNe (M$_{V}\geq-10$). The steeper power law for LRNe at high luminosities is consistent with the massive merger rates predicted by binary population synthesis models. We find that the rates of the brightest LRNe (M$_{r}\leq-13$ mag) are consistent with a significant fraction of them being progenitors of double compact objects (DCOs) that merge within a Hubble time. For ILRTs, we derive a volumetric rate of $2.6^{+1.8}_{-1.4}\times10^{-6}$\,Mpc$^{-3}$yr$^{-1}$ for M$_{\rm{r}}\leq-13.5$, that scales as dN/dL $\propto L^{-2.5\pm0.5}$. This rate is $\approx1-5\%$ of the local core-collapse supernova rate, and is consistent with theoretical ECSN rate estimates. \\ \\
\end{abstract}
\section{Introduction}
The advent of time domain surveys in the last decades has led to the discovery of ``gap transients" -- a new class of explosions that have $-16 \leq M_{V} \leq -10$ and occupy the luminosity gap between novae and supernovae (SNe) \citep{Kasliwal2011,Pastorello2019NatAs}. This class includes a diverse variety of transients such as faint core-collapse SNe \citep{Yang2021}, .Ia-like SNe \citep{Bildsten2007}, low-luminosity Iax SNe \citep{Karambelkar2021}, luminous red novae (LRNe, \citealt{Kulkarni07}), intermediate luminosity red transients (ILRTs, \citealt{Thompson09}) and outbursts in luminous blue variable (LBV) stars \citep{Smith11}. Among gap transients, there is a sub-class of hydrogen-rich explosions characterized by low expansion velocities and interaction with surrounding circumstellar material (CSM). This sub-class comprises of LRNe, ILRTs and LBV outbursts. 

LRNe are transients associated with the final stages of common envelope evolution (CEE) in a stellar binary system \citep{Ivanova13,Tylenda05a,Pastorello2019a}. The loss of angular momentum in a binary can initiate CEE that terminates with the inspiral of the binary on dynamical timescales. This can either lead to the merger of the two stars or the ejection of the CE and formation of a stable binary in a tighter orbit. Both cases are accompanied by energetic outbursts that are powered primarily by shocks or recombination in the ejected material \citep{Ivanova13,MacLeod17,Pejcha2017ApJ,Matsumoto2022}. The association of LRNe with CE-related outbursts was supported by the discovery of V1309Sco -- a Galactic LRN with archival photometric data showing a binary with a rapidly decaying orbital period in the years leading to the transient \citep{Tylenda11}. LRNe thus present an opportunity to probe the poorly understood physics of CEE, see \citet{Ivanova2013araa}. This is of particular importance because CEE is a crucial phase in the formation of double compact objects (DCOs, \citealt{VignaGomez2020}) that merge to radiate gravitaional waves, which are being detected regularly by LIGO \citep{LIGO2021a}.

LRNe generally have low expansion velocities ($<1000$\,km\,s$^{-1}$), a wide range of luminosities ($-3<M_{V}<-16$), long lasting ($\sim 100$ day) multi-peaked light curves that redden rapidly due to dust formation \citep{Kaminski2011AA,Kaminski2015AA}. Prior to 2021, only 4 Galactic and 11 extragalactic LRNe were known (\citealt{Kochanek14_mergers,Blagorodnova2021} and references therein). The extragalactic LRNe have $-9\leq M_{\rm{peak}}\leq -15$, while the Galactic LRNe in general have much lower luminosities. Progenitor primary stars have been identified for 7 LRNe so far and have revealed interesting correlations between the masses and peak luminosities \citep{Blagorodnova2021,Pastorello2019a}. Despite these advances, the volumetric rates of LRNe are largely unconstrained. The best estimate of the rate comes from \citet{Kochanek14_mergers}, who used 3 Galactic LRNe that had $-4<M_{V}<-10$ discovered over the last 30 years to determine their rate in the Milky Way. They find that very low luminosity events (M$_{r/V,\rm{peak}} \approx -3$ mag) are fairly common ($\sim 0.5$\,yr\,$^{-1}$), but the rate drops as $\approx L^{-1.4}$ with increasing luminosity. The rate of the more luminous, extragalactic LRNe has not been measured yet; and extrapolations based on the Galactic rate disagree by orders of magnitude with the expectations from population synthesis \citep{Howitt2020}. An accurate measurement of the rate and luminosity function of LRNe is needed to probe several CEE parameters (see \citealt{Howitt2020} for examples). 

ILRTs are an observationally related class of transients that are also characterized by low expansion velocities and reddening photometric evolution, but have single-peaked lightcurves and a narrower luminosity range ($-11<M_{V}<-15$) compared to LRNe \citep{Cai2021}. The origin of ILRTs still remains a mystery. They have been proposed to be electron-capture SNe in super-AGB stars \citep{Botticella09,Cai2021} or outbursts in dusty LBV stars \citep{Smith09,Andrews2021}. Both explanations are supported by the peculiar progenitors of ILRTs. The ILRTs SN\,2008S, NGC\,300\,OT, AT\,2019abn have been associated with extremely dusty, infrared-(IR) bright progenitors \citep{Botticella09,Prieto08, Jencson2019} consistent with super-AGB stars, while the proposed ILRT AT\,2019krl was associated with a blue supergiant or LBV progenitor \citep{Andrews2021}. About a dozen ILRTs have been studied extensively in the last decade (see \citealt{Cai2021} and references). Similar to LRNe, the volumetric rate has not been reliably measured. Based on two ILRTs SN\,2008S and NGC\,300\,OT, \citet{Thompson09} estimate the ILRT rate to be $\approx20\%$ of the CCSN rate. \citet{Cai2021} use a sample of 5 ILRTs reported by different surveys over the last decade and estimate a lower limit of $\approx8\%$ of the CCSN rate.

Reliable LRNe and ILRTs rate measurements have been hindered by their heterogeneous sample. The existing rate estimates have used transients reported by different surveys, and have not accounted for effects of survey completeness or selection biases \citep{Kochanek14_mergers,Cai2021}. A systematic sample of LRNe and ILRTs, preferably from a single survey is required to accurately constrain their rate. Such studies are now possible with experiments like the Census of the Local Universe (CLU, \citealp{De2020a}) on the Zwicky Transient Facility (ZTF, \citealt{Bellm2019,Graham2019,DeKany2020}). ZTF is an optical time domain survey with a 47\,sq.\,deg field-of-view that surveys the entire accessible northern sky at a cadence of $\approx2-3$ days in the \emph{g} and \emph{r} bands down to a depth of 20.5 mag. The CLU experiment aims to build a spectroscopically complete sample of transients detected by ZTF that are associated with galaxies in the CLU galaxy catalog \citep{Cook2019} to a depth of m$_{r}\approx20$ mag. The CLU experiment is ideal to detect low luminosity (M$_{r}\leq-16$) transients like LRNe and ILRTs out to large distances ($\approx 100$\,Mpc), consequently building a large sample of such rare transients. 

In this paper, we present a systematic sample of LRNe and ILRTs detected by the ZTF CLU experiment from June 1, 2018 to February 20, 2022. We utilize this sample and the actual observation history of ZTF to derive the rates of LRNe and ILRTs.  In Sec. \ref{sec:candidate_selection}, we describe the selection criteria used to construct our LRN and ILRT samples. Sec. \ref{sec:candidate_selection} also describes the photometric and spectrosocopic properties of LRNe and ILRTs identified in the CLU experiment. Sec. \ref{sec:rates} describes our methods to derive the luminosity function and volumetric rates of LRNe and ILRTs. Sec. \ref{sec:discussion} compares our results to previous measurements and theoretical predictions, and discusses the implications for progenitors of LRNe and ILRTs. We conclude with a summary of our results in Sec. \ref{sec:summary}.

\input{table_candidates_spec}

\section{Sample selection}
\subsection{Candidate filtering}
\label{sec:candidate_selection}
We focus our search on transients discovered as part of the ZTF CLU experiment \citep{De2020a}. Briefly, CLU aims to build a spectroscopically complete sample of transients associated with galaxies in the local universe ($<$200 Mpc). The CLU experiment uses alerts generated from all three (public, collaboration and Caltech) components of the ZTF survey (see \citealt{De2020a} for details). CLU uses the platform \texttt{skyportal} to save and coordinate followup of sources \citep{skyportal2019}. During ZTF Phase I (2018 June 01 -- 2020 October 30), the experiment was limited to all transients that 1) were within $100 \arcsec$ of known galaxies in the CLU galaxy catalog \citep{Cook2019}, 2) had more than two detections in ZTF-\emph{g} or \emph{r} filters, and 3) were brighter than $m_{r} = 20$ mag. Starting 2020 October 30 (ZTF Phase II), the filtering criteria were updated to select transients that were 1) within 100 kpc of CLU galaxies closer than 140 Mpc, 2) brighter than m$_{r} = 20.5$ mag, and 3) less luminous than M$_{r} = -17$ mag. A total of 3442 transients were saved by the CLU experiment since start of the ZTF survey. The experiment achieved a spectroscopic completeness of 88.5\% in ZTF Phase I and 79\% in ZTF phase II for sources brighter than 20 mag.  

Candidate LRNe and ILRTs were selected from their real-time ZTF alert photometry using the following selection criteria --
\begin{itemize}
    \item the transient must pass the CLU selection criteria, 
    \item the transient must be less luminous than M$_{r}=-16$ mag,
    \item the transient must have two ZTF alerts (i.e. two $>5\sigma$ detections) brighter than 20 mag in either \emph{g} or \emph{r} bands.  
\end{itemize}
The sample presented in this paper is restricted to events saved before 2022 February 20.

523 transients satisfied the selection criteria listed above. 109 of these were found to lie on top of faint underlying galaxies that are not present in the CLU catalog. These transients were excluded as they are likely supernovae in distant host galaxies. 55 additional candidates were excluded as they were flagged as image processing artifacts on visual inspection of the difference images. 73 candidates were eliminated as they showed small, long-term ($>1$ year) variations in brightness without significant colour changes, and were coincident with stars in the Milky Way or nearby galaxies  (M31 or M33). 

To further filter our candidates, we used follow-up spectroscopic observations (either from our followup campaigns or on TNS). In addition to the CLU experiment, some of our spectra were collected as part of the ZTF Bright Transient Survey (BTS, \citealt{Fremling2020,Perley2020}). The BTS classifications are already public, while all CLU classifications will be presented in separate papers focusing on different sub-samples (e.g. \citealt{De2020a}, Tzanidakis et al. in prep., Sit et al. submitted). Of the CLU transients that pass our criteria, seven sources were spectroscopically classified as active galactic nucleii (AGN), 65 as hydrogen-poor SNe (Ia, Iax, Ca-rich, Ib and Ic SNe), 11 as classical novae, 128 as Type II SNe -- based on broad ($v\geq5000$ km s$^{-1}$) H emission lines. 2 additional candidates were studied low-luminosity Type II SNe \citep{Reguitti2021,Yang2021} showing low velocity hydrogen lines. We were now left with 73 sources that did not match any of the above categories. 

We ran forced point-spread function (PSF) photometry at the location of these 73 sources on all ZTF difference images \citep{Masci2019}. This provides more accurate photometry than the real-time ZTF alerts, and also enables the recovery of sub-threshold ($>3\sigma$) detections. From this set, 14 candidates were ruled out as their forced photometry revealed them to have absolute magnitudes brighter than $-16$ mag. 18 additional sources were ruled out because their forced photometry lightcurves revealed slow, small amplitude long term ($>500$ day) variations. These slowly varying sources are likely foreground variable stars, but some could also be long duration giant outbursts in LBV stars such as $\eta-$Car or UGC~2773-OT \citep{Humphreys99,Smith2016}. One additional source at redshift z=0.002 showed fast erratic variations in ZTF data. Finally, five sources had M$_{r}\geq -10$ and fast fading lightcurves ($<10$ d), suggesting that they are classical novae. Of the remaining, one source showed a declining lightcurve with $g-r<0$ mag, similar to late-time SN lightcurves, unlike LRNe or ILRTs. The remaining 34 sources are promising LRNe and ILRT candidates, and are listed in Table \ref{tab:candidates}. 

\subsection{Classification}
The shortlist of 34 transients in Table \ref{tab:candidates} consists of LRNe, ILRTs, LBV outbursts and possibly some supernovae observed at late phases. Here, we discuss the classifications of these transients using their ZTF lightcurves, spectroscopic data and archival photometry. 

LRNe typically show multi-peaked lightcurves \citep{Pejcha2017ApJ,Matsumoto2022} which allows to photometrically differentiate them from ILRTs \citep{Cai2021}. Table \ref{tab:candidates} indicates the transients that show multiple peaks in their ZTF lightcurves.

Searching for previous outbursts in archival data can help distinguish LBV eruptions from LRNe/ILRTs, as LBV outbursts can be recurring. While LRNe are also known to show precursor emission in the years leading up to the merger, the precursors usually have much lower luminosities than the actual merger. The detection of historic outbursts comparable in brightness to the latest outburst is thus suggestive of an LBV eruption. We checked for historical activity at the locations of these sources using data from the ATLAS (\citealt{Tonry2018,Smith2020atlas}, depth $\approx$ 19.5 mag, dating back to $\sim$2014) and PTF (\citealt{Law09,Rau09}, depth $\approx$ 21.5 mag, dating back to $\sim$2009) surveys. All 34 sources have ATLAS data while 15 have PTF data. Table \ref{tab:candidates} indicates which transients were detected in the archival ATLAS or PTF data. The full forced photometry lightcurves will be available online (see Sec. \ref{sec:data_availability})

Spectroscopic data is available for 19 of the 34 transients. 11 of these sources only have medium resolution optical spectra, while the remaining 8 also have medium resolution NIR spectra. The optical spectra were taken with the Low-Resolution Imaging Spectrograph (LRIS, \citealt{Oke95}, R$\approx$750) on the Keck I Telescope, the Double Beam Spectrograph (DBSP, \citealt{Oke82}, R$\approx$1000) on the 200-inch Hale telescope (P200) on Mount Palomar, the Alhambra Faint Object Spectrograph and Camera (ALFOSC,R$\approx$360) on the 2.56m Nordic Optical Telescope (NOT), the Kast Double Spectrograph on the Shane 3-m telescope at the Lick Observatory (R$\approx1000$) and the Spectrograph for Rapid Acquisition of Transients (SPRAT, R$\approx$350) on the Liverpool Telescope. For ZTF\,21acpkzcc, we obtained a high resolution (R$\approx$15000) spectrum with the South African Large Telescope High Resolution Spectrograph (HRS, \citealt{Bramall2010}). The NIR spectra were taken with the Near-Infrared Echelle Spectrograph (NIRES, \citealt{Wilson2004}, R$\approx$2700) on the Keck II telescope and the Triplespec spectrograph (\citealt{Herter2008}, R$\approx$2500) on P200. The log of the spectroscopic observations is listed in Table \ref{tab:speclog}. In addition to these, we also obtained several low-resolution (R$\approx$100) spectra of these transients with the Spectral Energy Distribution Machine (SEDM; \citealt{Blagorodnova2018, Rigault2019, Kim2022}) spectrograph on the 60-inch telescope at Palomar Observatory. These spectra are not discussed here as their resolutions are too low for a useful analysis. All our spectra (including the low-resolution ones) will be made available online (Sec. \ref{sec:data_availability}).

LRNe, ILRTs and LBV outbursts can have similar spectroscopic properties especially at early times. However, some LBV eruptions have expansion velocities exceeding $\approx2000$ km s$^{-1}$, while LRNe and ILRTs have lower expansion velocities (typically $\sim1000$ km s$^{-1}$). The FWHM velocities of the H$\alpha$ emission line in the transients with spectroscopic coverage are listed in Table \ref{tab:candidates}. For transients with multiple spectra, the maximum velocity is listed. Early time spectra of both LRNe and ILRTs are characterized by narrow H$\alpha$ ($v_{\rm{FWHM}}\leq 1000$ km s$^{-1}$) emission. [\ion{Ca}{2}] emission was thought to be a defining feature of ILRTs, however it was recently discovered that some LRNe can show [\ion{Ca}{2}] emission \citep{Cai2019}, while some ILRTs do not show [\ion{Ca}{2}] at certain phases \citep{Andrews2021}. Late-time spectroscopic observations can help confidently distinguish between LRNe and ILRTs. In LRNe, H$\alpha$ emission grows narrower and weaker with time since peak, and eventually reemerges. Late-time spectra of LRNe also exhibit molecular absorption features. On the contrary, ILRTs show persistent narrow H emission throughout their evolution. Their late time spectra do not show molecular features, but have strong H$\alpha$ and \ion{Ca}{2} emission lines. 

\input{table_spectral_log}

Provided that not all of our sources have late-time spectroscopic coverage, we use the available spectra (from literature wherever necessary) together with the lightcurves to create a classification scheme for these transients. We classify the 34 transients into eight categories -- Possible LBV outbursts, LRN-gold, LRN-silver, LRN-bronze, ILRT-gold, ILRT-silver, ILRT-bronze, and Ambiguous. We describe these categories in detail below. Briefly, for LRNe and ILRTs, the gold sample comprises transients that we can confidently classify in either category, the silver sample comprises transients that have photometry and spectra indicative to their class, but lack smoking gun signatures associated with the class, and the bronze category comprises transients that have photometric evolution similar to their class, but do not have any spectroscopic data. The ``Ambiguous" category consists of 7 sources that have poor quality data, i.e. no spectra \emph{and} low-quality sparse photometric coverage owing to bad weather or solar conjunction of the transient, rendering us unable to place them in any of the other categories. Table \ref{tab:candidates} lists the classification of these sources based on this scheme. We describe the sources in each category below.

\subsubsection{Possible LBV eruptions}
This category includes sources that show broad ($v_{\rm{FWHM}}\geq2000$\,km\,s$^{-1}$) H$\alpha$ emission in their spectra, \emph{or} have historic outbursts in archival data. A total of 6 transients in our shortlist satisfy these criteria. Two of these -- ZTF\,21aagydmn and ZTF\,20abwilhb show broad H$\alpha$ emission with FWHM velocities of 2600 and 3300 km s$^{-1}$ and are likely LBV outbursts. The ATLAS and PTF data classifies four additional sources as potential LBVs -- ZTF\,19acpmbvd, ZTF\,21aantupk, ZTF\,21aclyyfm and ZTF\,21aaitlhy. We describe these sources in the Appendix \ref{app:lbvs}. 


\begin{figure*}[hbt]
    \centering
    \includegraphics[width=\textwidth]{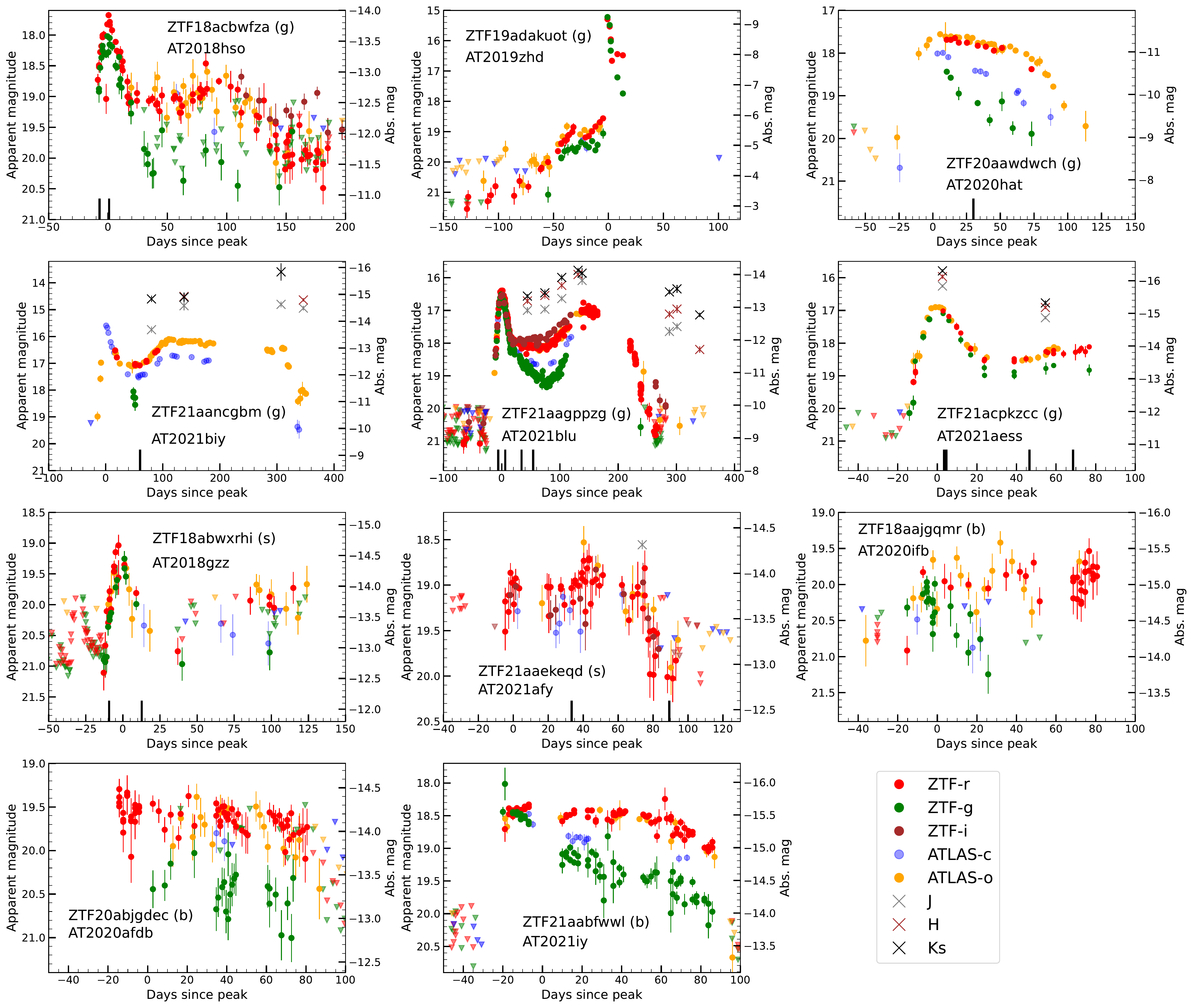}
    \caption{Forced photometry lightcurves of the 11 LRNe in our sample. The transients in gold, silver and bronze samples are marked in parantheses with g, s and b respectively. The ZTF-g, ZTF-r, ZTF-i, ATLAS-c and ATLAS-o band datapoints are plotted as green, red, brown, blue and orange points respectively. The J, H and Ks band magnitudes are plotted as gray, brown and black crosses respectively. Downward pointing triangles indicate 5-$\sigma$ upper limits. The days are in observer frame. The lightcurves have been corrected for extinction using the values listed in Table \ref{tab:phot_properties}. Solid black vertical lines indicate epochs at which the  spectra were obtained.}
    \label{fig:lrn_lcs}
\end{figure*}

\begin{figure*}[hbt]
    \centering
    \includegraphics[width=\textwidth]{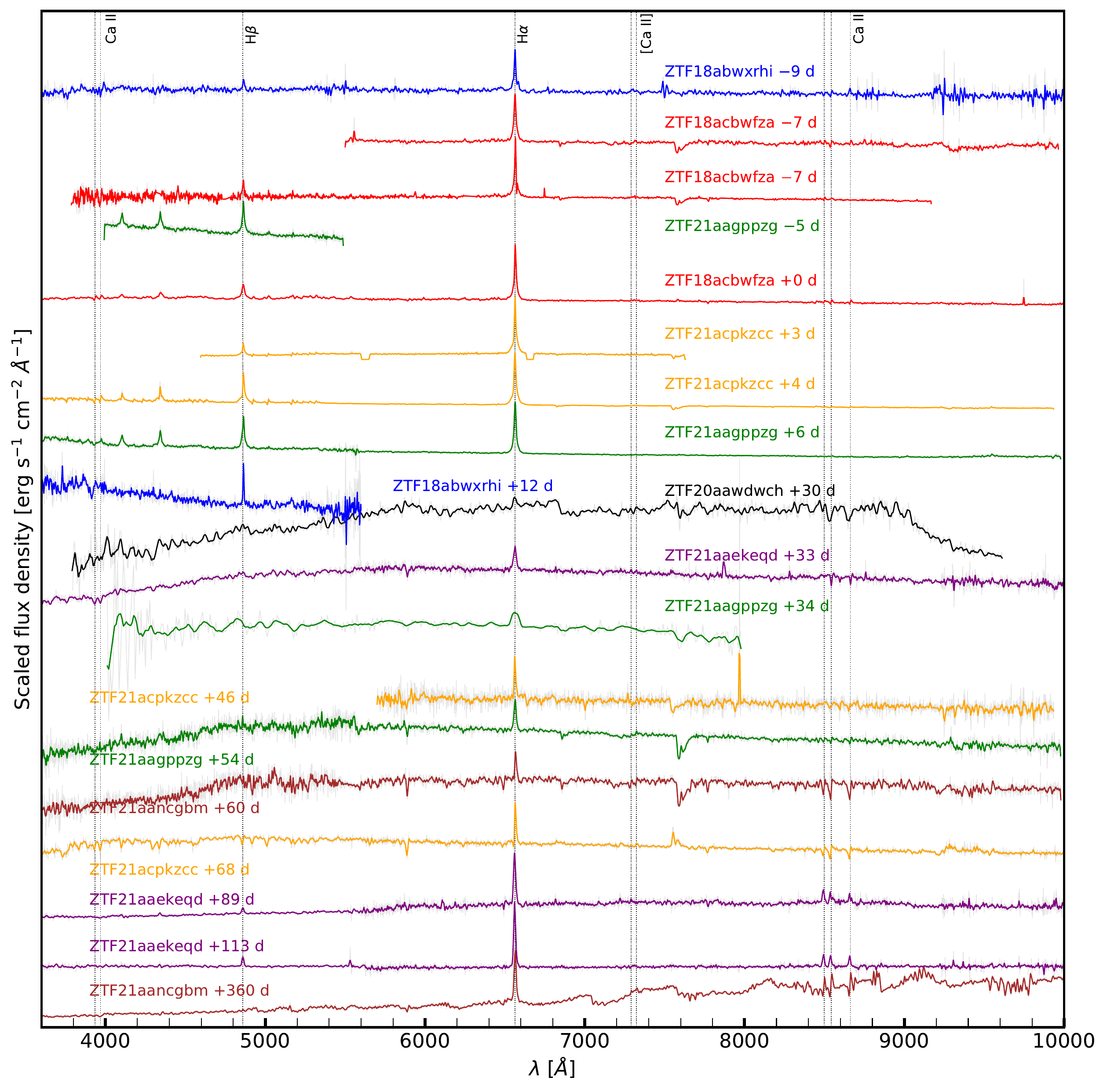}
    \caption{Optical spectra for LRNe presented in this paper.}
    \label{fig:lrn_spectra_optical}
\end{figure*}

\begin{figure*}[hbt]
    \centering
    \includegraphics[width=\textwidth]{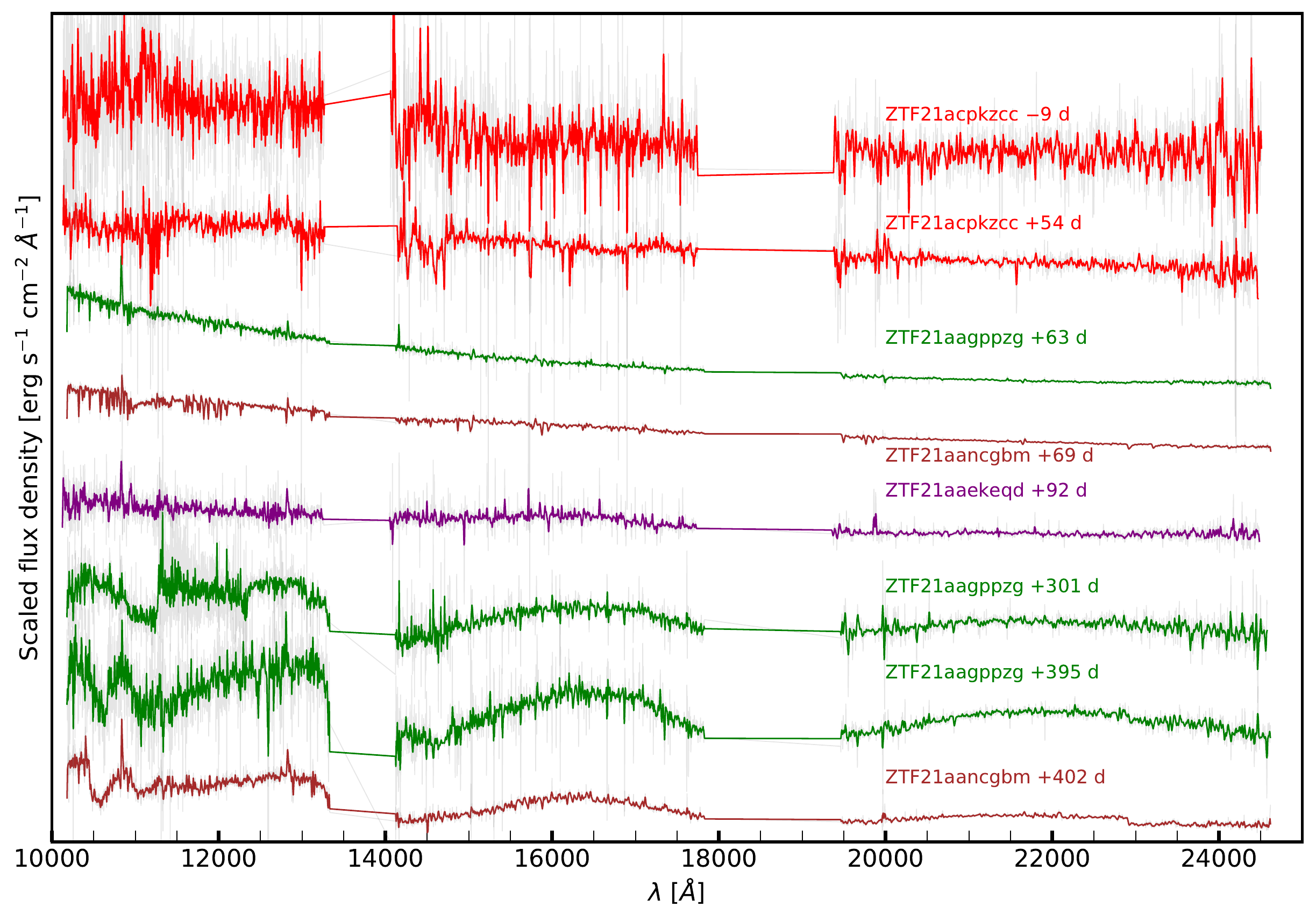}
    \caption{NIR spectra for LRNe presented in this paper.}
    \label{fig:lrn_spectra_nir}
\end{figure*}

\subsubsection{LRN-gold}
\label{sec:lrn_gold}
This category includes six spectroscopically confirmed LRNe detected in ZTF data in the last three years. Three of these have been studied in detail previously -- ZTF\,18acbwfza (AT\,2018hso, \citealt{Cai2019}), ZTF\,19adakuot (AT\,2019zhd, \citealt{Pastorello2021a}) and ZTF\,20aawdwch (AT\,2020hat, \citealt{Pastorello2021b}). Three other transients -- ZTF\,21aancgbm (AT\,2021biy, \citealt{Cai2022}), ZTF\,21aagppzg (AT\,2021blu, \citealt{Pastorello2022}) and ZTF\,21acpkzcc (AT\,2021aess, \citealt{DavisTNS2021}) were identified as possible LRNe in 2021. We initiated optical and NIR photometric and spectroscopic followup campaign for these objects, which confirmed their nature as LRNe. 

Our lightcurves of these transients are shown in Fig. \ref{fig:lrn_lcs}.  For all our calculations, we use host galaxy redshifts corrected for local velocity flows from the NASA Extragalactic Database (NED), and assume H$_{0} = 73$\,km\,s$^{-1}$\,Mpc$^{-1}$. We use Milky Way line-of-sight extinction values from \citet{Schlafly12}. We calculate the host galaxy extinctions for LRNe using \ion{Na}{1}\,D equivalent widths in their early time spectra (see discussion of individual objects below). We calculate the peak magnitudes and epoch of maximum brightness by fitting polynomials to the near-peak lightcurve. We calculate the pseudo-bolometric lightcurves by fitting blackbodies to the available optical data. The derived photometric properties and the adopted distances and extinctions for the LRNe in our sample are listed in Table \ref{tab:phot_properties}.

Our spectra of these transients are shown in Fig. \ref{fig:lrn_spectra_optical} and \ref{fig:lrn_spectra_nir}. We analyze the optical spectra and derive the hydrogen Balmer line velocities by fitting lorentzian profiles (which are better fits than gaussian profiles). We also measure the Balmer decrement ($\beta$) as the ratio of H$\alpha$ to H$\beta$ line fluxes in the spectra where both lines are detected. The spectroscopic properties are listed in Table \ref{tab:spec_properties}. We now briefly describe individual members of our sample.

\textbf{1.\,ZTF\,18acbwfza (AT\,2018hso)} is located in the starforming galaxy NGC\,3729. Our earliest spectrum (7 days before peak) shows \ion{Na}{1}\,D absorption with an equivalent width of $1.5\pm 0.4$\,\AA. Assuming the source of this absorption is dust in the interstellar medium and the correlation from \citet{Turatto2003} we derive $E(B-V)_{\rm{host}} = 0.23\pm0.06$ mag. This is consistent with the value used by \citet{Cai2019}. The lightcurve is characteristic of LRNe, with an early blue peak lasting for $\approx 20$ days, followed by a prolonged plateau where the transient evolves rapidly to redder colors. We note that the transient shows signs of evolving back to bluer colors at late times, when the \emph{r}-band lightcurve is declining but the \emph{g}-band lightcurve has plateaued. However, we caution that the \emph{g}-band detections at this phase from ZTF have low significance (3$\sigma$). Our spectra sample only the first blue peak, and show strong Balmer emission lines with v$_{\rm{FWHM}} \approx 500$\,km\,s$^{-1}$. These spectra also show narrow \ion{Ca}{2} NIR triplet lines with a P-cygni profile. The absorption of the P-cygni profile extends to a maximum velocity v$_{m}\approx400-500$\,km\,s$^{-1}$. \citet{Cai2019} report weak [\ion{Ca}{2}] in this transient. We find marginal evidence for [\ion{Ca}{2}] emission in only one of our spectra, taken 1 day post peak. Late-time spectra from \citet{Cai2019} show strong molecular absorption features.

\textbf{2.\,ZTF\,19adakuot (AT\,2019zhd)} is located in M31. Following \citet{Pastorello2021a}, we assume that the total extinction is dominated by the Milky Way. The transient has the lowest luminosity in our sample (peaking at M$_{r}\approx -9.5$ mag). We do not have any spectroscopic data for this transient, but late-time spectra presented in \citet{Pastorello2021a} show molecular absorption features. 

\begin{figure}[hbt]
    \centering
    \includegraphics[width=0.5\textwidth]{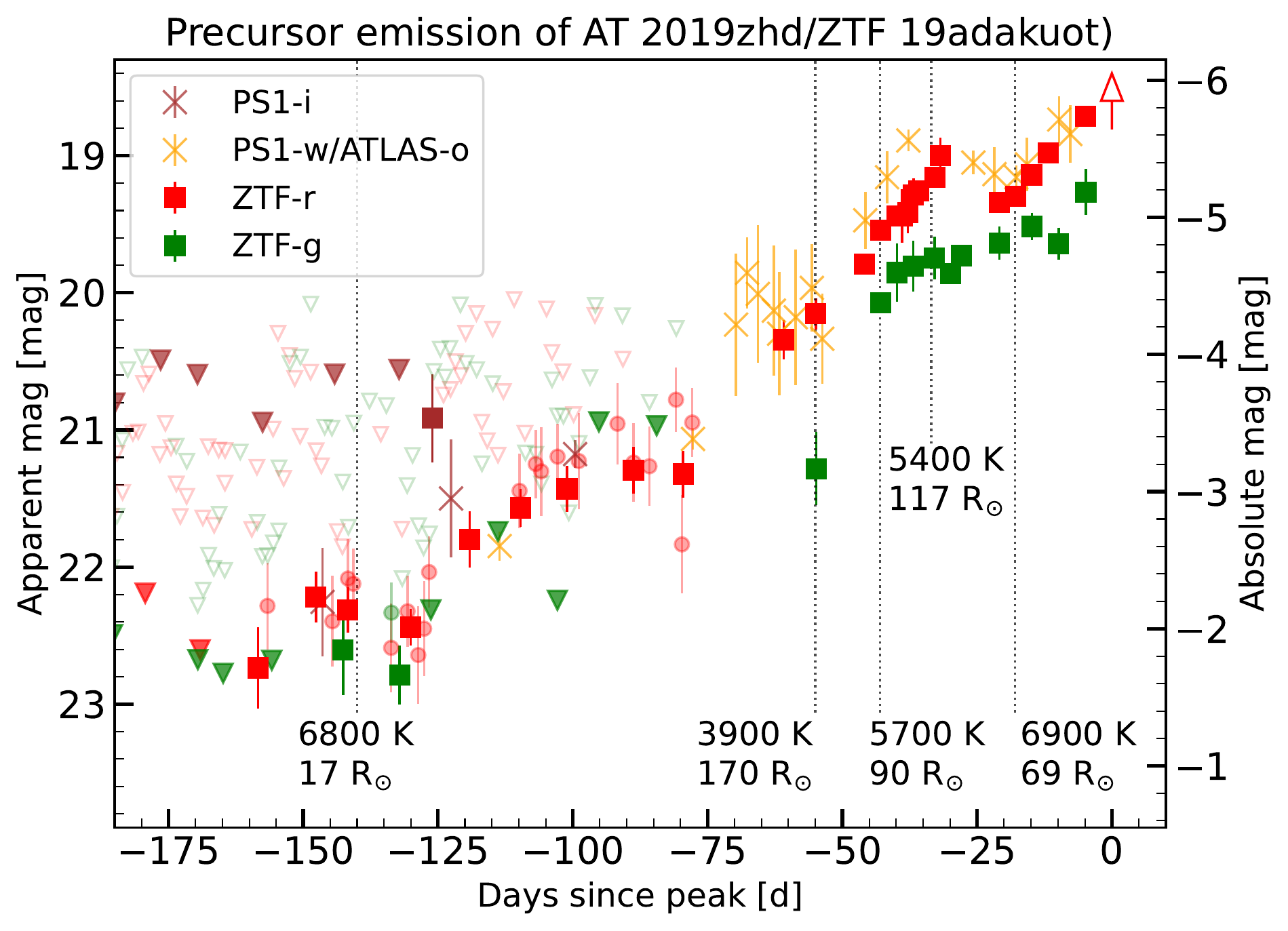}
    \caption{Pre-explosion \emph{g}, \emph{r} and \emph{i}-band detections for ZTF\,19adakuot. For phases $<-75$ days, we bin the ZTF flux measurements taken on the same day to increase the sensitivity. These points are plotted as faint background circles. We further bin these measurements in bins of 10 days to increase their significance. The binned measurements are showed in solid squares. For phases $>-75$ days, the squares represent ZTF forced photometry from single visits. The triangles denote 5-$\sigma$ upper limits. The crosses denote measurements reported previously in \citet{Pastorello2021a}. At the epochs marked by the dotted vertical lines, we indicate the effective temperatures and photospheric radii to demonstrate the dramatic variations exhibited by the precursor (see text). The initial pre-explosion behaviour is consistent with a photosphere L2-mass loss forming an optically thick photosphere detectable by ZTF at $\approx-150$ days, which expands and cools up to $\approx-70$ days. Further collisions in the L2 stream, or accretion onto the companion star can cause sudden increases in temperature over the next 70 days, until the transient brightens rapidly transitioning to the main explosion (marked by the red arrow).}
    \label{fig:ztf19adakuot_precursor}
\end{figure}

As noted in \citet{Pastorello2021a}, this transient shows precursor activity for a few months leading up to the explosion. In these months, the field of ZTF\,19adakuot was imaged with the ZTF camera several times per night. We stack the ZTF forced photometry from images taken on the same night, and recover several additional detections than the ones reported in \citet{Pastorello2021a}, which are plotted in Fig. \ref{fig:ztf19adakuot_precursor}. We further bin the photometry in bins of ten days to increase the significance of the detections. The ZTF detections date back to 160 days before the \emph{r}-band peak, when the transient was first detected at m$_{r}\approx22.7\pm0.3$ mag. The binned ZTF lightcurve shows that the transient went through a gradual, bumpy brightening phase for $\approx 160$ days before the main explosion. Three bumps can be identified in the \emph{r-}band lightcurve, each with a duration of $\approx 50$ days. During each of these bumps, the transient brightens by an increasing amount ($\approx$ 0.5, 1 and 2.5 mag respectively) before plateauing or declining by a modest amount ($\approx0.5$ mag) at the end of each bump. After the third bump, the transient brightens gradually by $\approx 1$ mag for 25 days before brightening rapidly by 4 mag as it transitions into the main explosion. This bumpy rise is reminiscent of the pre-outburst evolution of M31-LRN-2015 \citep{Blagorodnova2020}, V838\,Mon \citep{Munari2002AA,Tylenda05a} and AT\,2020hat \citep{Pastorello2019b}. 

At the end of the first bump (MJD 58759), fitting a single blackbody to the \emph{g} and \emph{r}-band detections suggests a source with T$_{\rm{eff}}=6600^{+2400}_{-1600}$ K, a radius R\,$=17^{+14}_{-7}$\,R$_{\odot}$ and a luminosity of 5.5$^{+1.3}_{-0.7}\times10^{2}$\,L$_{\odot}$. This photosphere is most likely formed by mass outflowing from the outer Lagrangian (L2) point (for example \citealt{Pejcha2017ApJ} or \citealt{Macleod2022}). On the second bump, the transient is only detected in \emph{r} band, with the most constraining \emph{g} band upper limit giving $g-r>0.8$ mag, suggesting the optically thick photosphere has cooled. This likely continues up to the beginning of the third bump on MJD 58837, where the temperature has cooled down to T$=3900^{+700}_{-700}$ K and the photosphere has expanded to R$=170^{+140}_{-70}$\,R$_{\odot}$ and L$=6.1^{+3.9}_{-1.7}\times10^{3}$\,L$_{\odot}$. This would imply a photospheric expansion speed of $\approx10-30$\,km\,s$^{-1}$. After this, the temperature increases quickly in the next 12 days on the third bump to 5700$^{+300}_{-300}$\,K with L\,$=7.5^{+0.3}_{-0.3}\times10^{3}$\,L$_{\odot}$, and reduced R\,$=90^{+12}_{-12}$\,R$_{\odot}$ on MJD 58849, suggesting the temperature increase was accompanied by a drop in opacity. This is followed by a surprisingly sharp decline in the redder (\emph{r}, \emph{o}) bands but a slow brightening in the \emph{g}-band. During this decline, the temperature increases to 6900$^{+1100}_{-1000}$\,K but the radius drops to 69$^{+27}_{-16}$\,R$_{\odot}$ with increased luminosity of 9.5$^{+0.8}_{-0.6}\times10^{3}$\,L$_{\odot}$ on MJD 58871. This is followed by an increase in luminosity and photospheric radius accompanied by a slight decrease in temperature, to L$=1.7^{+0.2}_{-0.1}\times10^{4}$\,L$_{\odot}$ on MJD 58887. In the next five days, the transient brightens rapidly to its peak luminosity of $\approx5.2\times10^{5}$\,L$_{\odot}$.

A full analysis of this complex precursor photometric evolution is outside the scope of this paper. A possible explanation for the abrupt temperature increases in this final phase is shocks due to collisions within the L2-stream, as suggested in \citet{Pejcha2017ApJ}. The total energy radiated during this $\approx40$ day duration is $\approx10^{44}$ erg. Assuming an L2 velocity of 30\,km\,s$^{-1}$, the mass required in the L2 stream is $\geq10^{-2}$M$_{\odot}$. 

\textbf{3.\,ZTF20aawdwch (AT\,2020hat)} is located in NGC\,5068. We follow \citet{Pastorello2021b} and assume that the extinction is dominated by the Milky Way. The ZTF lightcurve samples the post-peak decline, starting at m$_{r}$=17.90$\pm$0.06 mag. Our sole spectrum was taken at a phase of 30 days since peak, which shows a reddened continuum, extremely weak H$\alpha$ emission, some broad but weak molecular absorption features and \ion{Ca}{2} NIR triplet absorption lines -- characteristic of LRNe on the red plateau phase. Molecular absorption features are clearly detected in late-time spectra presented in \citet{Pastorello2021b}. 

\textbf{4.\,ZTF\,21aancgbm (AT\,2021biy, \citealt{Smith2021tns, Cai2022})} is located in the galaxy NGC\,4631. We do not have spectra at early phases, so we adopt the host extinction derived by \citet{Cai2022} using \ion{Na}{1}\,D absorption. The lightcurve is characteristic of LRNe, with an initial blue peak lasting for $\approx$ 50 days, followed by a rebrightening and a prolonged red plateau of $\approx$ 350 days. The field of ZTF\,21aancgbm is not part of the regular 2-day cadence ZTF survey, so the ZTF lightcurve samples only a small portion of the lightcurve. We combine the ZTF data with ATLAS data, which shows that the lightcurve lasts for more than 400 days, and is the longest duration LRN in our sample. 

We obtained \emph{JHKs} NIR imaging with P200/WIRC on several epochs on the plateau, which shows that the transient was significantly brighter in the NIR bands (by $>1$ mag) than in the optical (see Fig. \ref{fig:lrn_lcs}). The transient shows a late-time bump once it falls off the plateau, likely the result of collision with CSM as noted in \citet{Cai2022}. Our spectroscopic coverage comprises two optical spectra at +60 and +360 days since peak, and two NIR spectra at +69 and +402 days since peak. The +60 day spectrum shows a reddened continuum with weak, unresolved H$\alpha$ emission (v$_{\rm{FWHM}}\approx350$\,km\,s$^{-1}$) and \ion{Ca}{2} NIR triplet with P-cygni line profiles (with v$_{m}\approx400$\,km\,s$^{-1}$). We also detect weak [\ion{Ca}{2}] lines with P-cygni line profiles. The +69 d NIR spectrum shows several metallic absorption lines, as well as the CO absorption bandhead. The very late-time optical spectrum at +360 days shows a highly reddened continuum with several strong molecular absorption features of TiO and VO, which confirms the classification as a LRN. This spectrum also shows strong, narrow H$\alpha$ emission (with increased v$_{\rm{FWHM}}\approx500$\,km\,s$^{-1}$) and \ion{Ca}{2} NIR lines with P-cygni profiles (v$_{m}\approx600$\,km\,s$^{-1}$). 

The lightcurve shape and long duration of ZTF\,21aancgbm make it an interesting member of the LRN family. The 200 day long plateau resembles that seen in SNe IIP, and it is tempting to posit that it is powered by hydrogen recombination. \citet{Matsumoto2022} provide scaling relations between the plateau luminosity (L$_{\rm{pl}}$), duration (t$_{\rm{pl}}$) and ejected mass (M$_{\rm{ej}}$) for a recombination-powered LRN plateau, given by 
\begin{equation}
    \label{eq:mej}
    M_{\rm{ej}}\approx1.6 \,\rm{M}_{\odot}\times f_{\rm{ad, 0.3}}^{-1}\times(\frac{t_{\rm{pl}}}{100\rm{d}})\times (\frac{L_{\rm{pl}}}{10^{39}\rm{erg/s}})
\end{equation}
with f$_{\rm{ad},0.3}\approx1$ is a dimensionless factor quantifying inefficiencies in radiating the recombination energy. 
We define the plateau as the duration between the point where the transient begins rebrightening after the first blue peak to the time when it falls off the plateau to the same brightness. We calculate an average plateau luminosity L$_{\rm{pl}}\approx 2.3\times10^{40}$\,erg\,s$^{-1}$ and a plateau duration t$_{\rm{pl}}\approx274$ days. This suggests a recombining ejecta mass of $\approx$\,100\,M$_{\odot}$ for this LRN, implying an extremely massive progenitor. This seems unlikely, given that the correlation between peak luminosity and progenitor mass for LRNe \citep{Blagorodnova2021} predicts the primary progenitor mass in the range $\sim10-50$\,M$_{\odot}$. Additionally, from archival \emph{HST} imaging \citet{Cai2019} suggest a progenitor mass $\sim17-24$\,M$_{\odot}$. It is therefore very likely that the plateau luminosity is too high to be explained by hydrogen recombination alone. A plausible source of additional energy is shock interaction between the merger ejecta and pre-existing material around the binary that was ejected during the CE phase. Following \citet{Matsumoto2022}, assuming that only the timescale but not the luminosity of the transient is set by hydrogen recombination, the required recombining ejected mass is 
\begin{equation}
    \label{eq:tpl}
    t_{\rm{pl}}\approx 140\, \rho_{i,-11}^{-1/3} (\frac{M_{\rm{ej}}}{M_{\odot}})^{1/3} (\frac{v_{E}}{300\,\rm{km/s}})^{-1}\,\,\rm{days}
\end{equation}
where v$_{\rm{E}}$ is the mean ejecta velocity and $\rho_{i,-11} = \rho_{i}/10^{-11}$g\,cm$^{-3}\approx 1$ where $\rho_{i}$ is the characteristic density at which the recombination completes. We adopt v$_{E}\approx430$\,km\,s$^{-1}$ from \citet{Cai2022} which gives M$_{\rm{ej}}\approx22$\,M$_{\odot}$. This value is larger than that derived in \citet{Cai2022} due to the larger plateau duration that we have assumed. Assuming that this ejected mass runs into pre-existing mass (M$_{\rm{pre}}<<M_{\rm{ej}}$) with velocity v$_{\rm{pre}}<<\rm{v_{\rm{E}}}$, the generated shock luminosity can be written as \citep{Matsumoto2022}
\begin{equation}
    \label{eq:shock_intern}
    L_{\rm{pl}}^{\rm{sh}} \approx 7\times10^{39} \rho_{i,-11}^{-1/3} (\frac{M_{\rm{pre}}}{0.1M_{\rm{ej}}})  (\frac{v_{E}}{300\,\rm{km/s}})^{3}  (\frac{M_{\rm{ej}}}{M_{\odot}})^{2/3}\,\,\rm{erg/s}
\end{equation}
The observed plateau luminosity can be explained by shock interactions from Eq. \ref{eq:tpl} and \ref{eq:shock_intern} if M$_{\rm{pre}}\approx 0.3\,\rm{M}_{\odot}$. Given the very late-time bump in the lightcurve of this LRN, it is not unreasonable to expect significant CSM around the binary, providing a plausible explanation for its plateau.

\textbf{5.\,ZTF\,21aagppzg (AT\,2021blu, \citealt{Smith2021tns2, Pastorello2022})} is located on the outskirts of the galaxy UGC\,5829. It was first classified as an LBV outburst \citep{Uno2021tns}, however, our late-time NIR spectra show strong molecular absorption features, indicating that this is a LRN. We adopt the extinction to be dominated by the Milky Way, as our early optical spectra do not show any strong \ion{Na}{1D} absorption lines. 

The lightcurve is characteristic of LRNe and shows two pronounced peaks -- an initial blue peak for $\approx$ 50 days, followed by a second red peak lasting for $\approx 200$ days. Unlike ZTF\,21aancgbm, the second peak does not show a plateau but has a smooth rise and decline. The P200/WIRC NIR photometry shows that ZTF\,21aagppzg was also significantly brighter in NIR than in the optical. Particularly, at late times when the transient has faded below ZTF detection limits in the \emph{r}-band ($>21$ mag), it continues to be detected at $\approx$ 17 mag in the K band (see Fig. \ref{fig:lrn_lcs}). We stack the ZTF forced photometry prior to the outburst in 5-day bins and recover some archival detections. The first detection in \emph{r} band is $\approx 1040$ days prior to peak with m$_{r}=22.3\pm0.2$. The transient was detected again at four epochs in the \emph{r} and \emph{g} bands between 400 and 300 days before peak, at roughly constant magnitudes of m$_{r}=22.4\pm0.3$ and m$_{g} = 22.5\pm0.3$. Finally, it is detected again 95 days before peak at m$_{r}=21.8\pm0.3$, brightens to m$_{r}=20.9\pm0.2$ 65 days before peak and fades back to m$_{r}=21.9\pm0.3$ mag at 27 days before peak. It is also detected in the \emph{i} band during this time period, with m$_{i}=21.0\pm0.3$ mag. 

Our spectroscopic coverage comprises four optical spectra at $-5$, +6, +34 and +54 days since peak and three NIR spectra at +63, +301 and +395 days since peak. The two early-time optical spectra are characterized by a hot blue continuum with strong Balmer emission lines. The H$\alpha$ emission line has $v_{\rm{FWHM}} \approx 500-600$\,km\,s$^{-1}$. The H$\beta$ line profile appears to be a superposition of a broad component with a narrow component, however our spectral resolution is not sufficient to distinguish between the two. These spectra do not show the \ion{Ca}{2} NIR triplet lines. The third optical spectrum has lower resolution and a lower signal-to-noise ratio (S/N). It shows a reddened continuum, with H$\alpha$ being the only prominent feature. The final optical spectrum was obtained two months post peak, and shows a reddened continuum. The strength and velocity of the Balmer emission decreased, the H$\alpha$ line developed a double-peaked profile and the H$\beta$ is not detected. This spectrum also shows weak \ion{Ca}{2} NIR triplet absorption lines, along with a forest of metallic absorption lines. The NIR spectrum at +63 days shows hydrogen emission lines with some metallic absorption lines. Our two final late-time NIR spectra show strong molecular absorption features of TiO, VO and H$_{2}$O. 

Similar to ZTF\,21aancgbm, we can use scaling relations to determine the ejecta masses. We estimate the median luminosity on the red plateau  L$_{\rm{pl}}\approx1.6\times10^{40}$\,erg\,s$^{-1}$ and a plateau duration t$_{\rm{pl}}\approx180$ days. Using v$_{E}\approx500$\,km\,s$^{-1}$ and Eq. \ref{eq:tpl} gives the recombining ejecta mass M$_{\rm{ej}}\approx 10$\,M$_{\odot}$, although the estimate is very sensitive to the assumed ejecta velocity. Similar to ZTF\,21aancgbm, the recombination alone is not sufficient to explain the luminosity of the plateau, and Eq. \ref{eq:shock_intern} implies interaction with pre-existing mass M$_{\rm{pre}}\approx0.2$\,M$_{\odot}$. \citet{Pastorello2022} discuss archival \emph{HST} and ground-based imaging of this transient, and the properties of the putative progenitor. We note that in addition to the data presented in their paper, this transient also has archival \emph{Spitzer Space Telescope}/IRAC imaging from 2007-12-27. We detect a marginal source at the location of the transient with m$_{3.6} = 20.48 \pm 0.16$ mag and m$_{4.5} = 19.79\pm0.21$ mag (Vega system). We leave a more detailed analysis of this LRN and its progenitor properties to a future study.


\textbf{6. ZTF\,21acpkzcc (AT\,2021aess)} is the most luminous LRN in our sample, with peak M$_{\rm{r}} = -15.12 \pm 0.15$. We do not identify strong \ion{Na}{1}\,D in our early time spectra, although this is possibly due to low S/N. We assume that the extinction is dominated by the Milky Way, but caution that extinction due to the host galaxy could increase the peak luminosity estimates for this transient. Similar to the other LRNe described above, the lightcurve shows two distinct peaks -- a blue initial peak lasting $\approx40$ days followed by a reddened plateau. However, our photometric coverage stops at $\approx50$ days into the plateau as the transient went into solar conjunction. Similar to the previous two LRNe, ZTF\,21acpkzcc is also brighter in the NIR (by $>1$ mag) than in the optical bands. Stacking the archival ZTF lightcurve in three day bins shows some archival detections $\approx70$ days before peak at m$_{g}=21\pm0.3$ and m$_{r}=21.1\pm0.3$. Our spectroscopic coverage comprises three optical spectra at +4, +46 and +68 days since peak and one NIR spectrum at +54 days since peak. The early spectra shows narrow H$\alpha$ with $v_{\rm{FWHM}} \approx 500$\,km\,s$^{-1}$, which grows weaker and narrower with time (FWHM $< 350$\,km\,s$^{-1}$ at 46 days). The +68d spectrum also shows a weak P-cygni profile for H$\alpha$. The early time spectra also show a forest of \ion{Fe}{2} lines and the \ion{Ca}{2} NIR triplet lines with P-cygni profiles. The initial blue lightcurve peak with a red plateau together with weakening H$\alpha$ with time and NIR excess suggest that ZTF21\,acpkzcc is a LRN. The high luminosity of ZTF\,21acpkzcc is similar to the LRN AT\,2017jfs \citep{Pastorello2019b}, suggesting a massive binary origin.

We note that two additional LRNe were reported in the last three years -- AT\,2018bwo \citep{Blagorodnova2021} and AT\,2020kog \citep{Pastorello2021b}. AT\,2020kog was missed by ZTF because it landed in the chip-gaps of the ZTF detector. AT\,2018bwo was discovered on 2018-05-22, when ZTF was in the reference-building phase. The first ZTF visit to the field of AT\,208bwo was two months later on 2018-07-14. No alerts were generated as this field did not have a ZTF reference image. We ran post-facto image subtractions and measured forced PSF photometry and recover two detections of AT\,2018bwo in the \emph{r} band at $19.76\pm0.08$ and 20.15$\pm0.12$ mag on 2018-07-21 and 2018-07-24 respectively. 



\subsubsection{LRN-silver}
This category includes two sources that do not show all hallmarks of a LRN, but resemble LRNe in several aspects. Their lightcurves are plotted in Fig. \ref{fig:lrn_lcs} and their optical and near-infrared spectra are shown in Fig. \ref{fig:lrn_spectra_optical} and \ref{fig:lrn_spectra_nir} respectively. 

\textbf{1. ZTF\,18abwxrhi (AT\,2018gzz)} has peak M$_{\rm{r}}=-14.82\pm0.17$ and shows two pronounced lightcurve features : a blue first peak followed by a longer lasting red peak/plateau. Our photometric data is sparse for the red peak/plateau as the transient did not brighten significantly above the ZTF sensitivity limits. However, the lightcurve evolution inferred from the available data is strikingly similar to the LRNe in the gold sample. Additionally, a spectrum taken during the first blue peak shows narrow, marginally resolved Balmer emission lines with $v_{\rm{FWHM}} \approx 300$ km s$^{-1}$ ruling out a supernova origin, narrow, unresolved \ion{Ca}{2} NIR triplet lines with P-Cygni profiles and narrow \ion{Ca}{2} H \& K absorption lines. We do not detect \ion{Na}{1}\,D in this spectrum, and hence assume negligible host extinction. There is no spectroscopic coverage at late times. Given these similarities with LRNe, ZTF\,18abwxrhi is most likely a LRN, and we include it in the LRN-silver category. 

\textbf{2. ZTF\,21aaekeqd (AT\,2021afy)} is located on the outskirts of the galaxy UGC\,10043. While \citet{Pastorello2022} classify this transient as a LRN, their late-time spectra do not show obvious molecular absorption features (e.g. as those seen in ZTF\,21aancgbm). We also do not identify similar molecular features in any of our spectra. As noted in  \citet{Pastorello2022}, the lightcurve differs from other LRNe with similar luminosities. For these reasons, we include ZTF\,21aaekeqd in our LRN-silver sample. It is possible that obvious molecular features appear at later times, where spectroscopic coverage does not exist.

The transient has a 100-day long lightcurve with two low contrast peaks and is detected only in \emph{r} and \emph{i} bands by ZTF. In our earliest spectrum, we detect \ion{Na}{1}\,D absorption with equivalent width $1.8\pm0.6$\,\AA\,, consistent with \citet{Pastorello2022}. This corresponds to E(B--V)$_{\rm{host}}\approx 0.28\pm0.10$ mag, assuming that the interstellar medium is alone responsible for the \ion{Na}{1}\,D absorption. However, we caution that this is unlikely, given the remote location of the transient in the host. It is possible that the sodium absorption originates in circumstellar dust around the progenitor of the explosion, in which case the extinction estimate will be incorrect \citep{Poznanski2012}. Accounting for the Galactic component, we adopt a total E(B--V)$=0.33\pm0.10$ mag. With this, the transient reaches M$_{\rm{r,peak}}=-13.95\pm0.16$, and is one of the brighter members of our sample. We obtained an epoch of \emph{J-}band imaging at +60 days since peak with $J-r \approx 1$\,mag, similar to other LRNe. 

\begin{figure}
    \centering
    \includegraphics[width=0.5\textwidth]{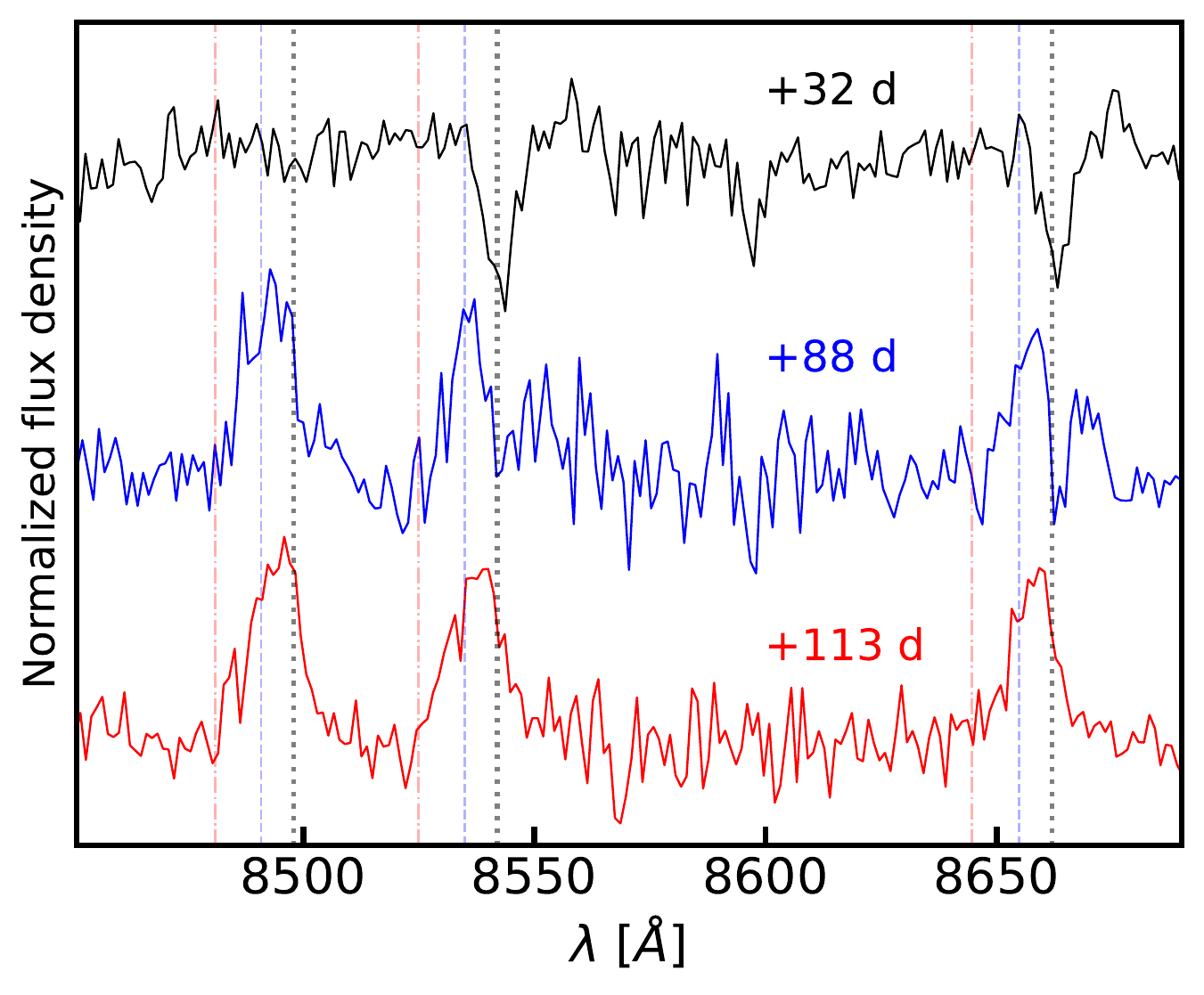}
    \caption{Evolution of \ion{Ca}{2} line profiles of ZTF\,21aaekeqd. In the earliest spectrum (+32 d, black), the lines are seen purely in absorption, and likely originate from a shell of slow-moving CSM around the binary. In the second spectrum, the profile has transitioned to emission superposed with a narrow absorption component. This is likely because the ejecta have crashed into the CSM, and swept up only part of the CSM. The shocks in the ejecta produce the emission, while the absorption comes from the unshocked, dense CSM shell outside the photosphere. In the final optical spectrum, the line profiles are pure emission, suggesting that all of the CSM has been swept up by the ejecta.}
    \label{fig:ZTF21aaekeqd_calcium}
\end{figure}

Our spectroscopic coverage comprises 3 optical spectra at phases of +33 d, +89 d, +114 d and one NIR spectrum at +92 d since first peak. The +33 day spectrum shows a reddened continuum with narrow H$\alpha$ ($v_{\rm{FWHM}} \approx 650$\,km\,s$^{-1}$) and \ion{Ca}{2} NIR triplet in absorption. Of the three \ion{Ca}{2} triplet lines, we detect only the 8542 and 8662\,\AA\, absorption lines. This is likely because the transition probability for the 8498\,\AA\, line is $\approx10$ times smaller than for the other two. Notably, the absorption lines are unresolved, with v$_{\rm{FWHM}}\leq200$\,km\,s$^{-1}$ (instrumental resolution in this wavelength range is v$_{\rm{inst}} \approx250$\,km\,s$^{-1}$, measured from sky emission lines). This velocity is much lower than the H$\alpha$ photospheric velocity at this phase ($\approx650$\,km\,s$^{-1}$). These absorption lines likely originate in a dense, slow-moving shell of circumstellar medium outside the photosphere, that was likely ejected prior to the explosion possibly during the CE phase. The +89 day spectrum shows a more reddened continuum with slower, marginally resolved H$\alpha$ ($v_{\rm{FWHM}} \approx 400$\,km\,s$^{-1}$) and narrow H$\beta$ ($v_{\rm{FWHM}} \approx 300$\,km\,s$^{-1}$). The \ion{Ca}{2} NIR triplet lines are now seen as a superposition of an emission component with v\,$\approx500$\,km\,s$^{-1}$, and a central narrow absorption component (see Fig. \ref{fig:ZTF21aaekeqd_calcium}). A possible explanation for this is that the ejecta have crashed into the CSM at this epoch, and the resulting shock has swept up only part of the CSM.  A photosphere forms at the shock interface from where the emission lines originate (\ion{Ca}{2} predominantly excited by collisions). The narrow, weak absorption component originates from the unshocked CSM shell lying outside the photosphere. The +80 day NIR spectrum shows similar narrow hydrogen emission lines. The spectrum also shows a weak broad feature in the \emph{H} band, which is similar to the H$_{2}$O absorption seen in ZTF\,21aagppzg. The final optical spectrum at +113 d shows a very weak continuum, but extraordinarily strong H$\alpha$ (v$_{\rm{FWHM}} \approx 300$\,km\,s$^{-1}$) and strong, unresolved H$\beta$ (v$_{\rm{FWHM}} < 400$\,km\,s$^{-1}$).  The \ion{Ca}{2} NIR triplet is seen purely in emission. This suggests that the shock has swept up all the CSM. The Balmer decrement in this spectrum is $\beta\approx7$, supporting that the emission originates in an interaction-dominated region. The absorption to emission transition of the \ion{Ca}{2} line profiles suggests that there is a dense, slow-moving shell of CSM around the progenitor. None of the other LRNe in our sample show such a transition. The late-time spectrum with very strong emission lines is unlike the late-time optical spectra of other LRNe in our sample (e.g. ZTF\,21aancgbm at 360 days), that show strong molecular absorption bands.  

\subsubsection{LRN-bronze}
This category includes three sources - ZTF\,18aajqkmr, ZTF\,20abjgdec and ZTF\,21aabfwwl that show multiple peaks in their lightcurves but do not have spectroscopic coverage (see Fig. \ref{fig:lrn_lcs}). These transients have peak absolute magnitudes of $-15.36$, $-14.37$ and $-15.72$, respectively. ZTF\,21aabfwwl was initially classified as a Type\,II SN \citep{Hinkle2021tns}, but the spectrum available on TNS is noisy and no H lines are clearly visible. We list these three transients as candidate LRNe.

\input{table_phot_properties}

\subsubsection{ILRT-gold}
\begin{figure*}[hbt]
    \centering
    \includegraphics[width=\textwidth]{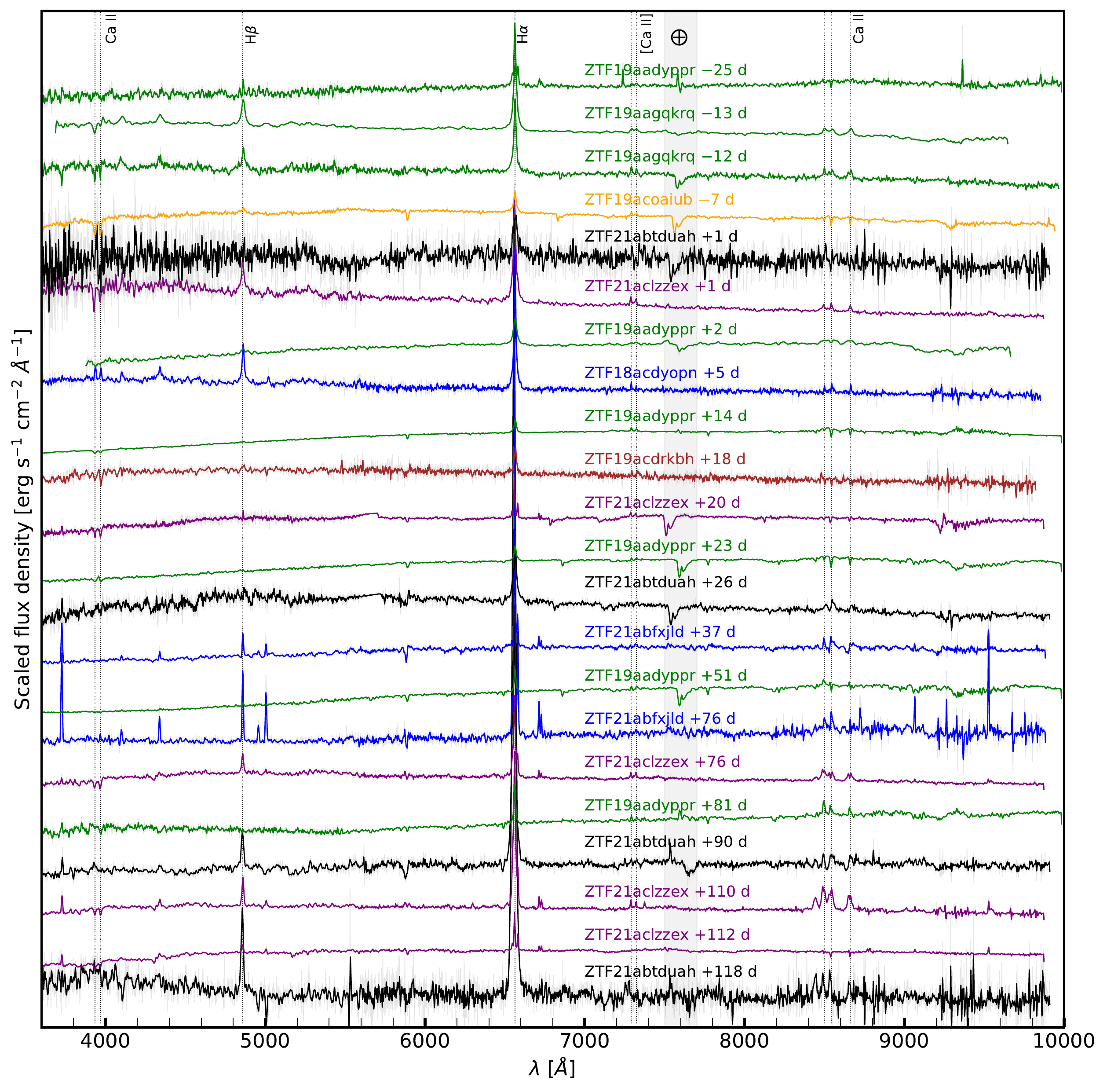}
    \caption{Optical spectra for ILRTs presented in this paper.}
    \label{fig:ilrt_spectra_optical}
\end{figure*}

\begin{figure*}[hbt]
    \centering
    \includegraphics[width=\textwidth]{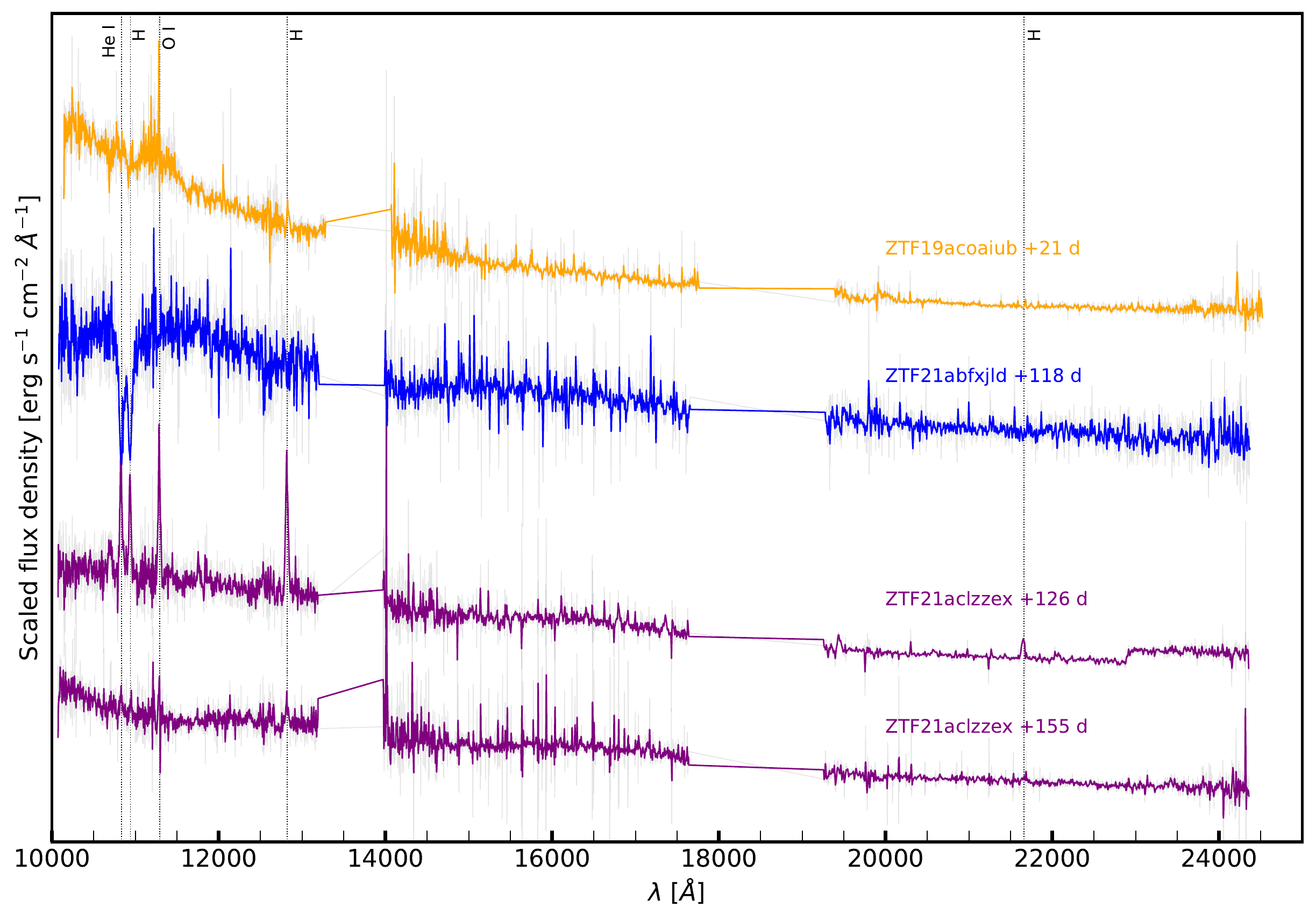}
    \caption{NIR spectra for ILRTs presented in this paper.}
    \label{fig:ilrt_spectra_nir}
\end{figure*}

This category includes six sources that we spectroscopically classify as ILRTs. Their lightcurves are shown in Fig. \ref{fig:ilrt_lcs}. We calculate the distances and extinction due to the Milky Way as was done for the LRNe (see Sec. \ref{sec:lrn_gold}). It is challenging to estimate the host galaxy extinctions because ILRTs have extremely dusty progenitors, and often exhibit variable \ion{Na}{1}\,D originating in outflows (\citealt{Cai2021, Smith09}, also see discussion of ZTF\,19acoaiub below). Instead, we use the peak optical $g--r$ colors of ILRTs to estimate the host extinction, similar to \citet{Humphreys11, Smith09, Jencson2019}. The spectra of ILRTs at peak show F-type absorption features, suggesting T $\approx7500$\,K. \citet{Cai2021} find that the ILRTs in their sample have peak B$-$V colors in the range $0.2-0.4$ mag, consistent with this interpretation. We use the peak colors of the ILRTs in our sample to estimate the value of E(B$-$V)$_{\rm{host}}$ required to bring the colors in this range. We estimate a non-negligible extinction for two ILRTs in our sample -- E(B$-$V) $\approx0.7$ for ZTF\,19aadyppr (consistent with \citealt{Jencson2019}) and E(B$-$V) $\approx$ 0.2 for ZTF\,19aagqkrq. For the other transients, we assume the extinction to be dominated by the Galactic component. We calculate pseudo-bolometric luminosities by fitting blackbodies to the available data. Table \ref{tab:phot_properties} lists the derived photometric properties of the ILRTs. Our optical and NIR spectra for these transients are shown in Figs. \ref{fig:ilrt_spectra_optical} and \ref{fig:ilrt_spectra_nir}. Similar to the LRNe, we derive line velocities by fitting Lorentzian line profiles and list the derived values in Table \ref{tab:spec_properties}. We briefly discuss the properties of the transients here.

Two of these transients -- ZTF\,19aadyppr and ZTF\,21aclzzex have spectroscopic coverage extending to late times. Of these, ZTF\,19aadyppr is in the galaxy M\,51 and has been studied in detail by \citet{Jencson2019}, who identified its red, dusty progenitor. The object shows a single-peaked red lightcurve, and its spectra show narrow H$\alpha$ and [\ion{Ca}{2}] emission, characteristic of ILRTs. 

ZTF\,21aclzzex is the brightest ILRT in our sample, with peak M$_{r}=-15.68\pm0.16$ mag.  Our lightcurve coverage is sparse and noisy, but the available data shows indications that the transient is red, and possibly has multiple peaks. However, our spectroscopic coverage extends to 163 days since peak, and resembles the evolution of ILRTs. An early time spectrum at 9 days since peak shows a blue continuum with narrow H$\alpha$, H$\beta$, \ion{Ca}{2} and [\ion{Ca}{2}] emission. The spectrum also shows \ion{Ca}{2} H \& K absorption. The next optical spectrum obtained at 84 d since peak shows a reddened continuum with narrow H$\alpha$, \ion{Ca}{2}, [\ion{Ca}{2}] emission and \ion{Ca}{2} H \& K absorption. [\ion{O}{1}] is detected as a weak emission line. The 118 d optical spectrum shows similar features, except the [\ion{O}{1}] has become stronger. Similar [\ion{O}{1}] emission was observed in the ILRT AT\,2013la \citep{Cai2021}. The 134 d NIR spectrum shows strong, narrow hydrogen emission lines, \ion{He}{1} and \ion{O}{1} emission. Although the lightcurve is not informative enough to classify the transient, the spectroscopic evolution is strikingly similar to several ILRTs, especially with [\ion{Ca}{2}] emission seen consistently in all spectra.  For this reason, we classify it as an ILRT.

The other four transients have only early-time spectroscopic coverage. Analogous to the other ILRTs, they all show narrow H$\alpha$ and [\ion{Ca}{2}] emission in their spectra. We note that while the LRN AT\,2018hso also showed [\ion{Ca}{2}] emission in its early spectra, it also had two pronounced peaks in its lightcurve. None of these four transients show multiple peaks and are not AT\,2018hso-like LRNe. We thus classify them as ILRTs based on narrow H$\alpha$, [\ion{Ca}{2}] emission and lack of multi-peaked lightcurves. 

Of these four, ZTF\,19acoaiub (AT\,2019udc) and ZTF\,19aagqkrq (AT\,2019ahd) have the best sampled lightcurves. ZTF\,19acoaiub is located in NGC\,718 and has peak M$_{r} = -14.68\pm0.04$ mag, a relatively fast declining single-peaked lightcurve. The ZTF lightcurve also samples the pre-peak rise. 7 days before peak, the transient was extremely red with $g-r\approx0.8$ mag. As it brightens, the transient also evolves to bluer colors of $g-r\approx0.4$ mag five days before peak and $g-r\approx0.2$ mag at peak. The transient then declines at a rate of $\approx$0.03 mag\,day$^{-1}$ for the next 60 days as it evolves back to redder colors. Our only optical spectrum was obtained 7 days before the peak and shows strong H$\alpha$ emission with v$_{\rm{FWHM}} \approx 1300$\,km\,s$^{-1}$. This spectrum also shows narrow \ion{Ca}{2} and [\ion{Ca}{2}] emission features, similar to several other ILRTs. This pre-peak spectrum shows very strong \ion{Na}{1} D absorption with EW\,$\approx4.8\pm0.7$\,\AA consistent with its pre-peak red colors. A likely explanation for this behaviour is that the progenitor of ZTF\,19acoaiub was dust enshrouded. This circumstellar dust was destroyed by the explosion, causing the transient to evolve to bluer colors as it brightened. This transient demonstrates the issue with using early-time \ion{Na}{1} D features to estimate the host extinction.

ZTF\,19aagqkrq (AT\,2019ahd) has peak M$_{r} = -13.72\pm0.15$ mag and has a blue color up to a few days before peak but rapidly transitions to red colors. Out spectroscopic coverage comprises two optical spectra obtained on successive nights at peak and 1 day post-peak. Both spectra show H$\alpha$ emission with v$_{\rm{FWHM}} \approx 700$\,km\,s$^{-1}$ and strong \ion{Ca}{2} and [\ion{Ca}{2}] emission lines. 

The remaining two -- ZTF\,18acdyopn and ZTF\,19acdrkbh -- have relatively sparse lightcurve sampling. ZTF\,18acdyopn has a peak absolute magnitude M$_{r} = -14.33\pm 0.19$ mag and declines by 1.5 mag in the \emph{r} band in 80 days. Although our lightcurve does not sample the peak, the available data shows that the transient had red colors a few days post-peak. An optical spectrum taken five days since peak shows narrow hydrogen Balmer emission lines (v$_{\rm{FWHM}} \approx 300$\,km\,s$^{-1}$), [\ion{Ca}{2}] , \ion{Ca}{2} H \& K as well as \ion{Ca}{2} NIR triplet emission lines. Similarly, ZTF\,19acdrkbh has peak M$_{r}=-14.62\pm0.15$, and was detected at the threshold of ZTF sensitivity, and has $g-r\approx0.6$ mag at peak. An optical spectrum at +19 days since peak shows narrow H$\alpha$ (v$_{\rm{FWHM}}\approx600$\,km\,s$^{-1}$) and possible \ion{Na}{1}\,D absorption, similar to other ILRTs. 


\input{table_spec_properties}

\subsubsection{ILRT-silver}
This category includes two sources -- ZTF\,21abtduah and ZTF\,21abfxjld -- that show luminosities and spectral evolution broadly similar to ILRTs, but do not have [\ion{Ca}{2}] emission. Although [\ion{Ca}{2}] emission in ILRTs is a sign of interaction with the surrounding circumstellar material (CSM), its strength is sensitive to the CSM density \citep{Humphreys11}. For example, the proposed ILRT AT\,2019krl did not show \ion{Ca}{2} emission in several of its spectra \citep{Andrews2021}.

\begin{figure*}[hbt]
    \centering
    \includegraphics[width=\textwidth]{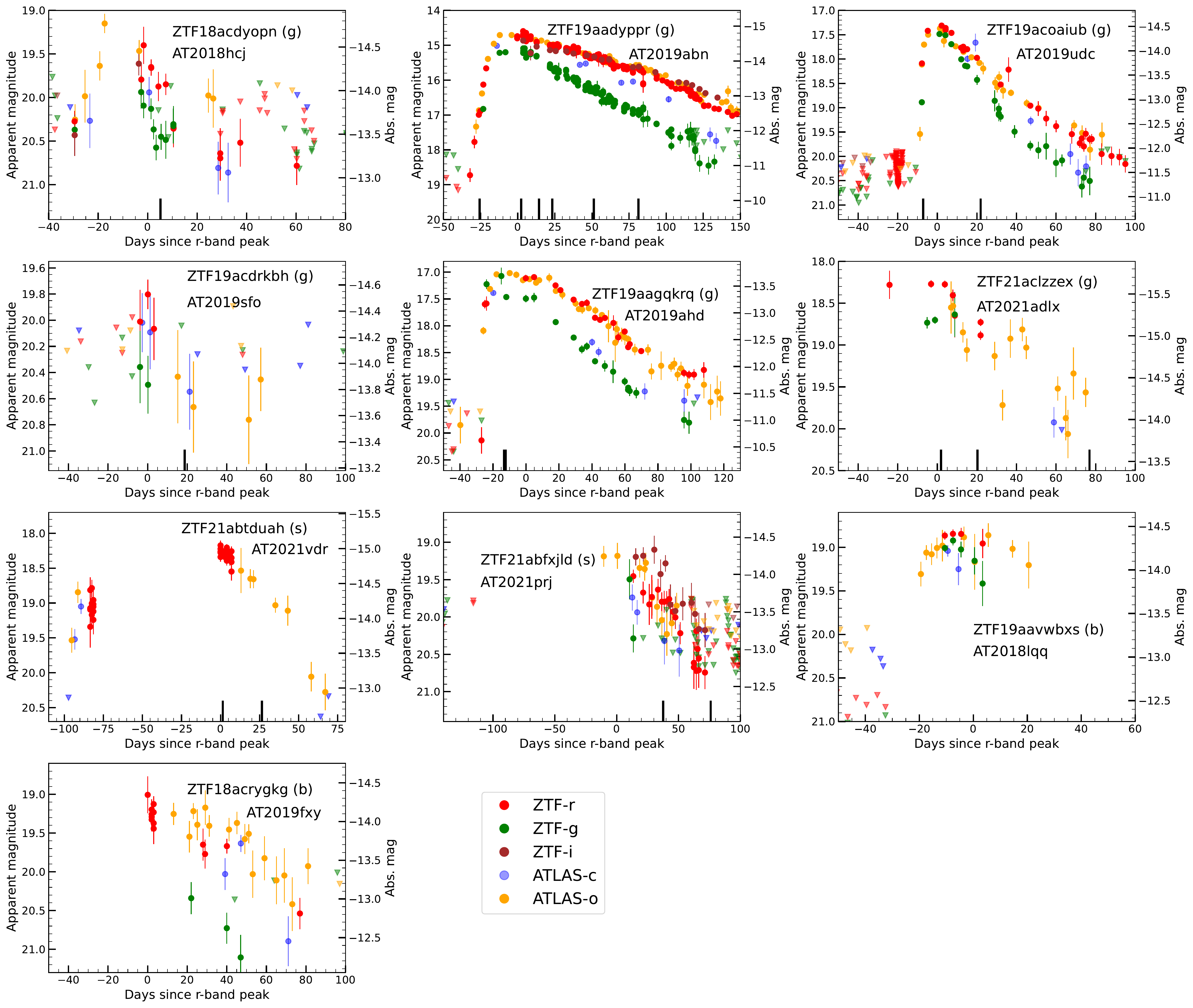}
    \caption{Forced photometry lightcurves of the 10 ILRTs in our sample. The transients in gold, silver and bronze samples are marked in parantheses with g, s and b respectively. The ZTF-g, ZTF-r, ZTF-i, ATLAS-c and ATLAS-o band datapoints are plotted in green, red, brown, blue and orange circles respectively. Downward pointing triangles indicate 5-$\sigma$ upper limits. The days are in observer frame. The lightcurves have been corrected for extinction using the values listed in Table \ref{tab:phot_properties}. Solid black vertical lines indicate epochs at which the  spectra were obtained.}
    \label{fig:ilrt_lcs}
\end{figure*}

For both transients, our lightcurve coverage does not sample the peak, and it is not clear whether they have multiple peaks. For ZTF\,21abtduah, our lightcurve samples the brightening and fading of the transient (see Fig. \ref{fig:ilrt_lcs}). We estimate a fiducial peak time by fitting a polynomial to these points. For ZTF\,21abfxjld, the ZTF lightcurve samples the fading of the lightcurve. However, the ATLAS \emph{o-}band lightcurve shows two measurements separated by 5 days that do not show significant evolution, suggesting that the transient peaked around then. We determine the peak time by fitting a polynomial to the ATLAS \emph{o-}band lightcurve. We have late-time spectroscopic coverage for both these transients. In a spectrum taken 24 days since the (fiducial) peak, ZTF\,21abtduah shows H$\alpha$ with v$_{\rm{FWHM}}\approx1200$\,km\,s$^{-1}$. A second spectrum obtained at 48 days since peak shows a double-peaked H$\alpha$ profile, suggesting interaction with slow moving CSM, or absorption in an external shell. This spectrum also shows narrow \ion{Ca}{2} NIR triplet in emission. The final spectrum taken at 140 days since peak shows strong H$\alpha$, H$\beta$, \ion{Ca}{2} NIR and \ion{O}{1} emission lines, similar to ZTF\,21aclzzex and AT\,2013la (however [\ion{Ca}{2}] lines are not seen). The H$\alpha$ line profile in this spectrum is also double-peaked. ZTF\,21abfxjld has spectra at +44d and 83 days since peak, both of which show H$\alpha$ emission with v$_{\rm{FWHM}}\leq300$\,km\,s$^{-1}$, and narrow [\ion{Ca}{2}] emission. We also have two NIR spectra taken at +26 d and +125 d since peak. The 26 day NIR spectrum shows narrow H emission lines. However, the 125 day spectrum shows strong but narrow H and He absorption in the \emph{J} band. Such features are not seen in any other ILRTs.

The low expansion velocities of these transients argue against a core-collapse SN origin for them. The lack of any molecular features at late times rules out LRNe. None of them show any significant outbursts in archival data. The nature of these transients is not completely clear. Their spectral features point towards CSM interaction. The lack of \ion{Ca}{2} emission suggests that the CSM is denser than in typical ILRTs. These transients could represent a peculiar variety of ILRTs. For these reasons, we classify these as ILRT-silver sources.


\subsubsection{ILRT-bronze}
This category includes two sources ZTF\,19aavwxbs and ZTF\,18acrygkg. ZTF\,19aavwxbs shows a single peak lasting 20 days in the ZTF and ATLAS lightcurve before it went into solar conjunction. A low resolution (R$\sim100$) spectrum of ZTF\,19aavwxbs taken with the Spectral Energy Distribution Machine (SEDM; \citealt{Blagorodnova2018, Rigault2019}) spectrograph on the 60-inch telescope at Palomar Observatory shows H$\alpha$ emission with v$_{\rm{FWHM}}\leq3000$\,km\,s$^{-1}$. No other features are discernible in the spectrum. ZTF\,18acrygkg has a lightcurve lasting for 40 days in the ZTF data. There is no spectroscopic data for ZTF\,18acrygkg. It is possible that both these transients are low-luminosity Type II SNe so we only list them as ILRT candidates. 

\subsubsection{Ambiguous sources}
Finally, we have seven sources that do not have spectroscopic coverage and the photometric data is not enough to determine a tentative classifications. Most of these sources are likely supernovae where the ZTF data samples the late-time phases. The 5-$\sigma$ alert lightcurves of these transients have durations shorter than 25 days.  

\section{Volumetric Rate}
\label{sec:rates}
To estimate the volumetric rates, we simulate the ZTF survey from 2018-06-01 to 2022-02-20 with the python package \texttt{simsurvey} \citep{Feindt_2019} using the ZTF pointing history and actual ZTF difference images to estimate the limiting magnitudes of each pointing. Using template lightcurves, we then inject LRNe and ILRTs for a range of possible rates and identify the number of transients that would be detected by the selection criteria of our experiment. Comparing simulations to the observed LRN/ILRT sample gives a first estimate of the rate. This estimate is then corrected for selection effects such as the spectroscopic completeness of the CLU experiment and the CLU galaxy catalog.  

\subsection{Template lightcurves}
LRNe exhibit a dichotomy in their lightcurves. The brighter LRNe (M$_{r} \geq -11$) are characterized by a double peaked lightcurve, while the lower luminosity events have a single peak and much shorter durations \citep{Blagorodnova2021, Pastorello2019a}. All except one of the ZTF events are brighter than $-11$ mag, and have double-peaked lightcurves lasting for $\sim 150$ days. ZTF\,19adakuot with M$_{\rm{r,peak}} \approx -9.5$ is the least luminous event in our sample and fades quicker than any of the other events. Owing to the different lightcurve shapes and durations of the brighter LRNe and the fact that most of the ZTF LRNe belong to this category, we restrict our rate analysis to events with M$_{\rm{r}}$ brighter than $-11$ mag.

\begin{figure*}[hbt]
    \centering
    \includegraphics[width=\textwidth]{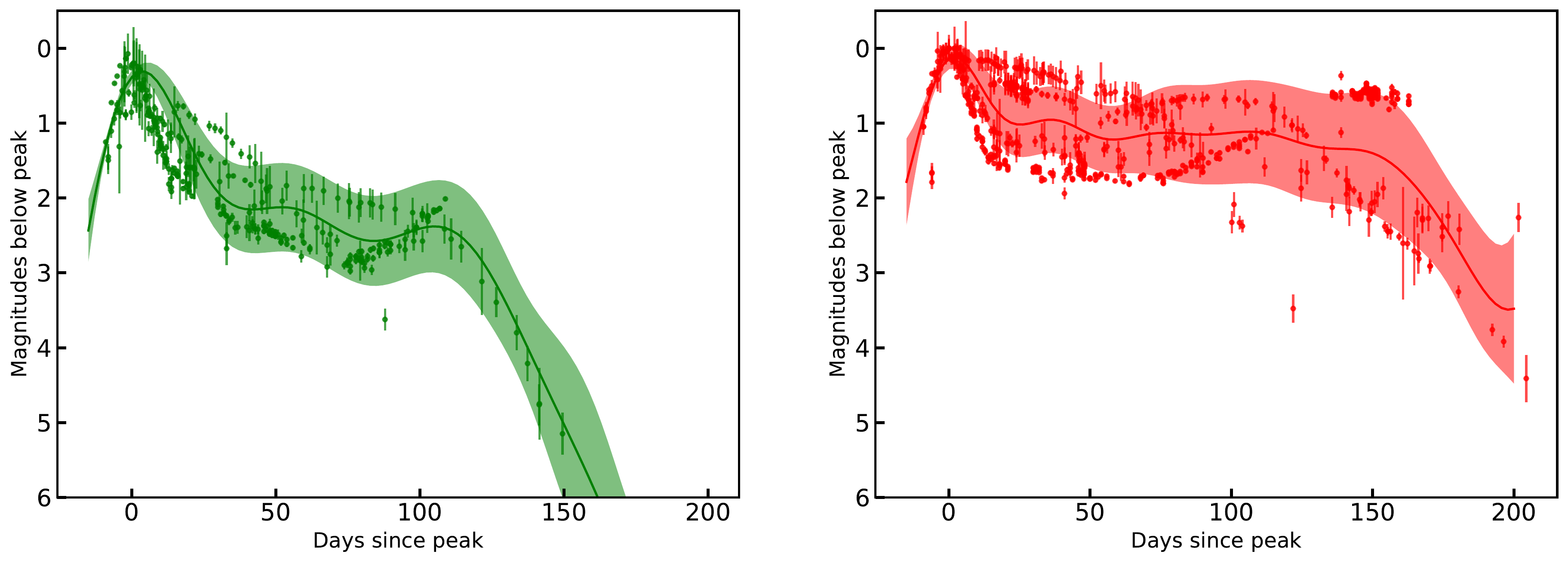}\\
    \includegraphics[width=\textwidth]{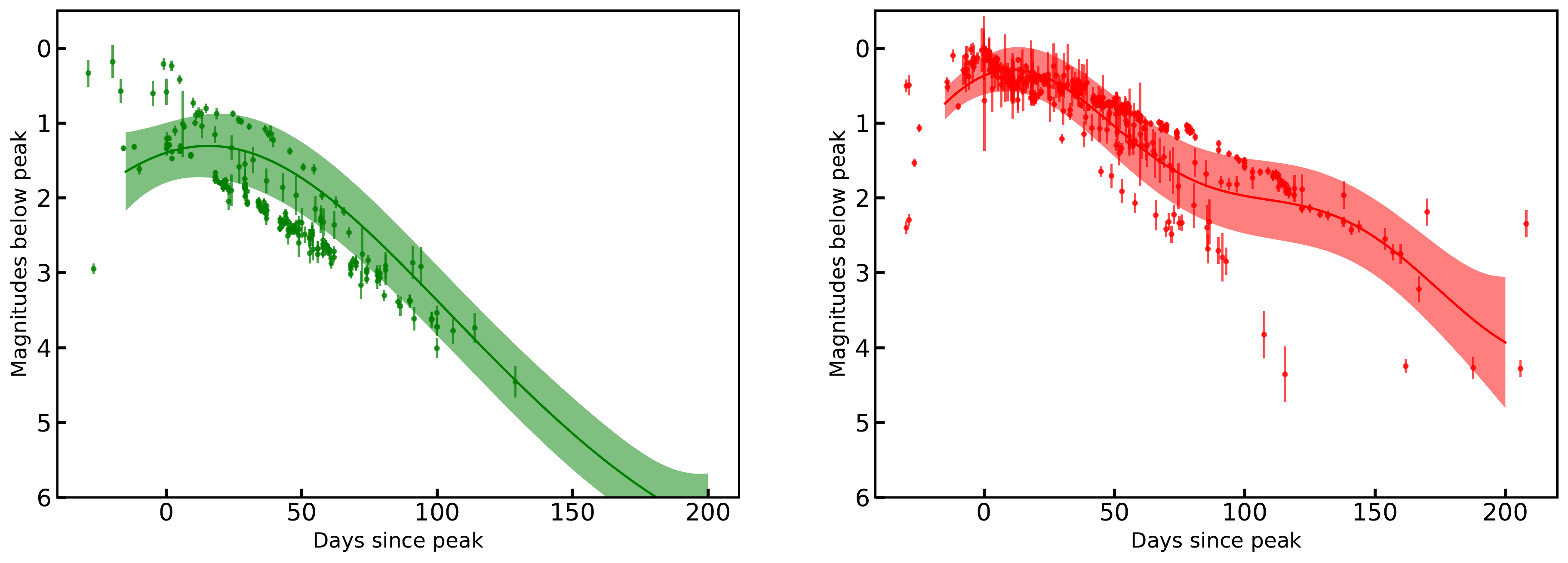}
    \caption{\emph{g} and \emph{r} band lightcurve templates (green and red colors respectively) for LRNe (top row) and ILRTs (bottom row). The solid points depict lightcurves of transients that were used to build the templates. The line and shaded regions show the templates and 68\% uncertainties derived from Gaussian Process fits to the transients. The templates capture the essence of the photometric evolution of LRNe (initial blue peak + long lived second peak/plateau) and ILRTs (single peak, red lightcurve). These templates are available online (see Sec. \ref{sec:data_availability}).}
    \label{fig:template}
\end{figure*}

We created an empirical LRN lightcurve template using the three well-sampled LRNe in our sample with M$_{\rm{r,peak}}\leq -11$ and two double-peaked LRNe from the literature AT\,2017jfs \citep{Pastorello2019b} and AT\,2020kog \citep{Pastorello2021b}. We first normalized the \emph{g} and \emph{r}-band lightcurves by their peak magnitude in the respective filter. We then fit a Gaussian Process model with a radial basis function (RBF) kernel to the normalized lightcurve. The \emph{g} and \emph{r}-template lightcurves of the LRNe are shown in the top row of Fig. \ref{fig:template}, and provided online.

The lightcurves of ILRTs are more homogeneous than LRNe. We construct lightcurve templates using \emph{g} lightcurves of 4 ILRTs (AT\,2018aes, AT\,2019abn, AT\,2019udc and AT\,2019ahd) and \emph{r-}band lightcurves of these and 4 additional ILRTs (AT\,2010dn, AT\,2012jc, AT\,2013la and AT\,2013lb). The ILRT templates are shown in the bottom row of Fig. \ref{fig:template}. Both LRN and ILRT template lightcurves are available online (Sec. \ref{sec:data_availability}).

\subsection{Luminosity Function}
\label{sec:lum_func}
In this paper, we have presented the first controlled sample of LRNe and ILRTs from a systematic survey. We use this sample to calculate the luminosity function of LRNe and ILRTs. 

\begin{figure*}
    \centering
    \includegraphics[width=0.5\textwidth]{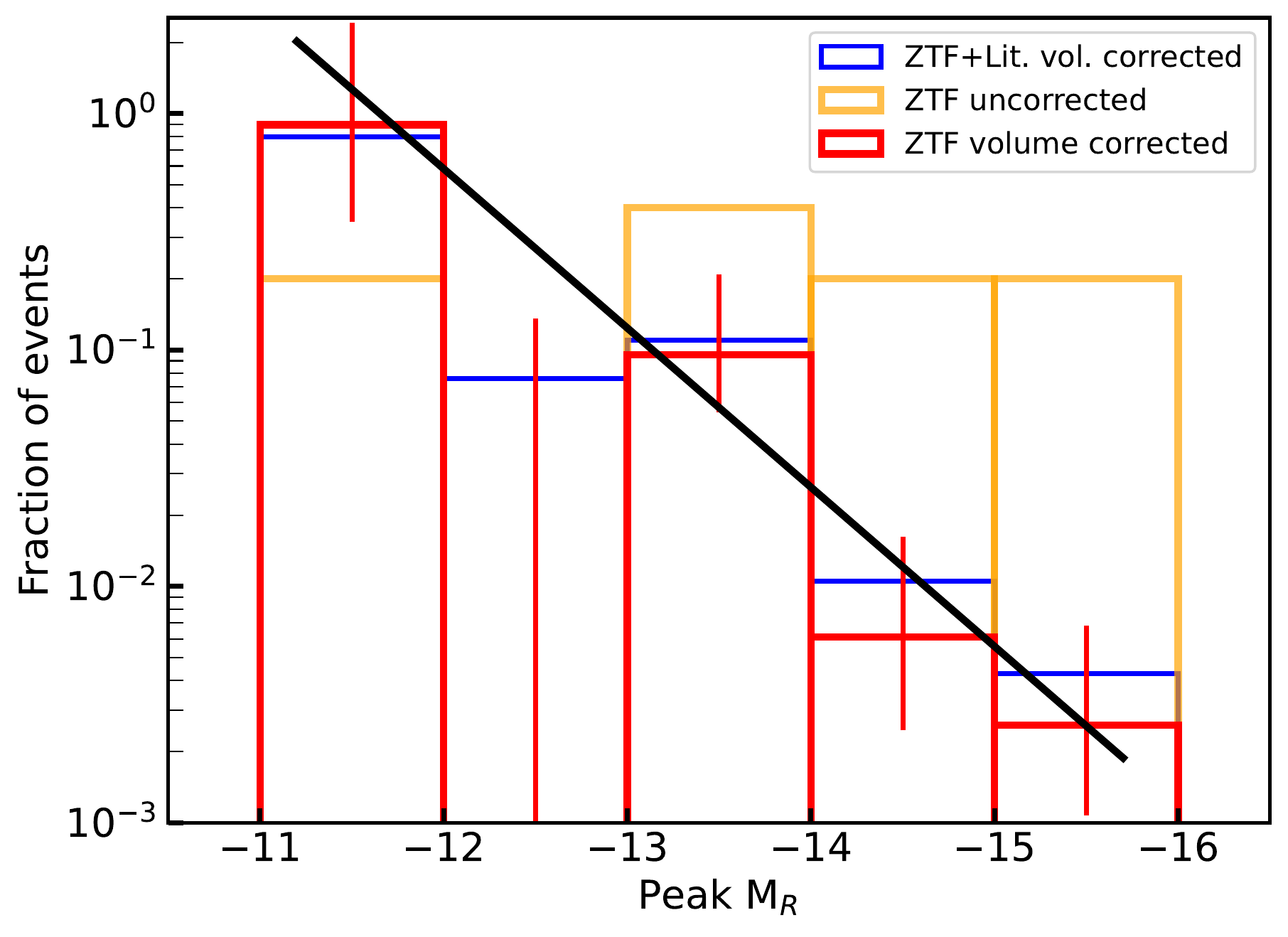}\includegraphics[width=0.5\textwidth]{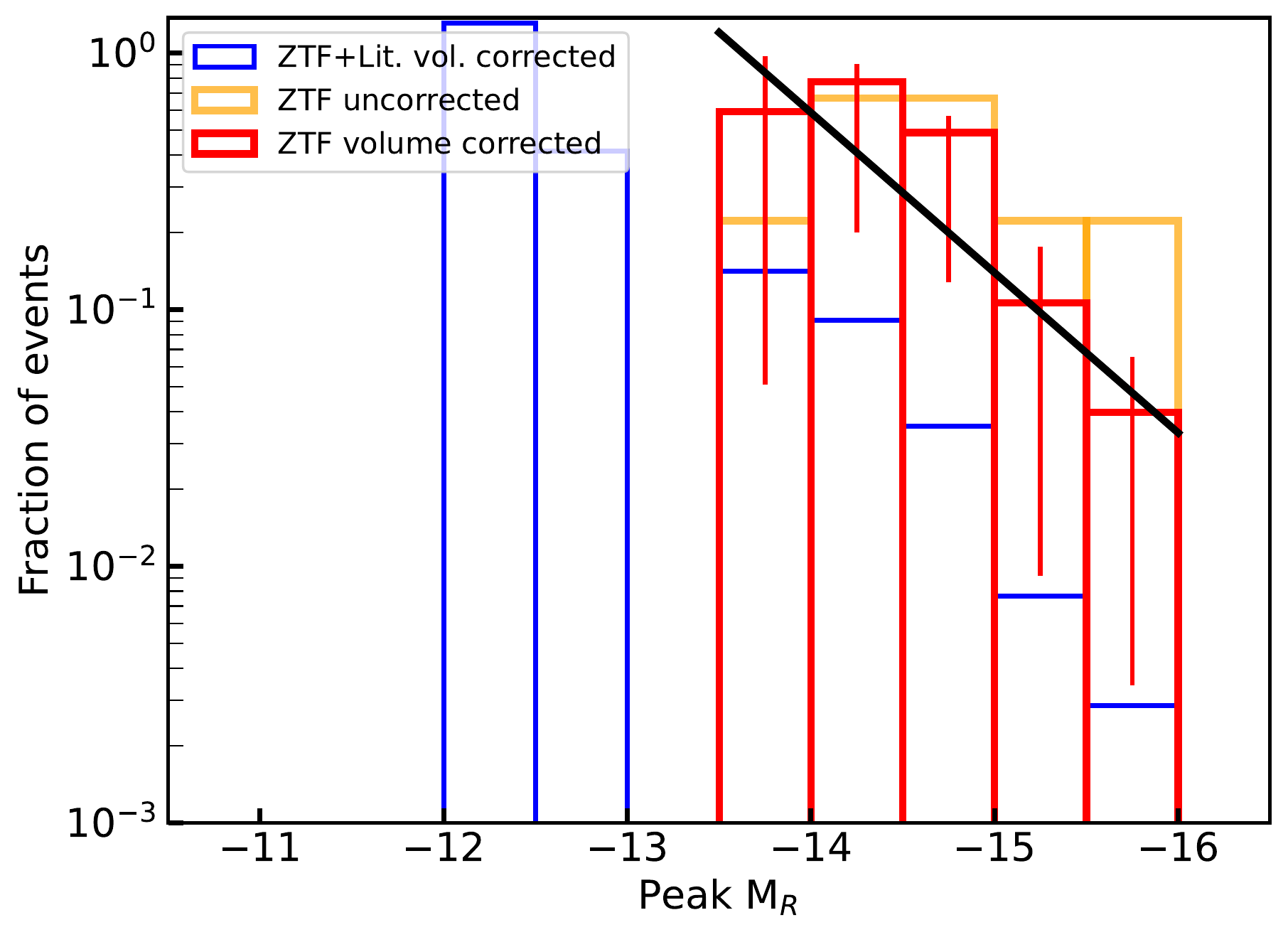}
    \caption{Distribution of the peak absolute magnitudes of LRNe (\emph{left}) and ILRTs (\emph{right}) in our sample. The volume corrected ZTF sample with Poisson errorbars is shown in red, and the volume corrected ZTF+literature sample is shown in blue. The black line shows a linear fit to this data (in log-space). We derive a luminosity scaling dN/dL $\propto L^{-2.5\pm0.3}$ for LRNe and dN/dL $\propto L^{-2.5\pm0.5}$ for ILRTs.}
    \label{fig:lrn_lum_fn}
\end{figure*}

We restrict ourselves to all gold and silver LRNe that have peak absolute magnitudes brighter than $-11$. Fig. \ref{fig:lrn_lum_fn} shows a histogram of the peak absolute magnitudes of the events in the ZTF sample. As the events are detectable out to different volumes (all smaller than the CLU volume limit of 200 Mpc), the histogram needs to be volume corrected to determine an accurate luminosity function. The volume corrected distribution of peak absolute magnitudes of LRNe in our sample is plotted in Fig. \ref{fig:lrn_lum_fn} with a different color. Each event has been weighted by $\frac{1}{\rm{V}_{\rm{max}}}$ where V$_{\rm{max}}$ is the maximum volume out to which that event can be detected ($V_{\rm{max}}\propto 10^{\frac{3}{5}(20 - \rm{M_{r,\rm{peak}}})}$). The volume corrected distribution shows a steep decline of the event rate with increasing peak luminosity. Fig. \ref{fig:lrn_lum_fn} also shows the distribution of the ZTF events combined with events from literature (taken from \citealt{Blagorodnova2021}). The distribution of the ZTF events is broadly consistent with that of the ZTF+literature events (although there are significant biases associated with the literature sample). The peak-luminosity distribution of ZTF events is well-fit by a straight line (in log-space) with a slope $0.6\pm0.1$. This corresponds to $\frac{dN}{dL} \propto L^{-2.5 \pm 0.3}$. This is significantly steeper than the luminosity function derived for low luminosity (M$_{V}$ $> -10$) LRNe \citep{Kochanek14_mergers}. The implications of these differences are discussed in Sec. \ref{sec:lrn_rate_disc}. 


Similarly, we calculate the ILRT luminosity function using the 8 gold and silver events from our sample. Fig. \ref{fig:lrn_lum_fn} (right panel) shows the volume-corrected distribution of the peak absolute magnitudes of ZTF ILRTs. We fit a straight line with slope $0.6\pm0.2$ to the distribution (in log-space), for M$_{r}<-13.5$. This corresponds to a luminosity function of $\frac{dN}{dL} \propto L^{-2.5\pm0.5}$. This scaling only samples the brighter end of the ILRT luminosity function. Fig. \ref{fig:lrn_lum_fn} also shows the distribution of previously known events from literature together with the ZTF sample, which extends to lower luminosities than the ZTF sample (M$_{r}\approx-12$). We note that for both LRNe and ILRTs, including the bronze or excluding the silver sample does not significantly change the derived luminosity function.

\subsection{Sub-sample for rate estimates}
\label{sec:rate_subsample}
We build a sub-sample from the candidates listed in Table \ref{tab:candidates} to use for rate calculations. All sources listed as potential LBVs are naturally excluded. To exclude the ``ambiguous" transients, we introduce an additional selection criterion -- a minimum threshold of 25 days on the duration of the ZTF lightcurve (i.e. the 5-$\sigma$ detections in \emph{g} or \emph{r} bands should span at least 25 days). Of the remaining sources, we only include those transients that have spectroscopic coverage (i.e. the gold and silver categories) to obtain an estimate of the rate, as this estimate is subsequently corrected for spectroscopic completeness. The sub-sample used for rate calculations is marked with an asterisk in Table \ref{tab:candidates}. 

The LRN sub-sample used for rate estimates comprises seven out of the eight gold and silver LRNe from Table \ref{tab:candidates}. ZTF\,19adakuot is excluded because of its low luminosity (Sec. \ref{sec:lum_func}). As noted in Sec. \ref{sec:lrn_gold}, the LRNe AT\,2018bwo and AT\,2020kog were not detected in the ZTF alert stream. Two detections of AT\,2018bwo were recovered post-facto, but it is excluded by the lightcurve-duration criterion. AT\,2020kog is excluded because it was in the chip-gaps of the ZTF camera, an effect that is accounted for by our survey simulations. 
The ILRT sub-sample comprises five out of the eight gold and silver ILRTs. The remaining three do not satisfy the lightcurve duration criterion. 

\subsection{Volumetric Rates}
The luminosity functions and template lightcurves were used to simulate LRNe and ILRTs, and the \texttt{simsurvey} simulations of the ZTF survey were used to count how many of them would be detected by our experiment. We apply the selection criteria described in Sec. \ref{sec:candidate_selection} and the lightcurve-duration criterion from Sec. \ref{sec:rate_subsample} to the simulated transients. We conduct 100 iterations of simulations for each rate to estimate the median and 1-$\sigma$ errors on the number of transients recovered for each rate.

\begin{figure*}[hbt]
    \centering
    \includegraphics[width=0.5\textwidth]{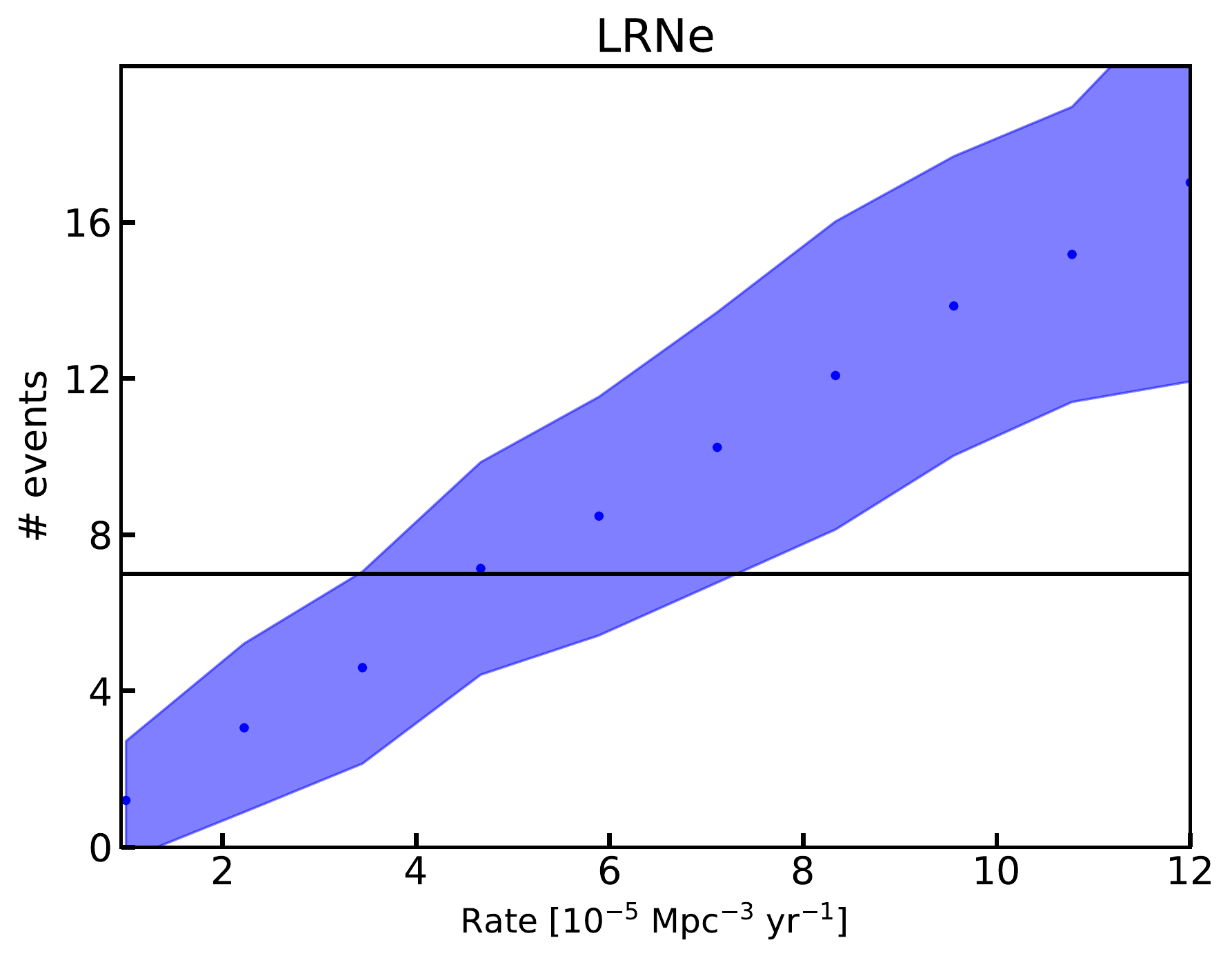}\includegraphics[width=0.5\textwidth]{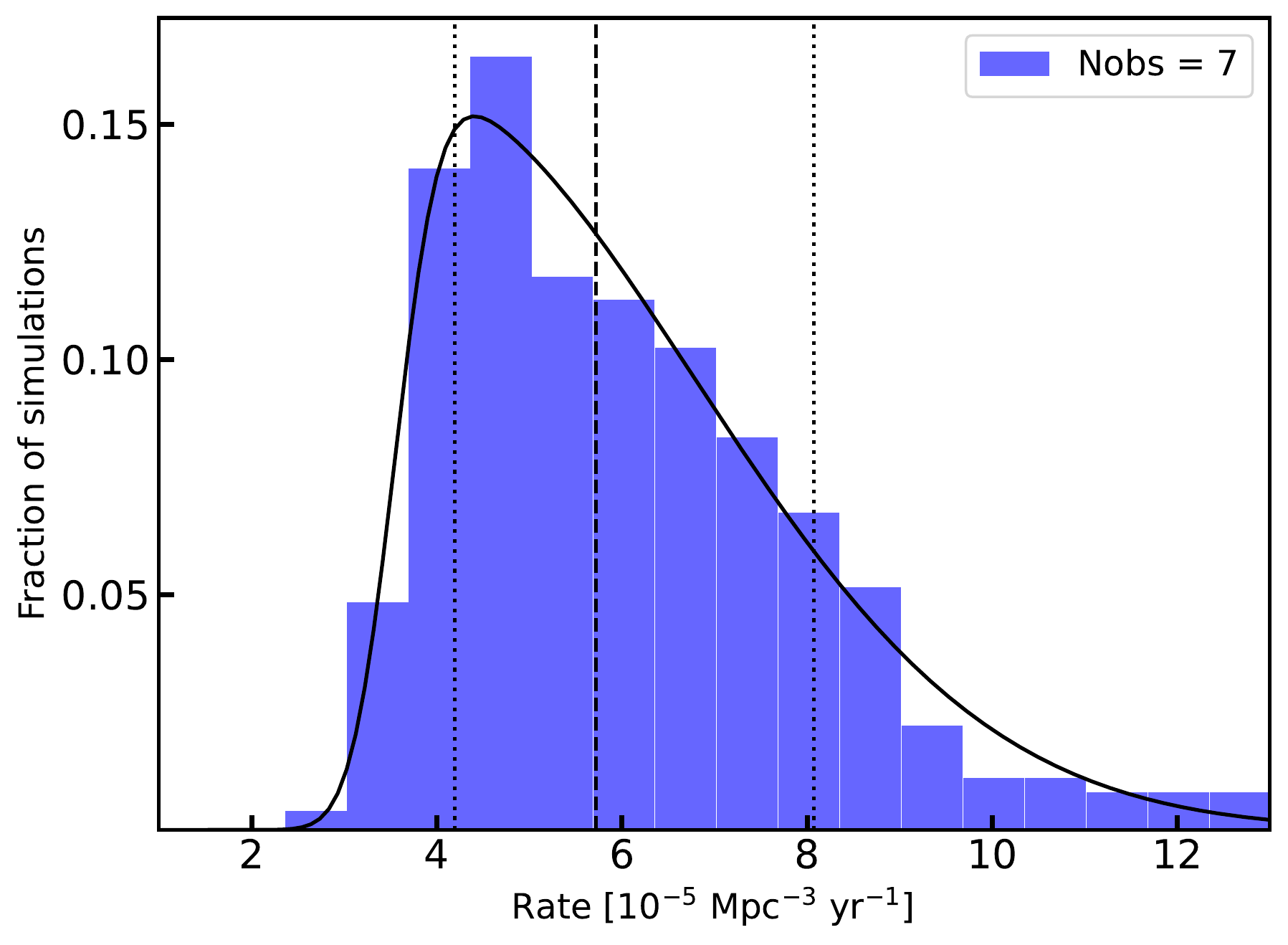}\\
    \includegraphics[width=0.5\textwidth]{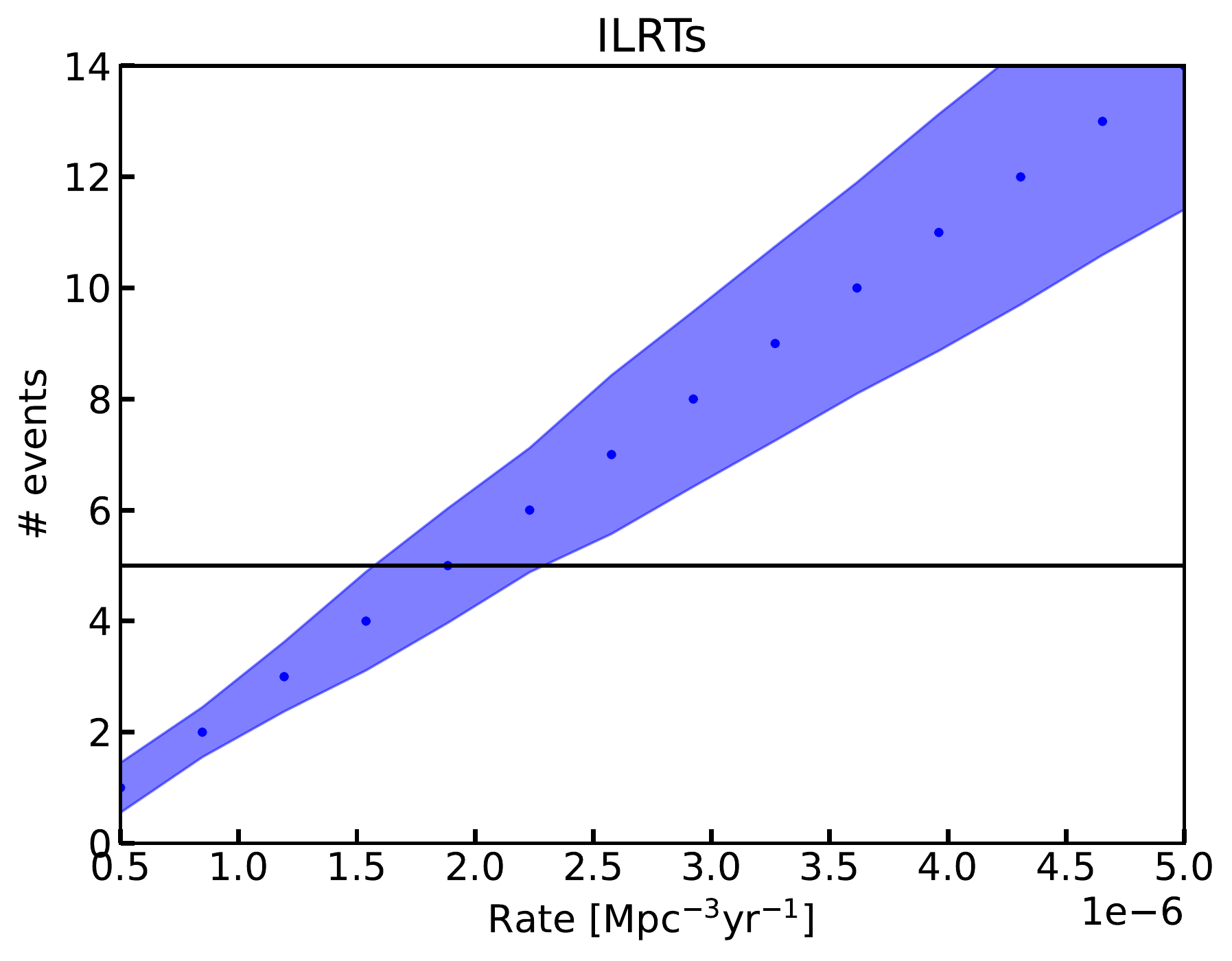}\includegraphics[width=0.5\textwidth]{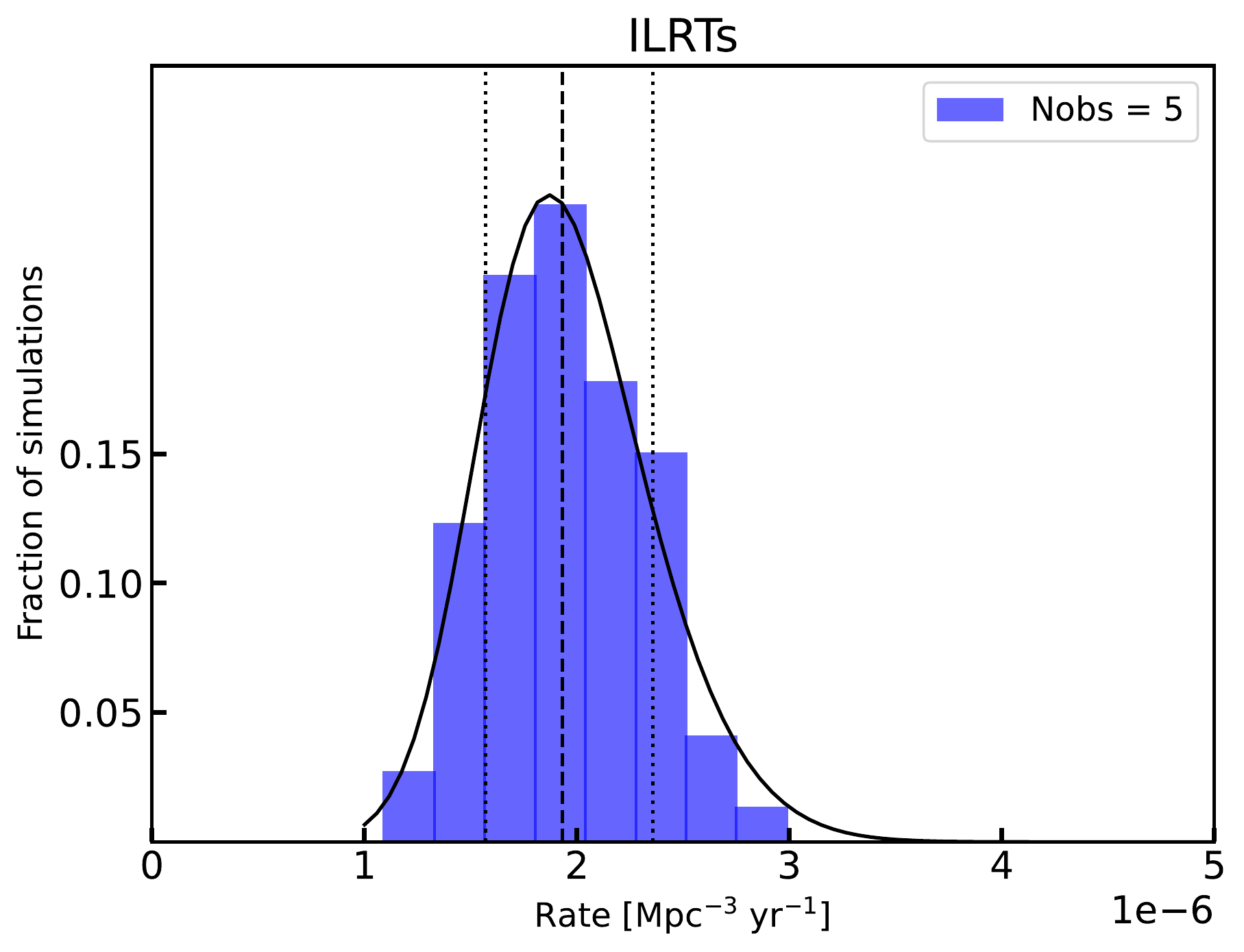}
    \caption{\emph{Left}: The number of simulated LRNe (\emph{top}) and ILRTs (\emph{bottom}) that pass our filtering criteria as a function of their volumetric rate. \emph{Right}: The fraction of simulations where the number of simulated transients that pass the filtering criteria LRNe equals the observed number of LRNe (\emph{top}) or ILRTs (\emph{bottom}). The distributions are fit by a skewed gaussian function, indicated as black solid curves.  The vertical dashed and dotted lines show the median and 68 percentile confidence limits of the distribution respectively.}
    \label{fig:lrn_rates}
\end{figure*}

The top-left panel of Fig. \ref{fig:lrn_rates} shows the number of LRNe that would be detected by our selection criteria as a function of their volumetric rate. The top-right panel of Fig. \ref{fig:lrn_rates} shows a histogram of the fraction of simulations where the number of detected simulated transients equals the observed number of seven LRN-gold and LRN-silver events. We fit the distribution with a skewed gaussian function to estimate the median and 68 percentile confidence limits. Accounting for an additional poisson uncertainty associated with the seven observed events we derive a volumetric rate of $r_{\rm{LRN},u} = 5.7^{+4.4}_{-2.7}\times10^{-5} $\,Mpc$^{-3}$\,yr$^{-1}$. %

This estimate does not account for four factors which may result in underestimation compared to the true LRN rate -- 1) the CLU experiment is limited to 100\arcsec or 30 kpc from nucleii of galaxies and will miss farther transients, 2) the CLU experiment is not 100\% spectroscopically complete,  3) the CLU experiment relies on the CLU galaxy catalog and is affected by its completeness, and 4) some LRNe will be missed due to inefficiencies of the ZTF image subtractions pipeline. First, we note that all literature events have been discovered within 100\arcsec or 140 kpc of their respective hosts. We searched through transients classified as part of the ZTF BTS -- an all-sky, magnitude limited survey with ZTF and did not find any additional LRNe. For this reason, we believe that the CLU offset criterion does not have a significant effect on the rate estimate. 
Second, the CLU experiment is $\approx80\%$ spectroscopically complete for m$_{r}<20$ mag transients. This suggests that the completeness corrected rate is $7.1^{+5.9}_{-3.4}\times10^{-5}$\,Mpc$^{-3}$\,yr$^{-1}$. However, the CLU completeness is a function of apparent magnitude, and varies from 100\% for m$_{r}<18$ mag, 95\% for m$_{r}<19$ mag. To account for this, we calculate the completeness as a function of apparent magnitude in bins of 1 mag. For each simulation of each of our rates, we bin the simulated transients by peak apparent magnitude, and count only the fraction of events that would be classified based on the CLU criteria. This exercise gives a value consistent with the simplified approach described above. This is because a majority of the events have faint apparent magnitudes, and the (large) uncertainties on this estimate are dominated by low number statistics. 

Third, to correct for the incompleteness of the galaxy catalog, we used the redshift completeness factor (RCF) derived from the BTS. As all events in our sample come from starforming galaxies, we calculate the RCF for starforming galaxies in the BTS sample as a function of redshift (z) and WISE 3.36\,\um absolute magnitude (M$_{\rm{W1}}$) as described in \citet{Fremling2020}. We then use the redshifts and M$_{\rm{W1}}$ magnitudes of the host galaxies of LRNe in our sample and weight each event by $\frac{1}{RCF(z,M_{W1})}$ to estimate the effect of galaxy catalog incompleteness. We find that this leads to an underestimation of the rate by $\approx 10\%$.

Finally, the ZTF image subtraction pipeline has two possible sources of inefficiency that are relevant for this calculation. In each science image, the pipeline actively masks pixels that are affected by quality cuts (e.g. saturation due to high brightness, cosmic rays, bad pixels). This introduces a time-dependent source of incompleteness. This dynamic masking does not have a significant effect as LRNe and ILRTs have long durations and are picked up by the pipeline eventually. A more serious issue pertains to the reduced efficiency of the image subtraction algorithm on bright galaxy backgrounds, as are common for the LRNe and ILRTs in our sample. The ZTF pipeline efficiency as a function of background brightness has not been studied to date, and this analysis is outside the scope of this paper. We therefore caution that our rates are possibly lower limits. Applying the corrections described above (except pipeline efficiency), we derive a corrected rate of $r_{\rm{LRNe}}=7.8^{+6.5}_{-3.7}\times10^{-4}$\,Mpc$^{-3}$\,yr$^{-1}$ for LRNe. 

Similarly, using the ILRT luminosity function and \texttt{simsurvey} simulations, we derive an uncorrected rate $r_{\rm{ILRT,u}}=1.9^{+1.3}_{-1.0}\times10^{-6}$\,Mpc$^{-3}$\,yr$^{-1}$ (see bottom panel of Fig. \ref{fig:lrn_rates}). Correcting for the incompleteness effects described above gives $r_{\rm{ILRT}}=2.6^{+1.8}_{-1.4}\times10^{-6}$\,Mpc$^{-3}$\,yr$^{-1}$. 


\section{Discussion}
\label{sec:discussion}
\subsection{LRNe}
\label{sec:lrn_rate_disc}
\begin{figure}[hbt]
    \centering
    \includegraphics[width=0.5\textwidth]{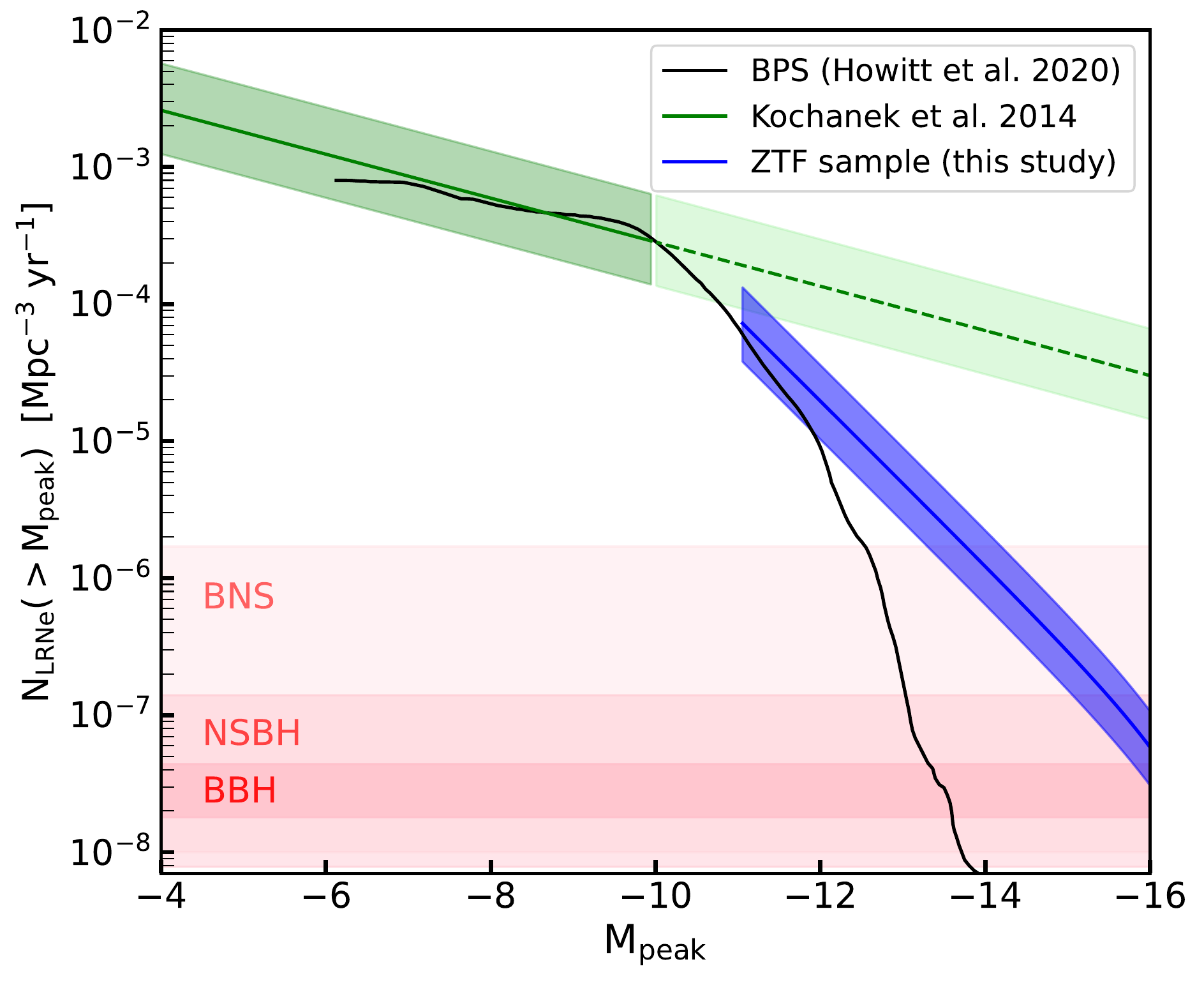}
    \caption{The rate of LRNe brighter than a given absolute magnitude as determined from our ZTF sample (blue), the Galactic sample of \citet{Kochanek14_mergers} sample (green) and binary population synthesis models of \citet{Howitt2020} (black line). The \citet{Kochanek14_mergers} sample was restricted to $-10\leq$M$_{V}\leq-4$, while the ZTF sample is restricted to $-16\leq$M$_{r}\leq-11$. The green dashed-line indicates the Galactic rate extrapolation to brighter luminosities. It is evident that the Galactic scaling overpredicts the observed ZTF rate by several orders of magnitude. The steeper drop of the LRN rate at higher luminosities is also seen in the BPS models. The BPS models underpredict the brightest (M$_{r}\leq-13$) rate of LRNe, which could be attributed to specific assumptions about CE physics in the models. The pink shaded regions indicate the constraints on the rates of BNS, NSBH and BBH mergers as measured by LIGO \citep{LIGO2021a}. The rates of the brightest LRNe are consistent with a significant fraction of them being progenitors of these compact merging systems. }
    \label{fig:lrn_rates_comparison}
\end{figure}

\citet{Kochanek14_mergers} used three Galactic mergers to estimate the rate of low luminosity stellar mergers with M$_{V}\geq-10$ mag. They find that the luminosity function of these mergers is roughly $\frac{dN}{dL} \propto L^{-1.4\pm0.3}$ and the rates of events brighter than M$_{V,\rm{peak}} = -3 (-10)$ is $\sim0.5$ (0.03) yr$^{-1}$. Our ZTF sample shows that their scaling does not extend to higher luminosities. For transients brighter than  M$_{r,\rm{peak}} = -11$, the luminosity function drops at a much steeper rate ($\frac{dN}{dL}\propto L^{-2.5}$), suggesting a broken power-law luminosity function for LRNe. Fig. \ref{fig:lrn_rates_comparison} shows the LRNe rate as a function of peak absolute magnitude derived from our ZTF sample and the \citet{Kochanek14_mergers} scaling. To convert the Galactic rate measurements from \citet{Kochanek14_mergers} to volumetric rates, we follow \citet{Howitt2020} and assume that the LRN rate scales with star-formation. We use a star-formation rate of 2\,M$_{\odot}$\,yr$^{-1}$ for the Milky Way \citep{Licquia2015} and an average cosmic star-formation rate of 0.015\,M$_{\odot}$\,yr$^{-1}$ \citep{Madau2014}. The ZTF and \citet{Kochanek14_mergers} rate estimates diverge at high luminosities. The volumetric rate of LRNe with brighter than M$_{r} = -11$ ($-13$) derived from the ZTF sample is lower by a factor of $\approx5$ (100) than that extrapolated from the \citet{Kochanek14_mergers} scaling.  
 


A steeper luminosity function at the brighter end has also been predicted by binary population synthesis models \citep{Howitt2020}. They follow the binary evolution of a population of binary systems with masses drawn from a Kroupa IMF and a binary fraction of unity, and find that 45\% of the simulated binaries undergo some form of unstable mass transfer. 38\% of these binaries result in stellar mergers while the remaining 62\% result in CE ejections. Qualitatively similar results were also obtained by the population synthesis study of \citet{Politano2010}, although they focused on the remnants of CEE rather than the associated transients. \citet{Howitt2020} used analytical expressions to approximate lightcurves associated with their simulated mergers and CE ejections, and determined the Galactic rate of LRNe as a function of their peak luminosity. Fig. \ref{fig:lrn_rates_comparison} also shows the volumetric LRN rate from these BPS simulations as a function of peak magnitude. We derive the volumetric rate using the Galactic and average cosmic star-formation rates as described above. As noted by \citet{Howitt2020}, the BPS rate agrees with the \citet{Kochanek14_mergers} value for low luminosities, but diverges for M$_{\rm{bol}} \leq -10$. The steep decline in the LRN rate seen in the BPS simulations is consistent with the rates derived from our ZTF sample. However, we note that the BPS simulations underpredict the rates for LRNe with M$_{r}<-13$. This could be a result of several assumptions about the LRN lightcurves, binary populations or CE physics used in the BPS simulations. Most importantly, \citet{Howitt2020} assumed that the LRN lightcurve is powered solely by hydrogen recombination and used analytical scaling relations from \citet{Ivanova2013araa} to estimate a LRN lightcurve. However, as noted in their analysis, this model is unable to reach the luminosities of the highest luminosity LRNe. Similar lightcurve modeling by \citet{Matsumoto2022} also shows that hydrogen recombination alone cannot explain the lightcurves of the highest luminosity LRNe. Additionally, hydrogen recombination is believed to power the plateau in LRNe, and not the initial blue peak (which is used for our and \citealt{Kochanek14_mergers} rate calculation). Using the recombination-powered lightcurve model thus underpredicts the LRN rate at the high luminosity end. More accurate models of LRN lightcurves are required to reconcile the predicted rate with the observed rate. Several other parameters about binary populations and CE physics can also contribute to the observed discrepancy. For example, a uniform binary fraction was assumed in the simulations, but the binary fraction increases with stellar mass \citep{Moe2017}. This would mean the number of massive binaries is higher, which would produce additional LRNe at the bright end. 

The broken power-law shape has interesting implications for our sample of LRNe. Observations as well as theoretical models of LRNe predict that the peak luminosities and durations of LRNe directly correlate with their progenitor masses \citep{Kochanek14_mergers, Blagorodnova2021, Ivanova13,Matsumoto2022, Cai2022}. Using the relation $L\propto M^{-2.2\pm0.3}$ from \citet{Cai2022}, our luminosity scaling implies a mass function of $\frac{dN}{dM}\propto M^{-4.3\pm1.1}$, different from standard IMF models. \citet{Kochanek14_mergers} found that the lower luminosity Galactic LRNe have progenitors consistent with the stellar IMF. An interesting possibility to explain this difference is to postulate that the low luminosity Galactic events are mergers while the more luminous extragalactic events are CE ejections. For a binary system with a given primary mass, the associated LRN is brighter and longer lived if it undergoes complete CE ejection rather than a merger where only a small fraction of the envelope is ejected. From their simulations, \citet{Howitt2020} find that the bright LRNe (M$_{\rm{bol}} \leq -10$) result almost exclusively from envelope ejections, while mergers result in lower luminosity transients. The broken power law would then suggest that the luminous CE ejections are much rarer than the less luminous stellar mergers. The range of peak luminosities and durations of the ZTF sample events presented here is consistent with the CE ejections in \citet{Howitt2020}. Thus in this picture, most if not all of the events in the ZTF sample analyzed here could be CE ejections. In this case, our derived rate would represent the rate of CE ejections in massive binaries -- an important step in the formation of double compact objects (DCOs, \citealt{VignaGomez2020}). It is interesting to compare our rate to the rate of DCO mergers detected by LIGO. \citet{LIGO2021b} determined the volumetric rate of binary neutron star (BNS), neutron star-black hole (NSBH) and binary black hole (BBH) mergers as $10-1700$\,Gpc$^{-3}$\,yr$^{-1}$, $7.8-140$\,Gpc$^{-3}$\,yr$^{-1}$ and $17-44$\,Gpc$^{-3}$\,yr$^{-1}$ respectively. The shaded pink regions in Fig. \ref{fig:lrn_rates_comparison} indicate these DCO merger rates. The plot shows that the brightest LRNe are consistent with a significant fraction of them being progenitors of DCO mergers.

Finally, we note another possible source of bias that could affect the LRN rate determined here. \citet{Pejcha2017ApJ} predict a population of LRNe originating in mergers involving giant primaries that should be completely dust enshrouded, and emit most of their radiation at IR wavelengths. \citet{Macleod2022} also show that dust formation in pre-merger outflows could obscure the binary system, causing the resulting LRNe to be observable only at IR wavelengths.  Some of the IR-only transients, dubbed SPRITES, that were discovered by the \emph{Spitzer Space Telescope} and had no optical counterparts are potential examples of such dust-enshrouded LRNe \citep{Kasliwal2017ApJ}. Searches for LRNe at infrared wavelengths are required to probe this dust-enshrouded population of LRNe. Upcoming NIR time-domain surveys such as the Wide-field Infrared Transient Explorer (WINTER, \citealt{Lourie2020}) at the Palomar Observatory and Dynamic REd All-sky Monitoring Survey (DREAMS, \citealt{Soon2019}) at the Siding Springs Observatory that are slated to commence operations in the second half of 2022 will help achieve this. Targeted surveys of nearby galaxies with these NIR telescopes will help discover and study IR-only LRNe, and shed light on the complete LRN landscape. Additionally, as all LRNe are brighter and longer lived at NIR wavelengths than the optical, the NIR surveys will also increase the population of optically-selected LRNe similar to those presented in this paper. Of particular importance will be additional discoveries of lower luminosity LRNe ($-9.5>\rm{M}_{\rm{r,peak}}>-11$) -- a luminosity range that has not been probed by our sample as they have much shorter optical lightcurves and can lack the early blue optical peak seen in their brighter counterparts \citep{Blagorodnova2021}. The upcoming NIR surveys will thus provide an unbiased and more precise measurement of the LRN luminosity function and volumetric rate.

\subsection{ILRTs}


We derive an ILRT rate of $\rm{R}_{<-13.5}\approx2.6\times10^{-6}$\,Mpc$^{-3}$\,yr$^{-1}$ for ILRTs that are more luminous than M$_{r}=-13.5$ mag. This rate is smaller than the LRN rate by two orders of magnitude, but we note that the luminosity threshold of our ILRT sample is brighter than that of our LRN sample. We cannot constrain the luminosity function of ILRTs with luminosities lower than M$_{r}=-13.5$, as none of our events lie in this luminosity range. In the literature, there are four ILRTs with $-12\geq \rm{M}_{r} \geq-13.5$, all discovered long before the start of ZTF \citep{Cai2021}. Assuming our luminosity function extends to M$_{r}\approx-12$, the total rate of ILRTs is $\rm{R}_{<-12}\approx2\times10^{-5}$\,Mpc$^{-3}$\,yr$^{-1}$. Additional discoveries of ILRTs in this low luminosity range are required to map out the lower end of the ILRT luminosity function.

\citet{Cai2021} derived a  lower limit on the ILRT rate of $9\times10^{-6}$\,Mpc$^{-3}$\,yr$^{-1}$. They used a sample of 12 ILRTs discovered in the last 12 years within 30 Mpc to estimate the ILRT rate, but their analysis did not include a luminosity function or completeness of the surveys. Additionally, their sample included two events with lower luminosities than our threshold, and their lower limit is consistent with the rough estimate $\rm{R}_{<-12}$ above. Comparing to the core-collapse supernova rate of $\approx1.01\times10^{-4}$\,Mpc$^{-3}$\,yr$^{-1}$ \citep{Perley2020}, we find that the rate of ILRTs ($\rm{R}_{<-13.5}$) is $\approx3^{+1.5}_{-1.5}\%$ of the CCSN rate. It remains to be seen how much the lower luminosity events contribute to this rate. 

Our rate is also lower than the estimate from \citet{Thompson09} who estimated the ILRT rate is $\sim20\%$ of the CCSN rate based on the two ILRTs SN\,2008S and NGC\,300OT. We note that NGC\,300OT is below the luminosity threshold of our sample, and the rough estimate $\rm{R}_{<-12}$ is consistent with their result.

Given the possible association of ILRT with ECSNe, it is worth comparing our rate to theoretical predictions. \citet{Poelarends2008} determine the range of stellar masses expected to undergo ECSN for different mass-loss and convective dredge-up prescriptions, and determine their rate to be $\approx4\%-24\%$ of the CCSN rate. The rate also depends sensitively on the metallicity, e.g \citet{Poelarends2007} find that for Z = 0.02, the ECSN rate is $\approx 3\%$ of the CCSN rate but for Z = $10^{-4}$ the rate is $\approx 25\%$. On the contrary, \citet{Doherty2015} use a different metallicity dependence on the mass-loss rate and find that the ECSN rate is $\approx2-5\%$ of the CCSN rate for all metallicities in the range Z $=$ [$10^{-5}$, 0.02]. Our measured ILRT rate ($\rm{R}_{<-13.5}$) is on the lower side, but still consistent with the wide range of theoretical calculations, and it is in line with the interpretation that several ILRTs could be ECSNe. 

We note however that we cannot rule out that some of the ILRTs presented here are LBV outbursts. Dusty LBV outbursts can also result in red, low luminosity transients with spectra showing narrow H with \ion{Ca}{2} and [\ion{Ca}{2}] features, similar to ILRTs \citep{Andrews2021}. While the transients in our sample do not show any archival activity for the last $\approx10$ years, it is possible that they experienced previous outbursts. Nevertheless,the connection of IRLTs to ECSNe is supported by strong evidence related to their dust-enshrouded progenitors with masses between 8--15 M$_{\odot}$ \citep{Thompson09,Botticella09,Jencson2019} and the fact that their remnants faded below their progenitor luminosities \citep{Adams16}.  Progenitors are not detectable for most of the ILRTs in our sample. As this will also be the case for a large number of ILRTs that will be detected by future large-scale deep surveys such as the Vera Rubin Observatory (VRO, \citealt{Ivezic2019}), it is important to identify additional ways of distinguishing ILRTs from LBV outbursts to confirm their nature as ECSNe. Possible ways to achieve this are extensive nebular-phase observations of ILRTs. On the one hand, late-time photometric observations would allow us to detect the presence, and estimate the amount of Ni generated in the explosion (see \citealt{Cai2021}). On the other hand, nebular spectroscopic observations would allow us to determine the composition of the ILRT ejecta where the presence of stronger Ni than Fe lines, weak O, Mg, C, Fe lines and a low [\ion{O}{1}]/[\ion{Ca}{2}] ratios would be evidence for the ECSN scenario (similar to AT\,2018zd, \citealt{Hiramatsu2021}).
\begin{figure*}[hbt]
    \centering
    \includegraphics[width=\textwidth]{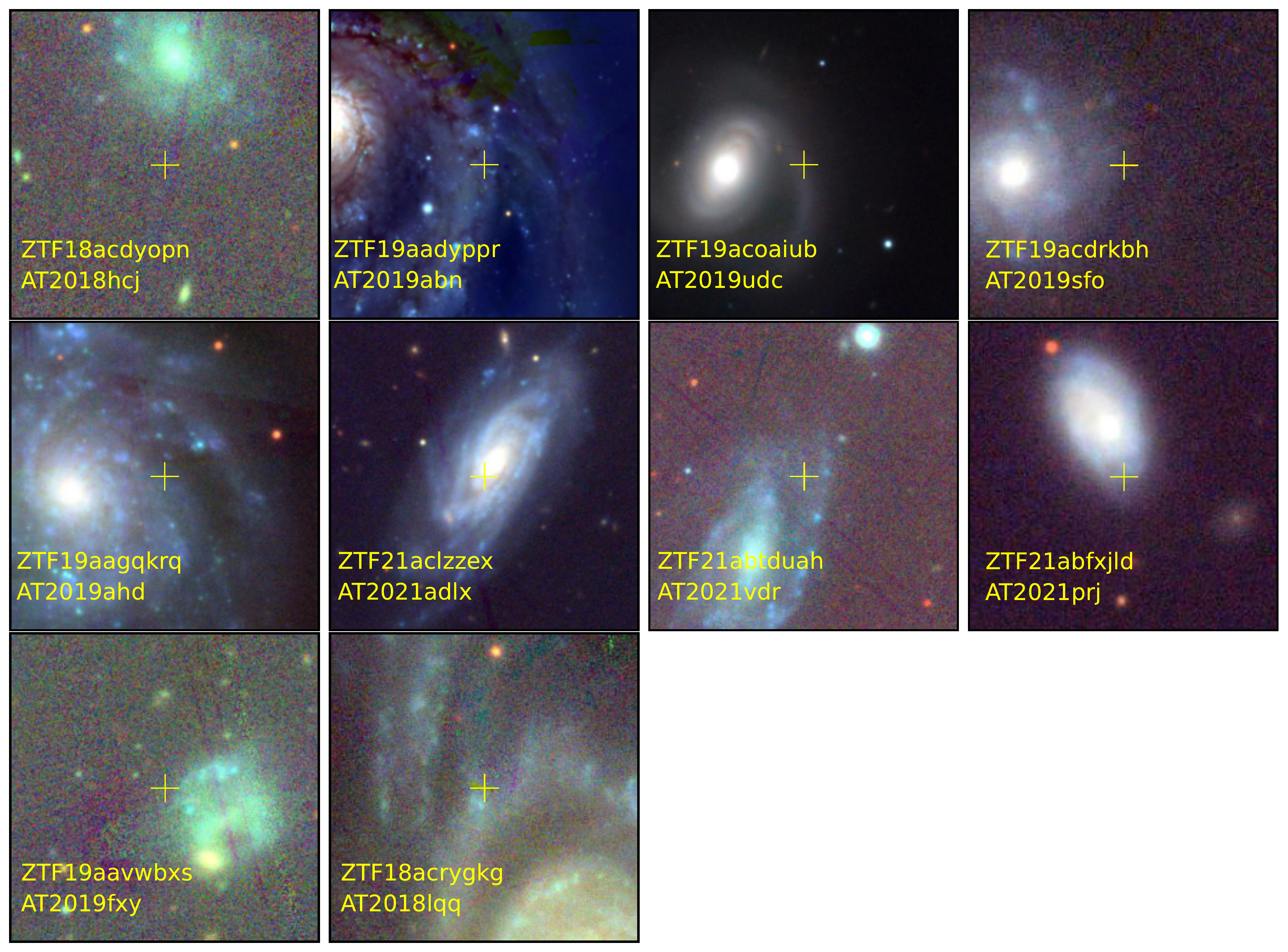}
    \caption{Host galaxies of the ILRTs in our sample. The positions of the ILRTs are indicated with a yellow cross, on top of color-coded PS1 images obtained from  \href{http://ps1images.stsci.edu/cgi-bin/ps1cutouts}{here}. Similar to LRNe, all ILRTs in our sample are located in star-forming galaxies.}
    \label{fig:ilrt_thumbnails}
\end{figure*}

\subsection{Host galaxies of LRNe and ILRTs}
Fig. \ref{fig:ilrt_thumbnails} shows thumbnails of the host galaxies of the ZTF ILRTs presented in this paper. The ZTF ILRTs belong to starforming galaxies, similar to all ILRTs discovered to date (see \citealt{Cai2021}). As neither ZTF nor the CLU galaxy catalog are biased towards star-forming galaxies \citep{Cook2019}, the sample confirms that ILRTs occur in predominantly young environments -- consistent with their ECSN or LBV interpretations.

\begin{figure*}[hbt]
    \centering
    \includegraphics[width=\textwidth]{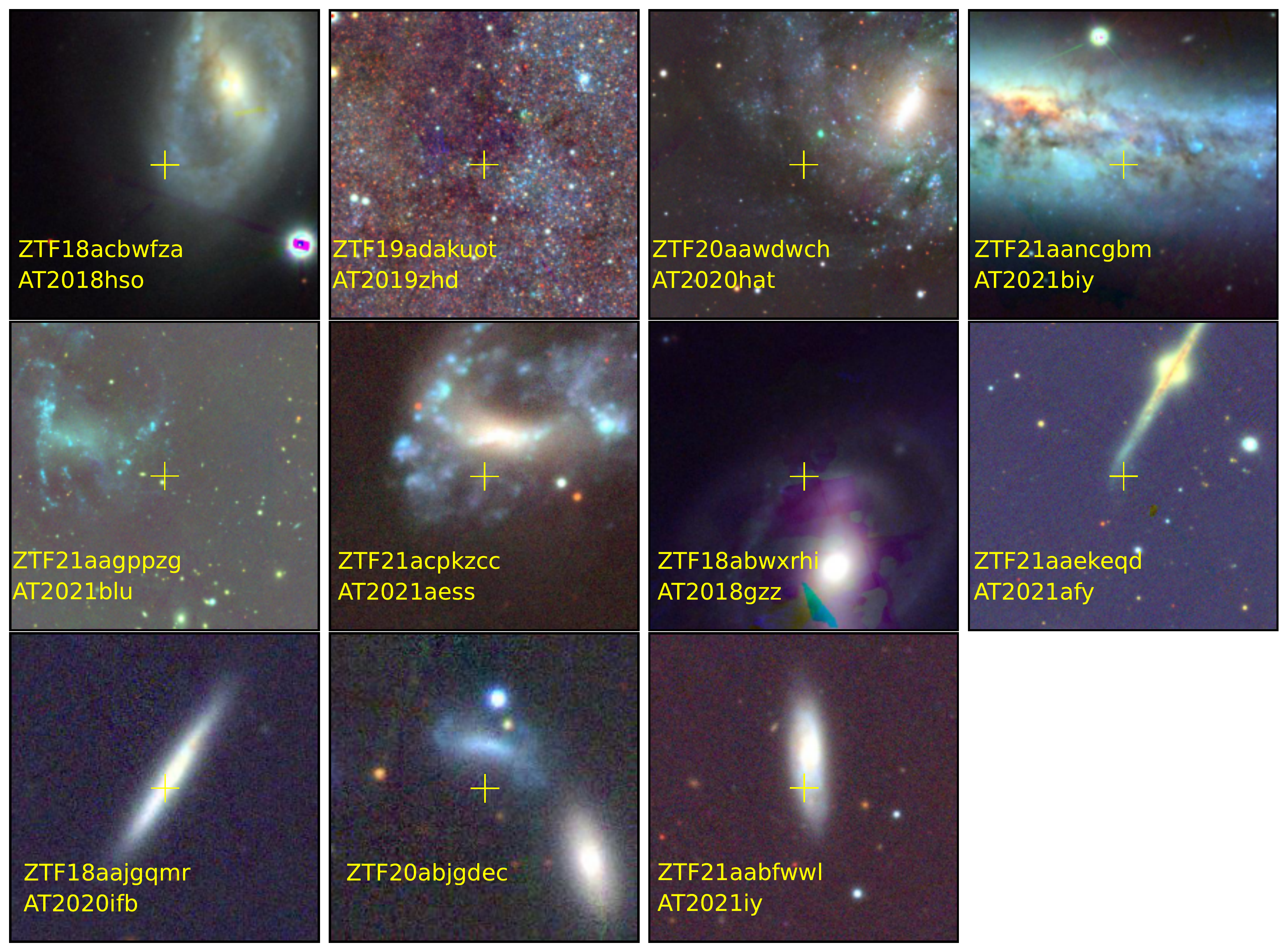}
    \caption{Host galaxies of the LRNe in our sample. The positions of the LRNe are indicated with a yellow cross. ZTF\,19adakuot is located in M31. All LRNe in our sample (including LRN-bronze events) are located in star-forming galaxies.}
    \label{fig:lrn_thumbnails}
\end{figure*}

Fig. \ref{fig:lrn_thumbnails} shows thumbnails of the host galaxies of the ZTF LRNe presented in this paper. All ZTF LRNe also belong to late-type starforming galaxies (see Table \ref{tab:candidates} for host galaxy morphologies), similar to all LRNe discovered in the last decade (see \citealt{Pastorello2019a,Blagorodnova2021}). This confirms that the luminous (M$_{r}\leq-11$) LRNe occur more commonly in star-forming environments. This is consistent with the expectation that LRNe in this luminosity range have massive ($\geq10$M\,$_{\odot}$) progenitors \citep{Blagorodnova2021}. Early-type galaxies are expected to have LRNe with lower mass progenitors and hence lower luminosities, making time-domain surveys less sensitive towards detecting them.


Only three LRNe have been discovered in old environments -- OGLE\,2002-BLG-360 in the Galactic bulge \citep{Tylenda13}, M31RV in the bulge of M31 \citep{Rich89} and M85\,OT in the S0-type galaxy M85 \citep{Kulkarni07}. Of these, M85\,OT does not follow the LRN peak luminosity-progenitor mass correlation \citep{Kochanek14_mergers,Blagorodnova2021}. The progenitor mass of M85\,OT has been constrained to $<7$M\,$_{\odot}$ \citep{Ofek08}. This is expected to produce a LRN with M$_{r,\rm{peak}}\geq-10$ -- much fainter than its actual peak M$_{r}\approx-12$. As it is not a canonical LRN, M85\,OT has been suggested to be an ILRT. However, ILRTs have starforming hosts unlike M85\,OT. Its progenitor mass is also incompatible with the ECSN or LBV scenario for ILRTs, suggesting that it is not a canonical ILRT either. An intriguing possibility is that M85OT is the merger of a WD with a companion evolved star. Such ``CV mergers" are believed to produce transients with observational characteristics similar to LRNe \citep{Metzger2021}. A WD progenitor would be consistent with the old environment of M85OT. As this is not a canonical merger of two stars, we would not expect the transient to follow the correlations observed in other LRNe. The Galactic slow nova CK\,Vul has been proposed to be a merger involving a white dwarf progenitor \citep{Eyres2018,Metzger2021}. M85OT could be an extragalactic member of this class of outbursts. Discoveries of additional LRNe in early-type galaxies will shed further light on the origin of this population.

\section{Summary and way forward}
\label{sec:summary}
Despite the discovery of a dozen LRNe and ILRTs in the last decade, their rate remained uncertain. In this paper, we compiled a systematic sample of LRNe and ILRTs using the Census of the Local Universe experiment on the Zwicky Transient Facility to address this issue. We present a sample of eight LRNe and eight ILRTs identified by the CLU experiment. We discuss the properties of these transients, and present new data for the precursor emission in AT\,2019zhd and line profile evolution in ZTF\,21aaekeqd (AT\,2021afy). We conduct simulations of the ZTF survey using actual ZTF observation history and correct for the completeness of CLU to derive a rate of $7.8^{+6.5}_{-3.7}\times10^{-5}$\,Mpc$^{-3}$\,yr$^{-1}$ for LRNe with absolute magnitudes M$_{r}$ in the range [$-11$,$-16$]; and $2.6^{+1.8}_{-1.4}\times10^{-6}$\,Mpc$^{-3}$\,yr$^{-1}$ for ILRTs with M$_{r}<-13.5$.

The rates of LRNe in this luminosity range are much lower than those extrapolated from Galactic measurements of low luminosity Galactic LRNe by \citet{Kochanek14_mergers}. Specifically, we find that the luminosity function of LRNe scales as $\propto L^{-2.5}$ in the range $-11<M_{r}<-16$, as opposed to L$^{-1.4}$ derived by \citet{Kochanek14_mergers} for $-4<M_{V}<-10$. This steeper decline at higher luminosities is broadly consistent with binary population synthesis models of \citet{Howitt2020}, however the BPS models underpredict the rates for M$_{r}<-13$ by as much as two orders of magnitude. This discrepancy is likely due to assumptions about the lightcurve models used in the BPS simulation. The rates of the brightest LRNe in our sample are consistent with a significant fraction of them being progenitors of double compact object systems. We note that all LRNe in our sample, and those discovered in the last decade belong to star-forming host galaxies. There have been no analogs of M85\,OT that was discovered in a S0-type galaxy. 


The ILRT rate corresponds to $2.6^{+1.8}_{-1.4}$\% of the local core-collapse SN rate. However, our sample only probes the high luminosity end of the ILRT luminosity function. Assuming our scaling extrapolates to lower luminosities, our measurement is consistent with previously measured constraints on the ILRT rate by \citet{Thompson09} and \citet{Cai2021}. The ILRT rate is also consistent with the wide range of theoretical predictions of rates of electron-capture SNe by \citet{Poelarends08} and \citet{Doherty2015}. Additional nebular phase observations of ILRTs and observations of their progenitors are required to confidently establish them as electron-capture SNe.

The future holds exciting prospects for studies of LRNe and ILRTs. Upcoming NIR time domain surveys such as WINTER and DREAMS will enable substantial progress in the next few years. These surveys will provide the first unbiased sample of the dusty, red transients and will be instrumental in uncovering hidden populations that could be missed by optical surveys. These results will set the stage for the VRO. Based on our rate estimates, we calculate that VRO will discover between 300 to 1500 LRNe and 200-700 ILRTs (with M$_{r,\rm{peak}}<-13.5$) per year (assuming a sensitivity of m$_{r}\approx24.2$ mag). While most of these will be too faint for spectroscopic classifications, these transients can be identified based on their low luminosities, long duration lightcurves and red photometric colors. An experiment similar to the Census of the Local Universe that keeps track of VRO transients in catalogued galaxies will be instrumental in the study of LRNe and ILRTs. 

\section{Acknowledgements}
Based on observations obtained with the Samuel Oschin Telescope 48-inch and the 60-inch Telescope at the Palomar Observatory as part of the Zwicky Transient Facility project. ZTF is supported by the National Science Foundation under Grants No. AST-1440341 and AST-2034437 and a collaboration including current partners Caltech, IPAC, the Weizmann Institute of Science, the Oskar Klein Center at Stockholm University, the University of Maryland, Deutsches Elektronen-Synchrotron and Humboldt University, the TANGO Consortium of Taiwan, the University of Wisconsin at Milwaukee, Trinity College Dublin, Lawrence Livermore National Laboratories, IN2P3, University of Warwick, Ruhr University Bochum, Northwestern University and former partners the University of Washington, Los Alamos National Laboratories, and Lawrence Berkeley National Laboratories. Operations are conducted by COO, IPAC, and UW. The ZTF forced photometry service was funded under the Heising-Simons Foundation grant 12540303 (PI: Graham). SED Machine is based upon work supported by the National Science Foundation under Grant No. 1106171. This work is also based on observations made with the Nordic Optical Telescope, owned in collaboration by the University of Turku and Aarhus University, and operated jointly by Aarhus University, the University of Turku and the University of Oslo, representing Denmark, Finland and Norway, the University of Iceland and Stockholm University at the Observatorio del Roque de los Muchachos, La Palma, Spain, of the Instituto de Astrofisica de Canarias.  SED Machine is based upon work supported by the National Science Foundation under Grant No. 1106171. The Liverpool Telescope is operated on the island of La Palma by Liverpool John Moores University in the Spanish Observatorio del Roque de los Muchachos of the Instituto de Astrof\'isica de Canarias with financial support from the UK Science and Technology Facilities Council. This work is part of the research program VENI, with project number 016.192.277, which is partly financed by the Netherlands Organisation for Scientific Research (NWO). A.V.F.’s group acknowledges generous support from the Christopher R. Redlich Fund, the U.C. Berkeley Miller Institute for Basic Research in Science, Sunil Nagaraj, Landon Noll, Sandy Otellini, and many additional donors. A major upgrade of the Kast spectrograph on the Shane 3 m telescope at Lick Observatory, led by Brad Holden, was made possible through generous gifts from the Heising-Simons Foundation, William and Marina Kast, and the University of California Observatories. Research at Lick Observatory is partially supported by a generous gift from Google.

\section{Data Availability}
\label{sec:data_availability}
Lightcurves and spectra of the LRNe, ILRTs and possible LBVs presented in this paper, and template LRN and ILRT lightcurves will be made available online after publication. The spectra will also be posted to WISeREP. The ZTF pointing history logs will be made available upon request to the corresponding author.



\bibliography{myreferences}
\bibliographystyle{apj}

\appendix
\section{Properties of sources classified as possible LBV outbursts}
\label{app:lbvs}
\begin{figure*}[hbt]
    \centering
    \includegraphics[width=\textwidth]{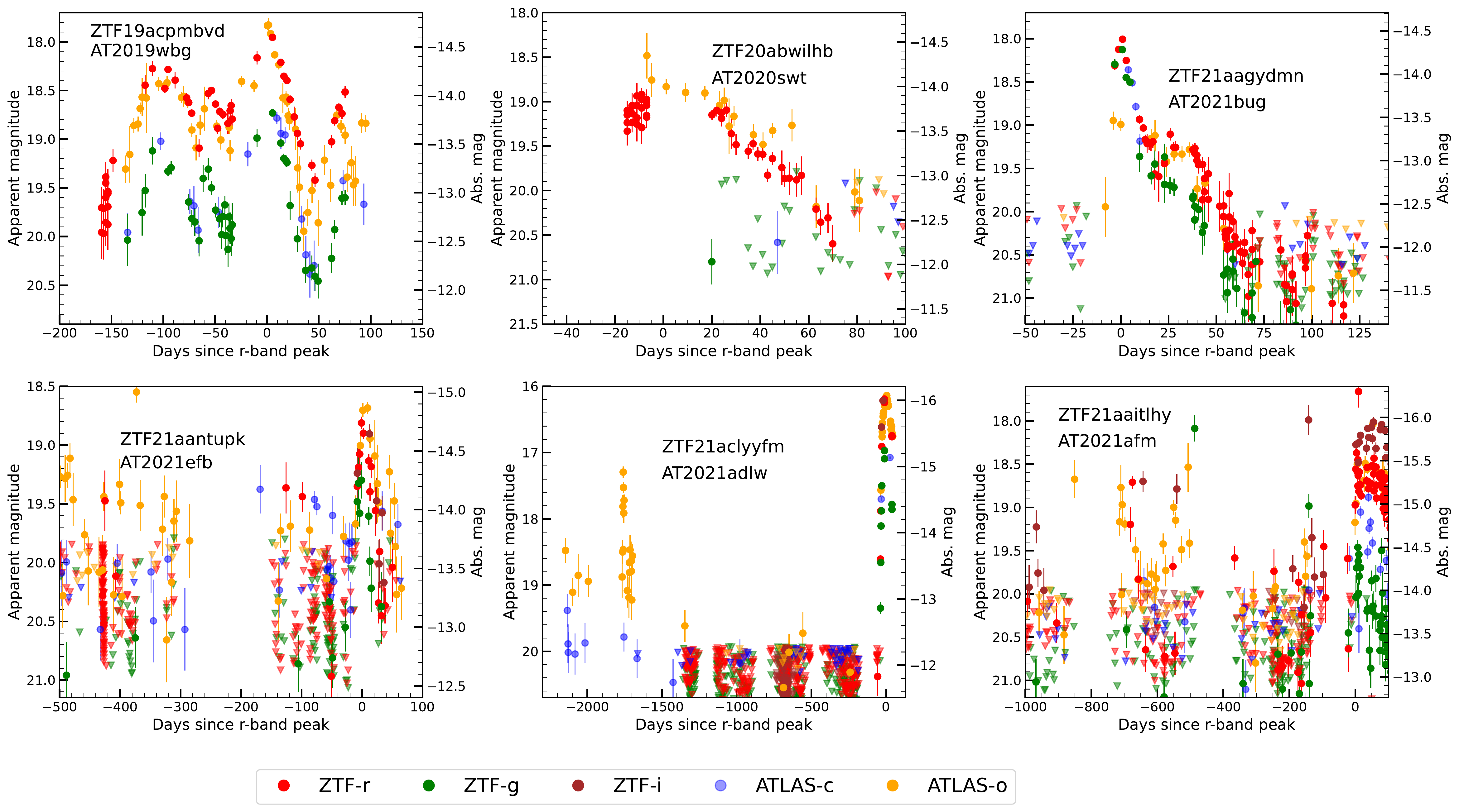}
    \caption{Lightcurves of transients that are possible LBV outbursts}
    \label{fig:lbv_lcs}
\end{figure*}

\begin{figure*}[hbt]
    \centering
    \includegraphics[width=\textwidth]{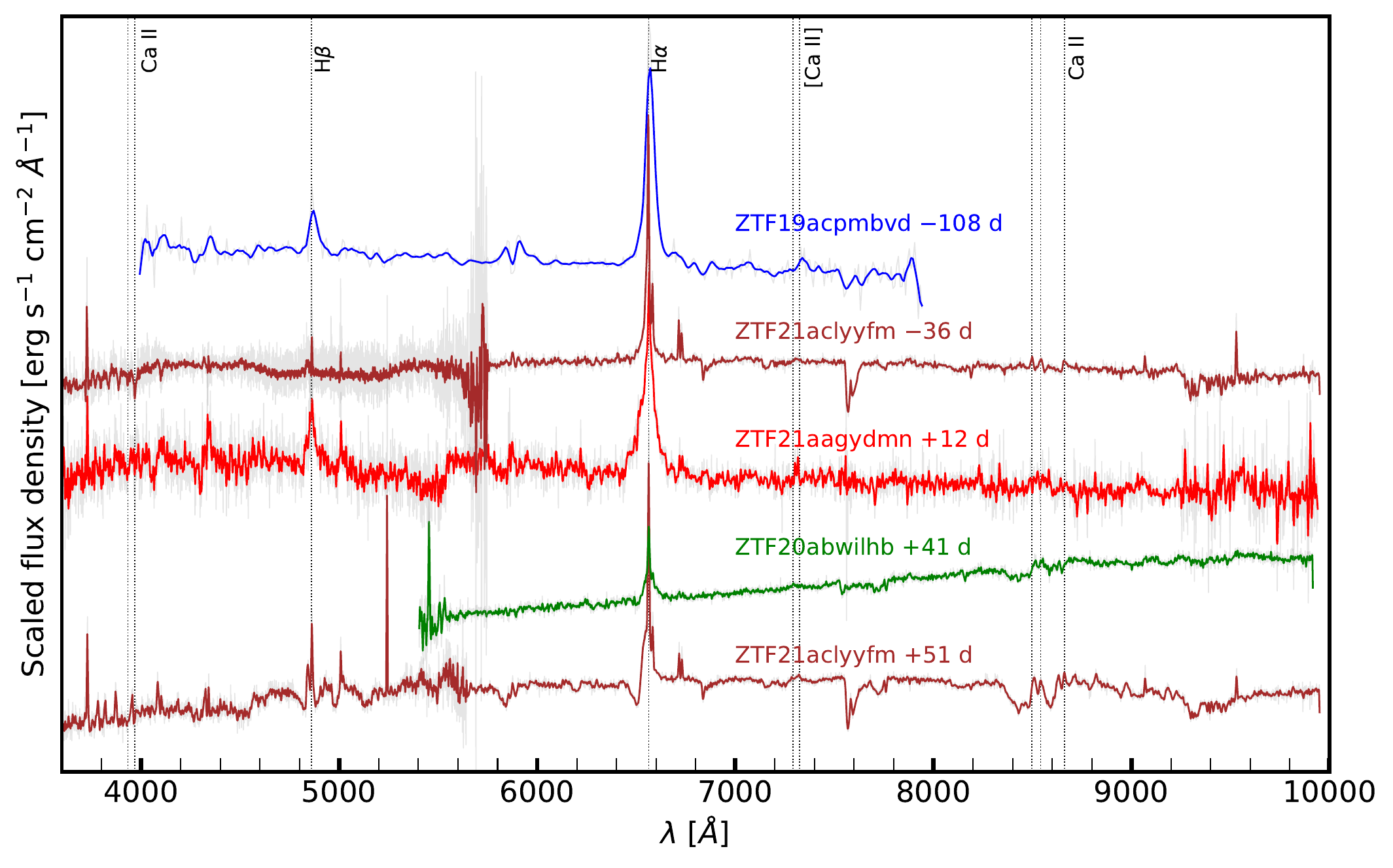}
    \caption{Optical spectra of transients that are possible LBV outbursts}
    \label{fig:lbv_spectra_optical}
\end{figure*}

The lightcurves of the six sources we classify as potential LBV outbursts are shown in Fig. \ref{fig:lbv_lcs} and their spectra are shown in Fig. \ref{fig:lbv_spectra_optical}. We discuss the individual objects below.

\begin{itemize}
    \item \textbf{ZTF\,19acpmbvd (AT\,2019wbg)} is located in the galaxy NGC\,4045, and shows a bumpy, red lightcurve with at least three peaks. The transient was detected in 2016 in iPTF data with m$_{g}\approx22$ mag. It was redetected in ZTF in 2022 at m$_{r}\approx19.5$ mag. A spectrum taken during this outburst shows H$\alpha$ with v$_{\rm{FWHM}}\approx 1700$\,km\,s$^{-1}$. 
    
    \item \textbf{ZTF\,20abwilhb (AT\,2020swt)} is located in the galaxy UGC\,03820. The lightcurve does not show multiple peaks, but shows signs of a $\approx40$day long plateau, after which the transient declines. The optical spectrum shows H$\alpha$ with v$_{\rm{FWHM}}\approx 3300$\,km\,s$^{-1}$, suggesting that this transient could be an LBV outburst or a low luminosity Type II SN.  
    
    \item \textbf{ZTF\,21aagydmn (AT\,2021bug)} is located in the galaxy NGC\,4533 and shows an unusual lightcurve that peaks at M$_{r}\approx-14.4$ and declines quickly. Once off the first decline, the lightcurve plateaus for $\approx10$ days, and then continues to decline. The spectrum shows H$\alpha$ with v$_{\rm{FWHM}} \approx 2600$\,km\,s$^{-1}$, however the lightcurve does not show a long-duration plateau as seen in low luminosity Type II SNe. The true nature of this source is unclear, but an outburst in a massive star is a possibility.
    
    \item \textbf{ZTF\,21aantupk (AT\,2021efb)} is located at the nucleus of the galaxy CGC\,003-005. The lightcurve shows erratic activity in the ATLAS and the ZTF data prior for several hundred days prior to the main explosion in 2021. We suggest that this source is either an LBV outburst or AGN activity. 
    
    \item \textbf{ZTF\,21aclyyfm (AT\,2021adlw)} is located in the galaxy NGC\,3813, and reached a peak absolute magnitude of M$_{r}\approx-16$ in the 2021 outburst. Archival ATLAS data shows that it underwent a similar outburst (M$_{o}\approx-15$) $\approx1800$ days prior to this outburst. An optical spectrum taken at an early phase during the 2021 outburst shows narrow H$\alpha$ with v$_{\rm{FWHM}}\approx1000$\,km\,s$^{-1}$, however late time spectra show ejecta with much larger velocities. This source is likely an LBV outburst or a ``weak" Type IIn supernova. 
    
    \item \textbf{ZTF\,21aaitlhy (AT\,2021afm)} is located close to the center of the galaxy NGC\,5657. It shows several detections in ATLAS and ZTF data in the $\approx$1000 days leading up to the 2021 explosion. We also detect the transient in the \emph{g}-band in PTF data taken during July 2011 and April 2016 at m$_{g}\approx 20$ mag. Owing to the extensive archival activity, we categorize this source as a possible LBV outburst.
    \end{itemize}

\label{lastpage}
\end{document}

%% file: journals.tex
\def\apjl{ApJ}%
\def\apjs{ApJS}%
\def\aap{A\&A}%

%% file: authors.tex
\author[0000-0003-2758-159X]{Viraj R. Karambelkar}
\email{viraj@astro.caltech.edu}
\affiliation{Cahill Center for Astrophysics, California Institute of Technology, Pasadena, CA 91125, USA}

\author{Mansi M. Kasliwal}
\affiliation{Cahill Center for Astrophysics, California Institute of Technology, Pasadena, CA 91125, USA}

\author[0000-0003-0901-1606]{Nadejda Blagorodnova}
\affiliation{Department of Astrophysics/IMAPP, Radboud University, Nijmegen, The Netherlands}

\author[0000-0003-1546-6615]{Jesper Sollerman}
\affiliation{The Oskar Klein Centre, Department of Astronomy, Stockholm University, AlbaNova, SE-10691 Stockholm, Sweden}

\author[0000-0003-2822-616X]{Robert Aloisi}
\affiliation{Department of Astronomy, University of Wisconsin-Madison, 475 North Charter Street, Madison, WI 53706, USA}

\author[0000-0003-3768-7515]{Shreya G. Anand}
\affiliation{Cahill Center for Astrophysics, California Institute of Technology, Pasadena, CA 91125, USA}

\author[0000-0002-8977-1498]{Igor Andreoni}
\affiliation{Joint Space-Science Institute, University of Maryland, College Park, MD 20742, USA}
\affiliation{Department of Astronomy, University of Maryland, College Park, MD 20742, USA}
\affiliation{Astrophysics Science Division, NASA Goddard Space Flight Center, Mail Code 661, Greenbelt, MD 20771, USA}

\author[0000-0001-5955-2502]{Thomas G. Brink}
\affiliation{Department of Astronomy, University of California, Berkeley, CA 94720-3411, USA}

\author{Rachel Bruch}
\affiliation{Department of Particle Physics and Astrophysics, Weizmann Institute of Science, 234 Herzl St, 76100 Rehovot, Israel}

\author[0000-0002-6877-7655]{David Cook}
\affiliation{IPAC, California Institute of Technology, 1200 E. California Blvd, Pasadena, CA 91125, USA }

\author[0000-0001-8372-997X]{Kaustav Kashyap Das}
\affiliation{Cahill Center for Astrophysics, California Institute of Technology, Pasadena, CA 91125, USA}

\author{Kishalay De}
\affiliation{MIT-Kavli Institute for Astrophysics and Space Research, 77 Massachusetts Ave., Cambridge, MA 02139, USA}

\author{Andrew Drake}
\affiliation{Cahill Center for Astrophysics, California Institute of Technology, Pasadena, CA 91125, USA}

\author{Alexei V. Filippenko}
\affiliation{Department of Astronomy, University of California, Berkeley, CA 94720-3411, USA}

\author[0000-0002-4223-103X]{Christoffer Fremling}
\affiliation{Cahill Center for Astrophysics, California Institute of Technology, Pasadena, CA 91125, USA }
\affiliation{Caltech Optical Observatories, California Institute of Technology, Pasadena, CA 91125, USA}

\author{George Helou}
\affiliation{IPAC, California Institute of Technology, 1200 E. California Blvd, Pasadena, CA 91125, USA }

\author[0000-0002-9017-3567]{Anna Ho}
\affiliation{Department of Astronomy, Cornell University, Ithaca, NY 14853, USA}

\author[0000-0001-5754-4007]{Jacob Jencson}
\affiliation{Department of Physics and Astronomy, Johns Hopkins University, 3400 North Charles Street, Baltimore, MD 21218, USA}

\author[0000-0003-3947-5946]{David Jones}
\affiliation{Instituto de Astrof\'isica de Canarias, E-38205 La Laguna, Tenerife, Spain} 
\affiliation{Departamento de Astrof\'isica, Universidad de La Laguna, E-38206 La Laguna, Tenerife, Spain}

\author[0000-0003-2451-5482]{Russ R. Laher}
\affiliation{IPAC, California Institute of Technology, 1200 E. California
             Blvd, Pasadena, CA 91125, USA}

\author[0000-0002-8532-9395]{Frank J. Masci}
\affiliation{IPAC, California Institute of Technology, 1200 E. California
             Blvd, Pasadena, CA 91125, USA}
             
\author{Kishore C. Patra}
\affiliation{Nagaraj-Noll-Otellini Graduate Fellow, Department of Astronomy, University of California, Berkeley, CA 94720-3411, USA}

\author[0000-0003-1227-3738]{Josiah Purdum}
\affiliation{Caltech Optical Observatories, California Institute of Technology, Pasadena, CA 91125, USA}

\author{Alexander Reedy}
\affiliation{Cahill Center for Astrophysics, California Institute of Technology, Pasadena, CA 91125, USA}

\author[0000-0001-8208-9755]{Tawny Sit}
\affiliation{Department of Astronomy, The Ohio State University, 140 West 18th Avenue, Columbus, OH 43210, USA}

\author[0000-0003-4531-1745]{Yashvi Sharma}
\affiliation{Cahill Center for Astrophysics, California Institute of Technology, Pasadena, CA 91125, USA}

\author[0000-0003-0484-3331]{Anastasios Tzanidakis}
\affiliation{Department of Astronomy, University of Washington, Seattle, WA 98195, USA}

\author{St\'efan J. van der Walt}
\affiliation{Berkeley Institute for Data Science, University of California, Berkeley}

\author[0000-0001-6747-8509]{Yuhan Yao}
\affiliation{Cahill Center for Astrophysics, California Institute of Technology, Pasadena, CA 91125, USA}

\author{Chaoran Zhang}
\affiliation{Center for Gravitation, Cosmology, and Astrophysics, Department of Physics, University of Wisconsin, Milwaukee, WI 53201, USA}

%% file: table_candidates_spec.tex
\begingroup
\renewcommand{\tabcolsep}{2pt}
\begin{table*}
\begin{center}
\begin{minipage}{18cm}
\caption{Properties of the 34 ZTF transients of interest}
\label{tab:candidates}
\begin{tabular}{llcccccccccc}
\hline
\hline
{ZTF }        & {AT}   & M$_{\rm{abs}}$      & multi-peak/ & Archival    &  Gal type$^{a}$  & H$\alpha$-FWHM$^{b}$ & [\ion{Ca}{2}] & Mol.      & Classification \\ 
{Name}        & {Name} & at peak             & plateau     & (PTF/ATLAS) &              & (km s$^{-1}$)        &        & feats.    &                \\ 
\hline 
ZTF18acbwfza* & 18hso  & $-13.82\pm0.15$     & mp          & --/no       & SB(r)a pec   & 500                  & em?    & yes$^{c}$ & LRN-gold       \\
ZTF19adakuot  & 19zhd  & $-9.60\pm0.15$     & pl          & no/yes      & SA(s)b       &                      &        & yes$^{c}$ & LRN-gold       \\
ZTF20aawdwch* & 20hat  & $-11.43\pm0.15$     & pl          & no/no       & SAB(rs)cd	& 130                  & no     & yes       & LRN-gold       \\
ZTF21aancgbm* & 21biy  & $-13.86\pm0.15$     & mp          & yes/no      & SB(s)d.      & 500                  & em?    & yes       & LRN-gold       \\ 
ZTF21aagppzg* & 21blu  & $-13.50\pm0.15$     & mp          & no/no       & starforming  & 500                  & no     & yes       & LRN-gold       \\
ZTF21acpkzcc* & 21aess & $-15.12\pm0.15$     & mp          & --/no       & SB(s)m?      & 500                  & no     & --        & LRN-gold       \\
\hline 
ZTF18abwxrhi* & 18gzz  & $-14.50\pm0.16$     & mp          & no/no       & (R')SB(rs)ab & 300                  & no     & no        & LRN-silver     \\ 
ZTF21aaekeqd* & 21afy  & $-13.95\pm0.16$     & pl          & --/no       & starforming  & 700                  & no     & no        & LRN-silver     \\
\hline
ZTF18aajgqmr  & 20ifb  & $-15.36\pm0.17$     & mp          &  --/no      & starforming  &                      &        &           & LRN-bronze     \\
ZTF20abjgdec  & 20afdb & $-14.37\pm0.16$     & mp          & --/no       & starforming  &                      &        &           & LRN-bronze     \\
ZTF21aabfwwl  & 21iy   & $-15.72\pm0.16$     & mp          & --/no       & starforming  &                      &        &           & LRN-bronze     \\
\hline
\hline
ZTF18acdyopn  & 18hcj  & $-14.33\pm0.19$     & no          & --/no       & SAdm.        & 300                  & em     & no        & ILRT-gold      \\
ZTF19aadyppr* & 19abn  & $-14.73\pm0.15$     & no          & yes/no.     & Sa           & 700                 & em     & no        & ILRT-gold      \\
ZTF19acoaiub* & 19udc  & $-14.62\pm0.15$     & no          & --/no       & SAB(s)a      & 1250                 & em     & no        & ILRT-gold      \\
ZTF19acdrkbh  & 19sfo  & $-14.62\pm0.20$     & --          & --/no       & starforming  & 600                  & em     & no        & ILRT-gold      \\
ZTF19aagqkrq* & 19ahd  & $-13.72\pm0.15$     & no          & --/no.      & SA(s)cd	    & 700                  & em     & no        & ILRT-gold      \\
ZTF21aclzzex* & 21adlx & $-15.68\pm0.16$     & no          & yes/no      & Sb           & 700                  & em     & no        & ILRT-silver    \\
\hline
ZTF21abtduah* & 21vdr  & $<-14.96$           & --          & no/no       & SB(s)d:      & 1300                 & no     & no        & ILRT-silver    \\
ZTF21abfxjld  & 21prj  & $<-14.21$           & --          & no/no.      & starforming  &$<300$                & no     & no        & ILRT-silver    \\
\hline
ZTF18acrygkg  & 18lqq  & $<-14.14$           & no          & --/no       & SA(s)bc      &                      &        &           & ILRT-bronze    \\
ZTF19aavwbxs  & 19fxy  & $-14.44\pm0.17$     & no          & --/no       & starforming  &                      &        &           & ILRT-bronze    \\
\hline
\hline
ZTF19acpmbvd  & 19wbg  & $-14.66\pm0.15$     & mp          & yes/yes     & SAB(r)a	    &    1700                 &        &           & LBV?           \\
ZTF20abwilhb  & 20swt  & $-13.87\pm0.16$     & no.         & no/no       & Scd          & 3300                 & no     & no        & LBV/Type II?   \\ 
ZTF21aagydmn  & 21bug  & $-14.36\pm0.15$     & no          & yes/no      & SAd          & 2600                 & no     & no        & LBV?           \\
ZTF21aantupk  & 21efb  & $-14.71\pm0.15$     & no          & --/yes      & -            &                      &        &           & LBV/AGN?       \\
ZTF21aclyyfm  & 21adlw & $-16.03\pm0.15$     & no          & --/yes      & SA(rs)b      &                      & no     & no        & LBV?                \\
ZTF21aaitlhy  & 21afm  & $-15.44\pm0.15$     & mp          & yes/yes     & SBb          &                      &        &           & LBV?                \\
\hline
ZTF18aawoeho  & 21ahuh & $-15.51\pm0.15$     & no          & no/no       & starburst    &                      &        &           & ?              \\
ZTF20ablmyzj  & 20afdc & $-15.95\pm0.30$     & no          & --/no       & S            &                      &        &           & ?              \\
ZTF20acfxnmv  & 20afdd & $-14.67\pm0.15$     & --          & --/no       & SB(s)cd      &                      &        &           & ?              \\ 
ZTF20acivtfy  & 20afde & $-13.72\pm0.16$     & --          & --/no       & starforming  &                      &        &           & ?              \\
ZTF21aapngrj  & 21gcg  & $-12.75\pm0.15$     & no          & --/no       &              &                      &        &           & ?              \\ 
ZTF21aakbdzz  & 21czz  & $-14.83\pm0.10$     & no          & --/no       & S            &                      &        &           & ?              \\
ZTF21aamwyxf  & 21dtz  & $-15.98\pm0.15$     & no          & no/no       & E            &                      &        &           & ?              \\
\hline
\hline
\end{tabular}

\begin{tablenotes} 
\item -- or unfilled entries indicate instances for which data is not available.
\item * : Events used for rate calculation, in Sec. \ref{sec:rates}; $a$ : Galaxy morphologies are taken from NED; $b$: Where multiple spectra are available, the maximum FWHM is reported; $c$: The molecular features are visible in spectra from the literature (see Sec. \ref{sec:lrn_gold}).
\end{tablenotes}

\end{minipage}
\end{center}
\end{table*}
\endgroup

%% file: table_spectral_log.tex
\begingroup
\renewcommand{\tabcolsep}{3pt}
\begin{table*}
\begin{center}

\caption{Observation log of spectra presented in this paper}
\label{tab:speclog}
\begin{tabular}{cccc|cccc}
\hline
\hline
{Name} & {Tel./ Inst.} & {Date} & {Resolution} & {Name} & {Tel./ Inst.} & {Date} & {Resolution}\\ 
\hline
ZTF18abwxrhi & P200/DBSP   & 2018-09-18 & 1500  & ZTF21aancgbm & Keck/NIRES  & 2021-04-18 & 1000  \\ 
ZTF18abwxrhi & P200/DBSP   & 2018-10-10 & 1000  & ZTF21aagppzg & Keck/NIRES  & 2021-04-18 & 1000  \\ 
ZTF18acbwfza & Gemini/GMOS & 2018-11-02 & 2000  & ZTF21aaekeqd & Keck/LRIS   & 2021-05-09 & 750   \\ 
ZTF18acbwfza & P200/DBSP   & 2018-11-02 & 1000  & ZTF21abfxjld & Keck/LRIS   & 2021-07-06 & 750   \\ 
ZTF18acbwfza & Keck/LRIS   & 2018-11-10 & 750   & ZTF21abfxjld & Keck/LRIS   & 2021-08-14 & 750   \\ 
ZTF18acdyopn & Keck/LRIS   & 2018-11-10 & 750   & ZTF21abfxjld & Keck/NIRES  & 2021-09-25 & 1000  \\ 
ZTF19aadyppr & P200/DBSP   & 2019-01-26 & 1000  & ZTF21abtduah & P200/DBSP   & 2021-11-06 & 1000  \\ 
ZTF19aagqkrq & NOT/ALFOSC  & 2019-02-11 & 350   & ZTF21aclyyfm & P200/DBSP   & 2021-11-06 & 1000  \\ 
ZTF19aagqkrq & P200/DBSP   & 2019-02-12 & 1000  & ZTF21aclzzex & Shane/KAST  & 2021-11-12 & 1000  \\ 
ZTF19aadyppr & NOT/ALFOSC  & 2019-02-23 & 350   & ZTF21acpkzcc & Keck/NIRES  & 2021-11-17 & 1000  \\ 
ZTF19aadyppr & Keck/LRIS   & 2019-03-07 & 750   & ZTF21acpkzcc & SALT/HRS    & 2021-11-29 & 15000 \\ 
ZTF19aadyppr & P200/DBSP   & 2019-03-16 & 1000  & ZTF21acpkzcc & P200/DBSP   & 2021-12-01 & 1000  \\ 
ZTF19aadyppr & P200/DBSP   & 2019-04-13 & 1000  & ZTF21abtduah & P200/DBSP   & 2021-12-01 & 1000  \\ 
ZTF19aadyppr & P200/DBSP   & 2019-05-13 & 1000  & ZTF21aclzzex & P200/DBSP   & 2021-12-01 & 1000  \\ 
ZTF19acdrkbh & Keck/LRIS   & 2019-10-27 & 750   & ZTF21aagppzg & P200/TSpec  & 2021-12-12 & 1000  \\ 
ZTF19acoaiub & P200/DBSP   & 2019-11-05 & 1000  & ZTF21acpkzcc & P200/DBSP   & 2022-01-12 & 1000  \\ 
ZTF19acoaiub & Keck/NIRES  & 2019-12-04 & 1000  & ZTF21acpkzcc & P200/TSpec  & 2022-01-20 & 1000  \\ 
ZTF19acpmbvd & LT/SPRAT    & 2019-12-23 & 350   & ZTF21aclzzex & Keck/LRIS   & 2022-01-26 & 750   \\ 
ZTF20aawdwch & NOT/ALFOSC  & 2020-05-15 & 300   & ZTF21aclyyfm & P200/DBSP   & 2022-02-02 & 1000  \\ 
ZTF20abwilhb & Keck/LRIS   & 2020-10-20 & 750   & ZTF21aancgbm & Keck/LRIS   & 2022-02-03 & 750   \\ 
ZTF21aagppzg & P200/DBSP   & 2021-02-08 & 1000  & ZTF21acpkzcc & Keck/LRIS   & 2022-02-03 & 750   \\ 
ZTF21aaekeqd & Keck/LRIS   & 2021-02-17 & 750   & ZTF21abtduah & Keck/LRIS   & 2022-02-03 & 750   \\ 
ZTF21aagppzg & P200/DBSP   & 2021-02-20 & 1000  & ZTF21aclzzex & Keck/LRIS   & 2022-03-01 & 750   \\ 
ZTF21aagydmn & P200/DBSP   & 2021-02-20 & 1000  & ZTF21abtduah & Keck/LRIS   & 2022-03-03 & 750   \\ 
ZTF21aagppzg & LT/SPRAT    & 2021-03-19 & 350   & ZTF21aclzzex & Keck/LRIS   & 2022-03-03 & 750   \\ 
ZTF21aancgbm & P200/DBSP   & 2021-04-09 & 1000  & ZTF21aagppzg & Keck/NIRES  & 2022-03-16 & 1000  \\ 
ZTF21aagppzg & P200/DBSP   & 2021-04-09 & 1000  & ZTF21aancgbm & Keck/NIRES  & 2022-03-17 & 1000  \\ 
ZTF21aaekeqd & Keck/LRIS   & 2021-04-14 & 750   & ZTF21aclzzex & Keck/NIRES  & 2022-03-17 & 1000  \\ 
ZTF21aaekeqd & Keck/NIRES  & 2021-04-18 & 1000  & ZTF21aclzzex & Keck/NIRES  & 2022-04-15 & 1000  \\ 

\hline
\hline
\end{tabular}

\end{center}
\end{table*}
\endgroup

%% file: table_phot_properties.tex
\begingroup
\renewcommand{\tabcolsep}{3pt}
\begin{table*}
\begin{center}
\begin{minipage}{15cm}
\caption{Photometric properties of LRN and ILRTs$^{a}$}
\label{tab:phot_properties}
\begin{tabular}{lcccccc}
\hline
\hline
{Name} & {DM$^{b}$} & {E(B--V)$^{c}$} & {m$_{\rm{r,peak}}$} &  {M$_{\rm{r,peak}}$} & MJD$_{\rm{peak}}$ & L$_{\rm{peak}}$\\ 
{} & {(mag)} & {(mag)} & {(mag)} &  {(mag)} & {(d)} & {(erg s$^{-1}$)}\\ 
\hline
ZTF18acbwfza & $31.60$ & 0.250 & $18.47 \pm 0.02$ & $-13.82 \pm 0.15$  & $58430.9 \pm 0.2$ & $9.3_{-0.3}^{+0.3} \times 10^{40}$   \\ 
ZTF19adakuot & $24.45$ & 0.055 & $15.00 \pm 0.01$ & $-9.60  \pm 0.15$  & $58891.6 \pm 0.1$ & $2.0_{-0.1}^{+0.1} \times 10^{39}$   \\ 
ZTF20aawdwch & $28.97$ & 0.093 & $17.79 \pm 0.02$ & $-11.43 \pm 0.15$  & $58963.0 \pm 1.2$ & $>1.2 \times 10^{40}$                \\
ZTF21aagppzg & $29.90$ & 0.020 & $16.45 \pm 0.01$ & $-13.50 \pm 0.15$  & $59258.9 \pm 0.1$ & $7.9_{-0.1}^{+0.1} \times 10^{40}$   \\
ZTF21aancgbm & $29.43$ & 0.261 & $16.28 \pm 0.03$ & $-13.86 \pm 0.15$  & $59252.9 \pm 0.5$ & $>8.9 \times 10^{40}$                \\
ZTF21acpkzcc & $32.09$ & 0.032 & $17.05 \pm 0.03$ & $-15.12 \pm 0.15$  & $59545.5 \pm 0.2$ & $3.3_{-0.2}^{+0.2}  \times 10^{41}$  \\
ZTF18abwxrhi & $33.70$ & 0.047 & $19.33 \pm 0.06$ & $-14.50 \pm 0.16$  & $58386.2 \pm 0.9$ & $1.9_{-0.3}^{+1.5}  \times 10^{41}$  \\
ZTF21aaekeqd & $32.87$ & 0.330 & $19.82 \pm 0.06$ & $-13.95 \pm 0.16$  & $59229.2 \pm 1.6$ & $6.0_{-4.6}^{+6.4}  \times 10^{41}$  \\
ZTF18aajgqmr & $35.00$ & 0.018 & $19.69 \pm 0.09$ & $-15.36 \pm 0.17$  & $58969.7 \pm 1.0$ & $3.3_{-0.4}^{+0.7}  \times 10^{41}$  \\
ZTF20abjgdec & $33.78$ & 0.141 & $19.80 \pm 0.06$ & $-14.37 \pm 0.16$  & $59045.6 \pm 2.1$ & $2.4_{-1.2}^{+13.3} \times 10^{41}$  \\
ZTF21aabfwwl & $33.99$ & 0.043 & $18.39 \pm 0.05$ & $-15.72 \pm 0.16$  & $59239.5 \pm 1.8$ & $5.2_{-0.7}^{+1.1}  \times 10^{42}$  \\
\hline
ZTF18acdyopn & $33.92$ & 0.064 & $19.76 \pm 0.11$ & $-14.33 \pm 0.19$  & $58426.8 \pm 2.4$ & $2.2_{-1.0}^{+21.7} \times 10^{41}$  \\
ZTF19aadyppr & $29.45$ & 0.705 & $16.65 \pm 0.01$ & $-14.73 \pm 0.15$  & $58534.7 \pm 0.5$ & $2.3_{-0.1}^{+0.1}  \times 10^{41}$  \\
ZTF19acoaiub & $31.83$ & 0.038 & $17.31 \pm 0.02$ & $-14.62 \pm 0.15$  & $58799.1 \pm 0.1$ & $2.0_{-0.2}^{+0.2}  \times 10^{41}$  \\
ZTF19acdrkbh & $34.33$ & 0.051 & $19.85 \pm 0.12$ & $-14.62 \pm 0.20$  & $58764.3 \pm 1.9$ & $1.8_{-0.4}^{+0.9}  \times 10^{41}$  \\
ZTF19aagqkrq & $30.76$ & 0.200 & $17.58 \pm 0.03$ & $-13.72 \pm 0.15$  & $58538.3 \pm 1.1$ & $8.0_{-0.4}^{+0.5}  \times 10^{40}$  \\ 
ZTF21aclzzex & $33.89$ & 0.019 & $18.26 \pm 0.06$ & $-15.68 \pm 0.16$  & $59528.6 \pm 1.7$ & $5.2_{-1.2}^{+3.7}  \times 10^{41}$  \\
ZTF21abtduah & $33.22$ & 0.022 & $< 18.31 $       & $<-14.96\pm 0.15$  & $<59522.5$        & $>1.3 \times 10^{41}$                \\
ZTF21abfxjld & $33.43$ & 0.106 & $< 19.51 $       & $<-14.21\pm 0.20$  & $<59359.8$        & $>8.0 \times 10^{40}$                \\
ZTF19aavwbxs & $33.26$ & 0.038 & $18.92 \pm 0.08$ & $-14.44 \pm 0.17$  & $58644.7 \pm 8.1$ & $2.2_{-0.3}^{+0.6}  \times 10^{41}$  \\
ZTF18acrygkg & $33.35$ & 0.036 & $< 19.31$.       & $<-14.14 \pm 0.16$ & $<58447.5$
&                                      \\
\hline
\hline
\end{tabular}
\begin{tablenotes} 
\item $a$ : The two classes LRN and ILRT are separated by a horizontal line. 
\item $b$ : The distance moduli have uncertainties of 0.15 mag and are taken from NED. 
\item $c$ : Galactic extinction values are taken from \citet{Schlafly11}; host extinction is calculated as described in the text.
\end{tablenotes}

\end{minipage}
\end{center}
\end{table*}
\endgroup

%% file: table_spec_properties.tex
\begingroup
\renewcommand{\tabcolsep}{1pt}
\begin{table}

\caption{Spectroscopic properties of LRN and ILRTs}
\label{tab:spec_properties}
\begin{tabular}{lccccc}
\hline
\hline
{Name} & {MJD} & {Phase} &  {v$_{\rm{H}\alpha, \rm{FWHM}}^{a}$} & {v$_{\rm{H}\beta, \rm{FWHM}}^{a}$}& {$\beta$}\\ 
{} & {(d)} & {(d)} &  {(km s$^{-1}$)} & {(km s$^{-1}$)} & {}\\
\hline
& & & LRNe & & \\
\hline
\hline
ZTF18acbwfza & 58424 & --7.1  & 510 $\pm$ 60 &               &               \\
ZTF18acbwfza & 58424 & --7.1  & 450 $\pm$ 30 & 750 $\pm$ 60  & 2.1 $\pm$ 0.1 \\
ZTF18acbwfza & 58432 & +0.9   & 350 $\pm$ 40 & 760 $\pm$ 120 & 2.2 $\pm$ 0.2 \\
\hline
ZTF18abwxrhi & 58379 & --9.3  & 280 $\pm$ 40 & 250 $\pm$ 120 & 3.8 $\pm$ 0.7 \\
\hline
ZTF21aaekeqd & 59262 & +33.4  & 650 $\pm$ 20 &               &               \\
ZTF21aaekeqd & 59318 & +89.3  & 340 $\pm$ 30 & 260 $\pm$ 70  & 2.1 $\pm$ 0.2 \\
ZTF21aaekeqd & 59343 & +113.8 & 300 $\pm$ 20 & $<$ 370       & 7.6 $\pm$ 0.6 \\
\hline
ZTF21aagppzg & 59253 & --5.9  & 490 $\pm$ 60 & 760 $\pm$ 100 & 2.5 $\pm$ 0.2 \\
ZTF21aagppzg & 59265 & +6.1   & 420 $\pm$ 40 & 700 $\pm$ 80  & 2.7 $\pm$ 0.2 \\
ZTF21aagppzg & 59313 & +54.1  & 500 $\pm$ 60 &               &               \\ 
\hline
ZTF21aancgbm & 59313 & +60.1  & $<350$       &               &               \\ 
ZTF21aancgbm & 59613 & +360.1 & 460 $\pm$ 40 &               &               \\ 
\hline
ZTF21acpkzcc & 59549 & +4.6   & 470 $\pm$ 50 & 540 $\pm$ 70  & 3.7 $\pm$ 0.3 \\
ZTF21acpkzcc & 59591 & +46.6  & $<$ 340      &               &               \\
ZTF21acpkzcc & 59613 & +68.6  & $<$ 300      &               &               \\
\hline
\hline
& & & ILRTs & & \\
\hline
\hline
ZTF18acdyopn & 58432 & +5.1   & 300 $\pm$ 40  & 630  $\pm$ 120 & 3.3 $\pm$ 0.4 \\
\hline
ZTF19aadyppr & 58509 & --25.7 & 330 $\pm$ 50  & $<$ 250        & 2.5 $\pm$ 0.3 \\
ZTF19aadyppr & 58526 & --8.7  & 640 $\pm$ 80  & 680  $\pm$ 180 & 2.4 $\pm$ 0.4 \\
ZTF19aadyppr & 58537 & +2.2   & 470 $\pm$ 170 & 2400 $\pm$ 850 & 0.9 $\pm$ 0.3 \\
ZTF19aadyppr & 58549 & +14.3  & 530 $\pm$ 120 &                &                \\
ZTF19aadyppr & 58558 & +23.2  & 710 $\pm$ 180 &                &                \\
ZTF19aadyppr & 58586 & +51.3  & 400 $\pm$ 70  &                &                \\
ZTF19aadyppr & 58616 & +81.3  & 330 $\pm$ 40  &                &                \\
\hline
ZTF19acoaiub & 58792 & --7.1  & 1250$\pm$ 300 & 2000 $\pm$ 850 & 3.4 $\pm$ 1.1 \\
\hline
ZTF19acdrkbh & 58783 & +18.7  & 590 $\pm$ 80  &                &               \\
\hline
ZTF19aagqkrq & 58525 & --13.4 & $<850$        & 1000 $\pm$ 200 & 2.3 $\pm$ 0.3 \\
ZTF19aagqkrq & 58526 & --12.4 & 670 $\pm$ 50  & 2000 $\pm$ 500 & 1.6 $\pm$ 0.2 \\
\hline
ZTF21aclzzex & 59530 & +1.9   & 670 $\pm$ 60  & 1200 $\pm$ 180 & 3.5 $\pm$ 0.5 \\
ZTF21aclzzex & 59605 & +76.9  & 500 $\pm$ 60  & 380  $\pm$ 110 & 5.5 $\pm$ 1.9 \\
ZTF21aclzzex & 59639 & +110.3 & 550 $\pm$ 40  & 300  $\pm$ 70  & 7.9 $\pm$ 0.9 \\
\hline
ZTF21abtduah & 59524 & +1.5   & 1170 $\pm$ 100 &                &               \\
ZTF21abtduah & 59549 & +26.5  & 1290 $\pm$ 100 &                &               \\
ZTF21abtduah & 59613 & +90.5  & 870 $\pm$ 30   & 690  $\pm$ 70  & 8.4 $\pm$ 0.6 \\
ZTF21abtduah & 59641 & +118.5 & 770 $\pm$ 30   & 780  $\pm$ 60  &10.6 $\pm$ 0.6 \\
\hline
ZTF21abfxjld & 59401 & +41.7  & $<$ 290       & $<$ 250        & 3.6 $\pm$ 0.5 \\
ZTF21abfxjld & 59440 & +80.2  & $<$ 270       & $<$ 260        & 4.6 $\pm$ 0.4 \\
\hline
\hline
\end{tabular}
\begin{tablenotes} 
\item The velocities have been corrected for instrumental resolutions listed in Table \ref{tab:speclog}. Upper limits are reported for unresolved lines.
\end{tablenotes}

\end{table}
\endgroup